\documentclass[prd,
              preprint,
              superscriptaddress,
              preprintnumbers,
              eqsecnum,
              showpacs,
              nofootinbib,
              nobibnotes,
              noeprint,
              nolongbibliography
              ]{revtex4-2}

\usepackage[linkcolor=blue,
            citecolor=blue,
            urlcolor=blue,
            colorlinks=true,
            breaklinks
            ]{hyperref}

\usepackage{booktabs}
\usepackage[italic]{hepnames}
\usepackage{amsfonts}
\usepackage{amsthm}
\usepackage{bm}
\usepackage{slashed}
\usepackage{dsfont}
\usepackage{tikz}
\usepackage[dvipsnames]{xcolor}
\usepackage[shortlabels]{enumitem}
\usepackage{xspace}
\usepackage{siunitx}
\usepackage{bookmark}
\usepackage{bbm}

\def\1eq#1{Eq.\nobreak\thinspace(\ref{#1})}

\def\2eqs#1#2{Eqs.\nobreak\thinspace(\ref{#1}) and\nobreak\thinspace(\ref{#2})}
\def\3eqs#1#2#3{Eqs.\nobreak\thinspace(\ref{#1}),\nobreak\thinspace(\ref{#2}) and\nobreak\thinspace(\ref{#3})}

\def\fig#1{\hyperref[#1]{Fig.\nobreak\thinspace\ref*{#1}}}
\def\figA#1{\hyperref[#1]{Fig.\nobreak\thinspace\ref*{#1}A}}
\def\figB#1{\hyperref[#1]{Fig.\nobreak\thinspace\ref*{#1}B}}

\def\tab#1{\hyperref[#1]{Tab.\nobreak\thinspace\ref*{#1}}}

\def\sect#1{\hyperref[#1]{Sec.\nobreak\thinspace\ref*{#1}}}
\def\appref#1{\hyperref[#1]{App.\nobreak\thinspace\ref*{#1}}}

\def\ie{{\it i.e.}, }
\def\eg{{\it e.g.}, }

\newcommand{\be}{\begin{equation}}
\newcommand{\ee}{\end{equation}}

\newcommand{\bea}{\begin{eqnarray}}
\newcommand{\eea}{\end{eqnarray}}

\def\is{S^{-1}}             

\def\g{\Gamma}              
\def\go{\Gamma_{\!1}}       
\def\ge{\Gamma_{\!2}}       
\def\got{\overline{\Gamma}_{\!1}} 
\def\get{\overline{\Gamma}_{\!2}} 

\def\ga{\Gamma_{\!5}}

\def\s#1{{\scriptscriptstyle #1}}

\def\gf{c_d}

\def\MOMt{$\widetilde{\text{MOM}}$}

\begin{document}

\title{\boldmath Pion physics with 
dressed quark-gluon vertices}

\author{M.N.~Ferreira}
\email{mnferreira@nju.edu.cn}
\affiliation{\mbox{School of Physics, Nanjing University, Nanjing, Jiangsu 210093, China
}}
\affiliation{{Institute for Nonperturbative Physics, Nanjing University, Nanjing, Jiangsu 210093, China
}}

\author{A.S.~Miramontes}
\email{angel.s.miramontes@uv.es}
\affiliation{\mbox{Department of Theoretical Physics and IFIC, University of Valencia and CSIC}, E-46100, Valencia, Spain}

\author{J.M.~Morgado}
\email{jose.m.morgado@uv.es}
\affiliation{\mbox{Department of Theoretical Physics and IFIC, University of Valencia and CSIC}, E-46100, Valencia, Spain}

\author{J.~Papavassiliou}
\email{joannis.papavassiliou@uv.es}
\affiliation{\mbox{Department of Theoretical Physics and IFIC, University of Valencia and CSIC}, E-46100, Valencia, Spain}

\author{J.M.~Pawlowski}
\email{j.pawlowski@thphys.uni-heidelberg.de}
\affiliation{\mbox{Institut f\"ur Theoretische Physik, Universit\"at Heidelberg}, Philosophenweg 16, Heidelberg, 69120, Germany}
\affiliation{\mbox{ExtreMe Matter Institute EMMI, GSI, Planckstrasse 1, Darmstadt, 64291, Germany}}

\begin{abstract}

Recently, a theoretical framework was set up in~\cite{Miramontes:2025imd}, which allows for the symmetry-preserving inclusion of full quark-gluon vertices in the 
description of the meson dynamics. In the present work, we develop a special truncation within this approach, which leads to a tractable set of functional equations that satisfy the fundamental chiral Ward-Takahashi identities. Specifically, the truncation allows us to simplify considerably the quark-gluon Schwinger-Dyson equation, without 
significant loss of quantitative accuracy. Importantly, this implies a substantial  
reduction of complexity of  
the renormalized Bethe-Salpeter 
equation: it is composed by a pair of one-loop diagrams 
that contain the full quark-gluon vertex, and 
a single two-loop diagram that is instrumental for 
the masslessness of the pion in the chiral limit. 
A detailed numerical analysis reveals that the incorporation of the aforementioned two-loop diagram is instrumental for the corresponding eigenvalue to reach unity. The key relation between the quark mass function and the pion wave function is shown to be satisfied to within the numerical precision of the loop integrals, which is at the level of about one percent or better.
The 
field-theoretic ingredients required for the 
extension of this 
analysis beyond the chiral limit are briefly discussed. 

\end{abstract}

\maketitle

\newpage 

\section{Introduction}\label{sec:Intro}

The self-consistent incorporation of QCD correlation functions into the 
dynamical equations that 
govern the physics of hadrons is particularly challenging, requiring the 
development of sophisticated truncation schemes, 
see, \eg  \cite{Munczek:1994zz, Matevosyan:2006bk, Fischer:2007ze, Fischer:2008wy, Fischer:2009jm,Chang:2009zb,Sanchis-Alepuz:2015tha,Williams:2014iea,Heupel:2014ina, Sanchis-Alepuz:2014wea, Williams:2015cvx, Sanchis-Alepuz:2015qra, Binosi:2016rxz, Williams:2018adr, Miramontes:2021xgn, Santowsky:2020pwd,Miramontes:2022mex, Gao:2024gdj, Miramontes:2025ofw, Fu:2025hcm, Huber:2025kwy,Miramontes:2025vzb}.
Recently, a theoretical framework was developed, 
which allows for the 
self-consistent inclusion of non-trivial quark-gluon vertices in the dynamical equations describing the physics of mesons~\cite{Miramontes:2025imd}. In particular, 
the set of Schwinger-Dyson equations (SDEs) satisfied by the quark propagator, $S$, the quark-gluon vertex, $\g^\mu$, and the (non-singlet) axial-vector vertex,
$\ga^{\mu}$, \cite{Alkofer:2000wg, Maris:2003vk, Fischer:2006ub, Binosi:2009qm, Maas:2011se, Bashir:2012fs,Cloet:2013jya, Eichmann:2016yit, Fischer:2018sdj, Huber:2018ned, Ferreira:2023fva}
are completely compatible 
with the constraints imposed 
by the chiral symmetry, 
in the form of Ward-Takahashi identities (WTIs).  

The analysis 
of~\cite{Miramontes:2025imd} 
singles out 
a practically unexplored correlation function~\cite{Bender:2002as, Chang:2009zb}, denominated ``gluon-axial-vector vertex'', and denoted by $G_5^{\mu\nu}$. 
The 
r\^ole of this vertex is 
instrumental, because 
its inclusion in the SDE 
of $\ga^{\mu}$
allows for a  symmetry-preserving
departure from the confines of the rainbow-ladder (RL) approximation \cite{Maris:1999nt, Maris:1999bh, Alkofer:2002bp, Eichmann:2008ae, Qin:2011dd, Hilger:2014nma, Heupel:2012ua, Eichmann:2015cra, Hilger:2015hka, El-Bennich:2016qmb, Mojica:2017tvh, Raya:2017ggu, Weil:2017knt, Serna:2017nlr, Gutierrez-Guerrero:2021rsx,Hernandez-Pinto:2023yin, Hernandez-Pinto:2024kwg,Chen:2019otg,Chang:2020iut,Xu:2024fun, Xu:2025hjf,Xu:2024vkn,Albino:2025bnr}. Note that the 
symmetry-restoring action of 
$G_5^{\mu\nu}$ hinges on the 
WTI that connects it to the quark-gluon vertex $\g^\mu$. 
However, as illustrated in~\cite{Miramontes:2025imd},   
the dynamical equation that determines $G_5^{\mu\nu}$
has a rather complicated structure, 
being composed 
by graphs that 
contain $G_5^{\mu\nu}$ explicitly
(``$G$-{\it dependent}\,''), 
and those containing 
$\ga^{\mu}$
instead 
(``$G$-{\it independent}\,'').
Therefore, 
a truncation that 
reduces the complexity of this equation without 
compromising its symmetry properties  
would be particularly useful, allowing for a preliminary 
numerical exploration of this 
novel approach. 

In the present work we show that such a truncation is indeed feasible,
as long as the corresponding 
quark-gluon vertices are also appropriately modified.
In particular, one may 
drop the $G$-dependent diagrams 
entirely, by setting $G_5^{\mu\nu}=0$
inside them, such that the 
gluon-axial-vector vertex is exclusively 
computed 
from the $G$-independent graphs.
The symmetry remains intact
(\ie the key WTIs are still satisfied) 
provided that, at the same time, a simplified version of the SDE that governs the quark-gluon vertex is used~\cite{Alkofer:2008tt,Williams:2015cvx,Aguilar:2024ciu}. In particular, denoting by 
$q$ the momentum of the gluon,   
one sets 
\mbox{$\g^\mu(q,r,-p)= V(q) 
\gamma^\mu$}
\textit{inside} the relevant diagrams, where the function 
$V(q)$ corresponds to the 
classical form factor of the quark-gluon vertex 
in the so-called 
``symmetric'' kinematic configuration
defined as \mbox{$q^2=r^2=p^2$}. 

We stress that, although the aforementioned SDE takes $V(q)$ as input, it returns a quark-gluon vertex that displays the full kinematic content. It possesses eight transverse form factors, and in the Landau gauge the longitudinal ones do not take part in the dynamics. These form factors depend non-trivially on three kinematic variables. We also note, that a variant of this approximation has been used in the detailed study of the quark-gluon vertex presented in~\cite{Gao:2021wun}.

Within this truncation, the 
dynamical equation for 
$G_5^{\mu\nu}$ is drastically 
simplified, and we readily obtain the Bethe-Salpeter equation (BSE) that controls the formation of the bound states (pions) \cite{Salpeter:1951sz,PhysRev.84.350,Bethe1957,Nakanishi:1969ph,Jain:1993qh,Munczek:1994zz}. This BSE consists of four diagrams, 
one corresponding 
to the standard RL graph, but now with \textit{dressed vertices}, and three
additional graphs that originate 
precisely from the vertex $G_5^{\mu\nu}$. Once the multiplicative renormalization
has been carried out, see~\cite{Fischer:2003rp,Aguilar:2010cn,Aguilar:2018epe}, 
the final BSE is 
described by three diagrams,
two that are ``one-loop dressed''
and one that is ``two-loop dressed''. The full quark-gluon vertex enters both in the 
quark gap equation 
and the 
one-loop dressed diagrams of the BSE.

Importantly, the numerical solution of the BSE for massless pions in the chiral limit reveals that the dominant term of the pion amplitude satisfies the axial WTI \cite{Miransky:1994vk,Maris:1997hd} within the numerical accuracy of the loop integrals, see \fig{fig:chi_2B}. 

The article is organized as follows. In \sect{sec:summ} 
we review the theoretical framework, introducing the key quantities, together with the most salient relations. Next, in \sect{sec:spt}, we discuss in detail the symmetry-preserving truncation of the dynamical equation 
satisfied by $G_5^{\mu\nu}$.
In \sect{sec:bse} the 
truncation is employed 
to obtain the BSE that controls 
the formation of pions.
In \sect{sec:ren} the 
multiplicative renormalization 
of the SDE-BSE system is 
carried out.
Then, in \sect{num} we solve the system to obtain the Bethe-Salpeter amplitude (BSA) of the massless pion. 
In \sect{sec:Disc} we present our discussion and conclusions. 
Finally, in \appref{app:DiagrammaticProof} we offer a diagrammatic demonstration of the WTI satisfied by the vertex 
$G_5^{\mu\nu}$ within the truncation employed,
while in \appref{app:identity} we discuss an interesting integral identity connecting the chirally symmetric and chiral symmetry breaking terms of the quark-gluon vertex.

\section{Review of the theoretical framework}\label{sec:summ}

In this section we present a 
brief account of the
main ingredients, 
key relations, and general notation  
employed 
in the present work. 

We start by pointing out that 
the calculations will be carried out in 
Minkowski space, and the final results will be passed to Euclidean space, in order to implement the numerical treatment. The main elements of our analysis may be summarized as follows.\\[-2ex]

(${\it i}$) 
The central component of this approach is 
the axial-vector vertex, $\ga^{a \mu} 
=t^a\ga^{\mu}$,
where $t^a$  
are the generators of the flavour $SU(N_f)$ algebra;
for $N_f=2$, $t^a = \sigma^a/2$, 
while for $N_f=3$, 
$t^a = \lambda^a/2$,
 where $\sigma^a$ 
and $\lambda^a$
are the Pauli and Gell-Mann matrices, respectively. 
Note that
this vertex is associated with the flavour non-singlet current 
$j^{a\mu}_5(x)=\bar{\psi}(x)\gamma_5 \gamma^\mu t^a{\psi}(x)$, 
 see discussion in Appendix~B of~\cite{Miramontes:2025imd}.
 
In the limit of vanishing current quark masses ($m=0$), the vertex $\ga^{\mu}$
satisfies the well-known WTI \cite{Itzykson:1980rh,Miransky:1994vk}
\be
\label{eq:wtig5}
-P_\mu\ga^\mu(P,p_2,-p_1)=\is(p_1)\gamma_5+\gamma_5\is(p_2)\,,
\ee
where  $\is(p)$ is the inverse quark propagator.\\[-2ex]

(${\it ii}$)
According to the usual decomposition, 
\be\label{eq:invS}
S^{-1}(p)=A(p)\slashed{p}-B(p)\,,
\ee
where $A(p)$ and $B(p)$ are the dressings of the (Dirac) vector and scalar structures,  respectively, and 
\mbox{${\mathcal M}(p) = B(p)/ A(p)$} 
is the constituent quark mass.  

\begin{figure}[!t]
    \hspace*{-1cm}
    \includegraphics[scale=1.25]{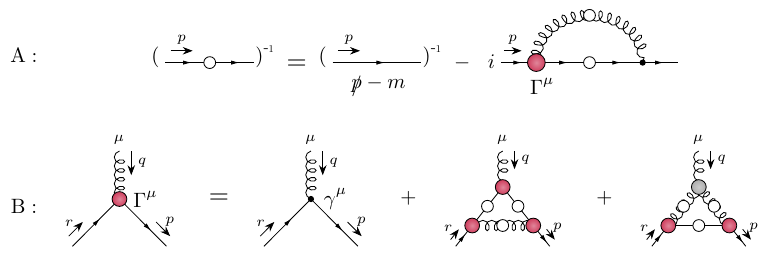}
    \caption{\textbf{\textit{Panel A:}} Diagrammatic representation of the SDE 
    (gap equation) that determines the quark propagator $S(p)$. White 
    circles denote full propagators, 
    red circles stand for full quark-gluon vertices, while  
    small black circles indicate tree-level quark-gluon vertices. 
    \textbf{\textit{Panel B:}} 
    The SDE of the quark-gluon vertex (red circle) obtained within the 3PI formalism. The gray circle denotes the fully-dressed three-gluon vertex.}
    \label{fig:summ}
\end{figure}

The momentum evolution of $A(p)$ and $B(p)$ is controlled by the gap equation, shown in \figA{fig:summ}, 
given by 
\begin{align}
\label{eq:GapEq}
\is(p)= \slashed{p} - m 
+ ig^2C_f\int_q\gamma^\nu S(q)\g^\mu(q-p,p,-q)\Delta_{\mu\nu}(q-p) \,,
\end{align}
which involves the fully-dressed quark-gluon vertex, denoted 
by $\g^\mu(q,r,-p)$, as its main ingredient. 
In \1eq{eq:GapEq}, $g$ is 
the QCD gauge coupling, $C_f=4/3$ the Casimir eigenvalue of the fundamental $SU(3)$ representation, and $\Delta_{\mu\nu}$ stands for the full gluon propagator in the 
Landau gauge, 
\begin{align}
\Delta_{\mu\nu}(q) =  P_{\mu\nu}(q) \Delta(q) 
\,, \qquad\qquad P_{\mu\nu}(q)= g_{\mu\nu} - 
\frac{q_{\mu}q_{\nu}}{q^2} \,.
\label{eq:gluoprop}
\end{align}
The term $m$ denotes the 
current quark mass; throughout this analysis it will be considered to be 
vanishing, $m=0$ (chiral limit). 
Moreover, 
\begin{align}
\int_q :=  \int_{\mathbbm{R}^4}\frac{{\rm d}^4 q}{(2\pi)^4} \,,
\end{align}
where the use of a symmetry-preserving regularization scheme is implicitly assumed.

After taking appropriate traces, 
the quark gap equation in \1eq{eq:GapEq} is
reduced to a set of coupled nonlinear integral equations that determine the dressing functions $A(p)$ and $B(p)$.
In particular, we have in Minkowski space 
\begin{subequations} 
\label{eq:QuarkGapEq}
\begin{align} \nonumber 
    B(p) =&\,- \displaystyle\frac{ig^2 C_f}{4}\int_q~a(q)\textrm{Tr}\left[\gamma^\nu\slashed{q}\ge^\mu(q-p,p,-q)\right]\Delta_{\mu\nu}(q-p)\\[1ex]\nonumber 
    &\,- \frac{ig^2 C_f}{4}\int_q~ b(q)\textrm{Tr}\left[\gamma^\nu\go^\mu(q-p,p,-q)\right]\Delta_{\mu\nu}(q-p)\,,\\[1ex]\nonumber 
    p^2A(p) =&\,  p^2+ \frac{ig^2 C_f}{4}\int_q~a(q)\textrm{Tr}\left[\slashed{p}\gamma^\nu\slashed{q}\go^\mu(q-p,p,-q)\right]\Delta_{\mu\nu}(q-p)\\[1ex]
    & \hspace{.5cm}+\frac{ig^2 C_f}{4}\int_q~ b(q)\textrm{Tr}\left[\slashed{p}\gamma^\nu\ge^\mu(q-p,p,-q)\right]\Delta_{\mu\nu}(q-p)\,,
\end{align}
with
\begin{align} 
a(p) :=c(p)A(p) \,,\qquad
b(p) := c(p)B(p) \,,\qquad
c(p):= \frac{1}{A^2(p)p^2-B^2(p)}\,.
\label{eq:bcquark}
\end{align}
\end{subequations}
In \1eq{eq:QuarkGapEq} we have separated 
the quark-gluon vertex into the ``odd'' ($\go$)
and ``even'' ($\ge$) components,
\begin{align}
\g^\mu(q,r,-p)=
\underbrace{\go^\mu(q,r,-p)}_{{\rm odd\,\#\,of\,\gamma}} \,\,+ \,\, \underbrace{\ge^\mu(q,r,-p)}_{{\rm even\,\#\,of\,\gamma}}  \,.
\label{eq:QgOddEven}
\end{align}

(${\it iii}$)
In the Landau gauge, it is natural to consider the ``transversely 
projected'' quark-gluon vertex, 
$\overline{\Gamma}_\mu(q,r,-p)= P_{\mu\nu}(q)\Gamma^{\nu}(q,r,-p)$. 
In general kinematics, $\overline{\Gamma}_\mu(q,r,-p)$ can be spanned by eight independent tensors, $\bar\tau_{i}$, 
namely  
\begin{align} 
\label{decomp}
    \overline{\Gamma}^\mu(q,r,-p)=\sum_{i=1}^{8}\lambda_i(q,r,-p)\bar\tau_{i}^\mu(r,-p) \,, 
    \qquad\qquad  
 \bar\tau_{i}^\mu(r,-p)=  P^\mu_{\nu}(q)\tau_{i}^\nu(r,-p) \,,
\end{align}
where the $\tau^\nu_{i}$ in Minkowski space are given by~\cite{Mitter:2014wpa, Cyrol:2017ewj, Gao:2021wun, Ihssen:2024miv} 
\begin{align}
\label{Taus}
    &\tau^\nu_{1}(r,-p) =\gamma^\nu\,, \quad &&\tau^\nu_{2}(r,-p) =  (p+r)^\nu\,, \nonumber \\
    &\tau^\nu_{3}(r,-p) = (\slashed{p}+\slashed{r})\gamma^\nu\,, \quad &&\tau^\nu_{4}(r,-p) = (\slashed{p}-\slashed{r})\gamma^\nu\,,\nonumber\\
    &\tau^\nu_{5}(r,-p) =  (\slashed{p}-\slashed{r})(p+r)^\nu\,, \quad &&\tau^\nu_{6}(r,-p) =(\slashed{p}+\slashed{r})(p+r)^\nu\,,\nonumber\\
    &\tau^\nu_{7}(r,-p) = -\frac{1}{2}[\slashed{p},\slashed{r}]\gamma^\nu\,, \quad &&\tau^\nu_{8}(r,-p) = -\frac{1}{2}[\slashed{p},\slashed{r}](p+r)^\nu \,,
\end{align}
and the $\lambda_i(q,r,-p)$ denote form factors that depend on three Lorentz scalars.
Within this basis, the component $\got^\nu$ is spanned by the elements 
$\{\bar\tau^\nu_{1}, \bar\tau^\nu_{5},
\bar\tau^\nu_{6},\bar\tau^\nu_{7}\}$, while  
$\get^\nu$ by   
$\{\bar\tau^\nu_{2}, \bar\tau^\nu_{3},
\bar\tau^\nu_{4},\bar\tau^\nu_{8}\}$. 

Within the three-particle-irreducible (3PI) scheme~\cite{Alkofer:2008tt,Williams:2015cvx,Aguilar:2024ciu},
the one-loop dressed SDE that controls the 
evolution of the form factors 
$\lambda_i(q,r,-p)$ is 
shown diagrammatically in 
\figB{fig:summ}. 

\begin{figure}[!t]
    \hspace*{-1cm}
    \includegraphics[scale=1]{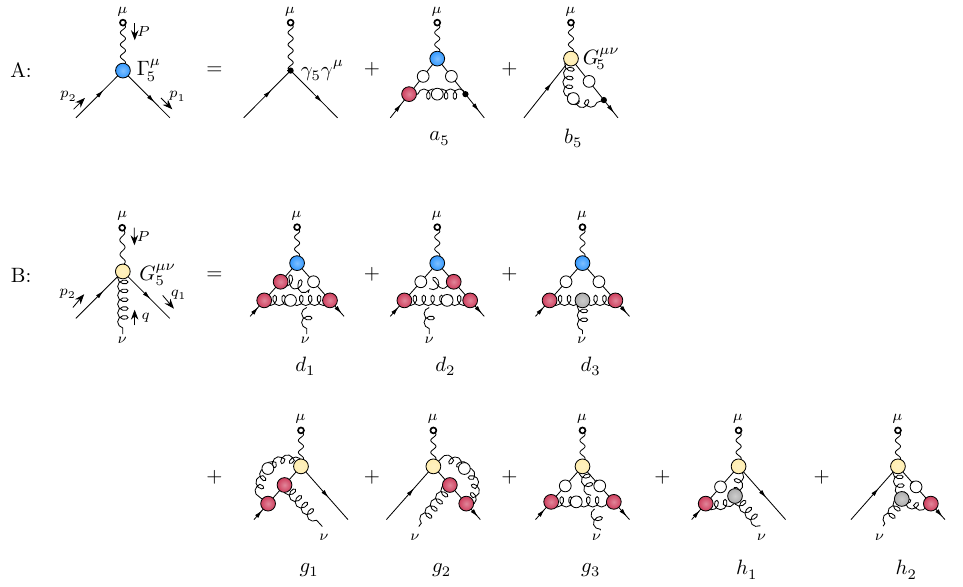}
    \caption{\textbf{\textit{Panel A:}} The SDE of the axial-vector vertex, $\ga^\mu$ (blue circles). The graph ($b_5$) contains the ``gluon-axial-vector'' vertex, $G_5^{ab \mu\nu}$ (yellow circle), which is crucial for the preservation of the axial WTI.
    \textbf{\textit{Panel B:}} The one-loop dressed representation of the gluon-axial-vector vertex; we refer to the graphs $d_1,~d_2,~d_3$ as $G$-{\it independent}, while to the  $g_1,~g_2,~g_3,~h_1,~h_2$ as $G$-{\it dependent.}}
    \label{fig:1ld}
\end{figure}

(${\it iii}$) Instrumental in this entire analysis is 
the SDE satisfied by the vertex $\ga^\mu$, 
shown in \figA{fig:1ld}. 
Note that the diagram ($a_5$) 
contains the fully-dressed quark-gluon vertex $\Gamma_{\nu}$, depicted as 
the red circle; 
this graph 
provides the standard RL description, after the substitution $\Gamma_{\nu} \to \gamma_{\nu}$.
A key component of this 
SDE is the ``gluon-axial-vector'' vertex \cite{Bender:2002as, Chang:2009zb, Miramontes:2025imd}, \mbox{$G_5^{ab \mu\nu} = ig 
t^a \frac{\lambda^b}{2} G_5^{\mu\nu}$}, 
represented 
by the yellow circle in diagram ($b_5$) of \figA{fig:1ld}.  Crucially, $G_5^{\mu\nu}$
is related to the quark-gluon vertex by the WTI \cite{Chang:2009zb, Miramontes:2025imd}
\be
\label{eq:wtiG5}
-iP_\mu G_5^{\mu\nu}(P,q,p_2,-q_1)=\g^\nu(q,p_1,-q_1)\gamma_5+\gamma_5\g^\nu(q,p_2,-q_2)\,,
\ee
where $q_i:=p_i+q$.
By virtue of \1eq{eq:wtiG5}, 
when the 
diagrams contributing to the SDE 
of $\ga^\mu$ are contracted by
$P_\mu$, the axial WTI of \1eq{eq:wtig5} emerges precisely,
with the inverse quark propagators 
satisfying the gap equation 
with the {\it full} quark-gluon vertex. 

(${\it iv}$) When the chiral symmetry is dynamically broken, a nonvanishing 
mass function $B(p)$ 
is generated as a solution of  
\1eq{eq:GapEq}. 
As a result, in the 
limit $P \to 0$, $p_1=p_2 :=p$, the WTI of \1eq{eq:wtig5} 
yields 
\be
\label{eq:WTIP0-I}
\lim_{P\to 0}P_\mu\ga^\mu(P,p_2,-p_1)=
2 B(p) \gamma_5\,.
\ee
This result 
forces $\ga^\mu$ 
to contain a pole term, 
$\ga^\mu(P,p_2,-p_1)|_{\textrm{pole}}$,
associated with the attendant Nambu-Goldstone boson (pion), of the form 
\be\label{eq:g5Pole}
\left.\ga^\mu(P,p_2,-p_1)\right|_{\textrm{pole}}=\frac{P^\mu}{P^2} \,\chi(P,p_2,-p_1)\gamma_5 \,.
\ee
The quantity $\chi(P,p_2,-p_1)$ is the BSA of the pion, which, in the limit $P \to 0$, is composed by two form factors \cite{Miransky:1994vk},
\begin{align}
\chi(0,p,-p) = \chi_1(p)
+ \chi_3(p)\slashed{p} \,.
\label{chi0pp}
\end{align}
The comparison of 
\2eqs{eq:g5Pole}{chi0pp}
furnishes the well-known symmetry-induced relations 
\begin{align}\label{eq:chi12}
\chi_1(p)&=2B(p)\,,& \chi_3(p)&=0\,.
\end{align}
The first relation is particularly 
powerful, 
connecting the dominant component of the pion amplitude to the quark mass function. 

(${\it v}$)
In the limit $P \to 0$,
the SDE that governs the vertex $\ga^\mu$ furnishes the BSEs 
for both 
$\chi_1(p)$ and $\chi_3(p)$.
Quite importantly, as was explicitly shown in~\cite{Miramontes:2025imd},
these BSEs admit the 
relations 
given by \1eq{eq:chi12}
as their exact solutions.
In that sense, symmetry and dynamics are harmoniously intertwined.
Crucial in the demonstration of this key property
is the WTI of \1eq{eq:wtiG5}, which, in the limit $P \to 0$,
becomes
\be
\lim_{P\to 0} P_\mu G_5^{\mu\nu}(P,q,p_2,-q_1)=2 i\ge^\nu(q,p,-p-q)\gamma_5\,.
\label{eq:G5poleG2}
\ee

(${\it vi}$)
The one-loop dressed approximation
of $G_5^{\mu\nu}$
is given by the diagrams 
shown in \figB{fig:1ld}. It turns out 
that the contraction of this 
set of diagrams 
by $P_{\mu}$
generates the r.h.s. of 
the WTI in \1eq{eq:wtiG5}, where the corresponding 
quark-gluon vertices 
$\g^\nu(q,p_1,-q_1)$ and 
$\g^\nu(q,p_2,-q_2)$
satisfy the SDE shown 
in 
\figB{fig:summ}~\cite{Miramontes:2025imd}.

\section{Symmetry-preserving truncation}\label{sec:spt}

There is a practical difficulty associated with the 
truncation described 
in item (${\it vi}$) of the previous section. Specifically, 
a subset of the diagrams 
that define the vertex $G_5^{\mu\nu}$ in \figB{fig:1ld}
contain the $G_5^{\mu\nu}$ itself as their ingredient 
(second row, $G$-dependent graphs), thus  
converting 
the dynamical equations for $\ga^\mu$ and $G_5^{\mu\nu}$ into a coupled system.
Even though such a system is 
in principle tractable, it is certainly useful to introduce an 
operationally simpler approach, which does not compromise the validity of the essential WTIs. In particular, as we 
demonstrate in this section, 
one may omit the 
aforementioned subset of graphs entirely,  at the 
expense of introducing a compensating adjustment at the level of 
the quark-gluon SDE in \figB{fig:summ}.

The starting point of our truncation is to set 
$G_5^{\mu\nu} =0$ on the r.h.s. of the equation shown in \figB{fig:1ld},
thus eliminating the $G$-dependent graphs.
After doing so, the equation that furnishes 
$G_5^{\mu\nu}$ reduces to the three 
diagrams in the first row of \figB{fig:1ld} ($G$-independent graphs). However, these remaining diagrams no longer reproduce 
the WTI of \1eq{eq:wtiG5}, with the 
full quark-gluon vertices (red circles) satisfying the SDE 
shown \figB{fig:summ}.

\begin{figure}[!t]
    \centering
    \hspace*{-1.5cm}\includegraphics[scale=1]{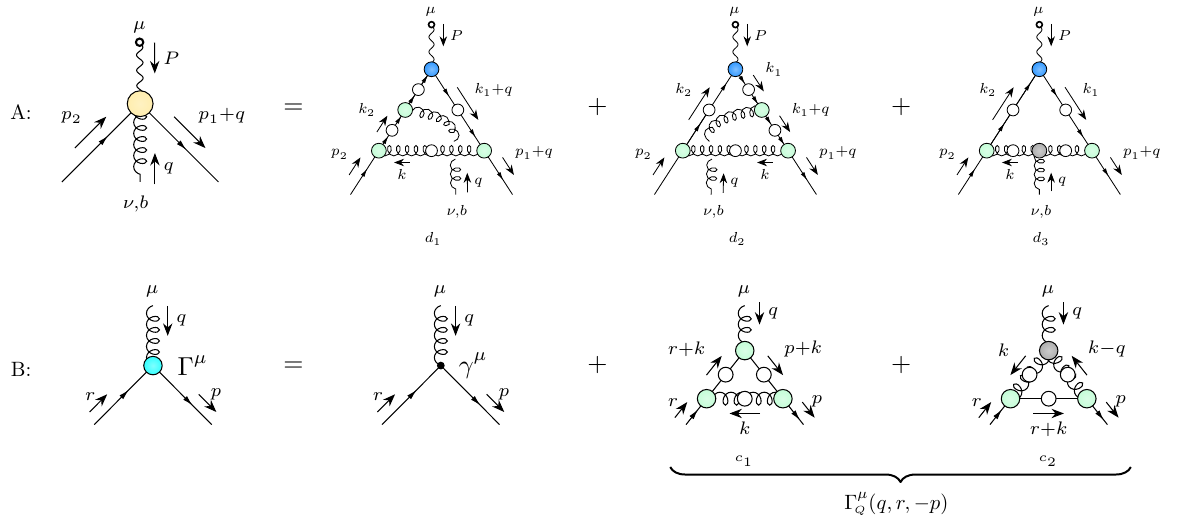}
    \caption{\textbf{\textit{Panel A:}} 
    The truncated version of the SDE 
    for the vertex $G_5^{\mu\nu}$, 
    composed only by $G$-independent graphs. The green circles denote the 
    component $V_{\mu}$, defined in \1eq{eq:qgsym}.
    \textbf{\textit{Panel B:}} The SDE 
    of the quark-gluon vertex (cyan circle), which is compatible with the truncation 
    of the $G_5^{\mu\nu}$ shown in the panel above.}
    \label{fig:1ld_green}
\end{figure}

Nonetheless, it turns out that a simplification implemented on  
the $\g^\mu$ appearing {\it inside}  
all one-loop dressed diagrams
in \figB{fig:summ} and \figB{fig:1ld}
restores the key WTI. 
Specifically, upon inspection, one recognizes that the global replacement 
\be\label{eq:qgsym}
 \g_\mu(q,r,-p) \to V_\mu(q)\,, \qquad \qquad    V_\mu(q):=\gamma_\mu V(q)=\hspace{-4.5cm}\raisebox{-1.25cm}{\includegraphics[scale=1]{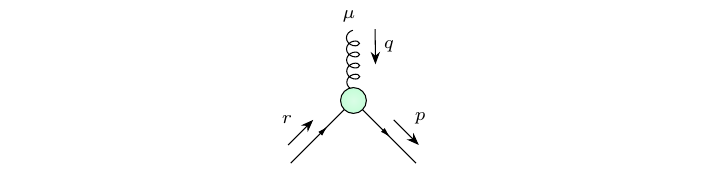}}\hspace{-4.5cm}\,,
\ee
gives rise to a symmetry-preserving 
approximation, defined by 
the system of equations depicted 
diagrammatically in the 
two panels of \fig{fig:1ld_green}.

As will become evident from the demonstration leading to \1eq{qWTI}, the WTI is preserved regardless of the precise form chosen for $V(q)$;
the only formal requirement is that 
$V(q)$ be function of 
a single kinematic variable, 
namely the momentum carried by the gluon.
This need arises because, in going from 
to \1eq{qWTI} to \1eq{eq:Pd1d2}, 
certain key cancellations take place 
among various terms upon shifting appropriately the 
integration variable;
these cancellations go through provided that the $V(q)$ has the aforementioned momentum dependence.
The actual expression used for 
$V(q)$ in the numerical analysis will 
be discussed in \sect{qgnum}.

We emphasize 
that even though the $V_\mu(q)$ 
used as input in the r.h.s. of the SDE 
in \figB{fig:1ld_green} 
contains only the classical tensor, 
the output obtained,  
represented by the cyan circle, 
displays the {\it full} kinematic 
structure associated with a quark-gluon vertex. In particular, 
in the Landau gauge, the resulting vertex $\g^\mu(q,r,-p)$
is composed by eight tensorial structures, see \1eq{decomp}, 
multiplied by the corresponding 
nonvanishing form factors, $\lambda_i$, which depend on three kinematic variables, \eg $q^2$, $r^2$ and $q\cdot r$. Note that this full vertex will enter in the 
graph ($a_5$) of the SDE for the axial-vector vertex, thus effectuating the 
departure from the RL approximation. 
 
Given the above discussion, the closed form of the 
dynamical equation displayed in 
\figB{fig:1ld_green} is given by 
\be\label{eq:qgsde}
\g^\mu(q,r,-p) =
\gamma^\mu+c_1^\mu+c_2^\mu\,,
\ee
with
\bea\label{eq:c1c2}
c_1^\mu & = & c_a\int_k V^\beta(k)S(p+k)V^\mu(q)S(r+k)V^\alpha(k)\Delta_{\alpha\beta}(k)\label{eq:c1}\,,\\
\nonumber\\
c_2^\mu & = & c_b\int_k V^\beta(k-q)S(r+k)V^\alpha(k)\Delta_{\rho\beta}(k-q)\Delta_{\alpha\delta}(k)\g^{\mu\delta\rho}(q,-k,k-q)\label{eq:c2}\,,
\eea
where $\g^{\mu\delta\rho}$ is the full three-gluon vertex, $c_a=-ig^2(C_f-C_A/2)$, $c_b=-ig^2C_A/2$, and $C_A$ is the Casimir eigenvalue of the adjoint representation [$N$ for $SU(N)$].

It is convenient for what follows 
to introduce the ``quantum'' part, $\g^\mu_{\!\scriptscriptstyle{Q}}(q,r,-p)$, of the 
quark-gluon vertex, defined 
and diagrammatically represented as 
\be\label{eq:qgquant}
 \g^\mu_{\!\s{Q}}(q,r,-p) = c_1^\mu+c_2^\mu
 \,, \qquad \qquad    \g^\mu_{\!\s{Q}}(q,r,-p) := \hspace{-0.75cm}
 \raisebox{-1.5cm}{\includegraphics[scale=1.2]{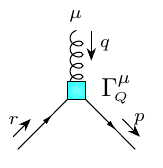}}\,.
\ee

Having set the stage, we next 
derive the WTI satisfied by 
the vertex $G_5^{\mu\nu}$, 
given by the diagrams 
$d_1$, $d_2$ and $d_3$
in \figA{fig:1ld_green}, 
\be\label{eq:G5}
G_5^{\mu\nu}(P,q,p_2,-q_1)=d_1^{\mu\nu}+d_2^{\mu\nu}+d_3^{\mu\nu}\,.
\ee
Contracting the first two graphs 
by $P_\mu$, and  employing
\1eq{eq:wtig5} under the integral sign, we get
\begin{align}\label{eq:Pd1d2}
    P_\mu d_1^{\mu\nu} & = -L_1^\nu-L_3^\nu, &P_\mu d_2^{\mu\nu} & = -L_2^\nu-L_4^\nu\,,
\end{align}
with
\bea
    L^\nu_1 & = & g^2\kappa_a \int_k V^\beta(k)\gamma_5 S(p_2+k+q)V^\nu(q) S(p_2+k)V^\alpha(k)\Delta_{\alpha\beta}(k)\,,\nonumber\\
    \nonumber\\
    L^\nu_2 & = & g^2\kappa_a\int_k V^\beta(k) S(p_1+k+q)V^\nu(q) S(p_1+k)\gamma_5 V^\alpha(k)\Delta_{\alpha\beta}(k)\,,\nonumber\\
    \nonumber\\
    L^\nu_3 & = & g^2\kappa_a\int_k V^\beta(k) S(p_1+k+q)\gamma_5 V^\nu(q) S(p_2+k) V^\alpha(k)\Delta_{\alpha\beta}(k)\,,\nonumber\\
    \nonumber\\
    L^\nu_4 & = & g^2\kappa_a \int_k V^\beta(k) S(p_1+k+q) V^\nu(q)\gamma_5 S(p_2+k) V^\alpha(k)\Delta_{\alpha\beta}(k)\,,
\eea
where $\kappa_a=C_f-C_A/2$. Clearly, since $V^\nu(q)\gamma_5+\gamma_5 V^\nu(q)=0$, the contributions $L_3^\nu$ and $L_4^\nu$ cancel upon addition, giving
\bea
-P_\mu\left(d_1^{\mu\nu}+d_2^{\mu\nu}\right) & = & g^2\kappa_a \int_k V^\beta(k)\gamma_5 S(p_2+k+q)V^\nu(q) S(p_2+k)V^\alpha(k)\Delta_{\alpha\beta}(k)\nonumber\\
\nonumber\\
&+ & g^2\kappa_a\int_k V^\beta(k) S(p_1+k+q)V^\nu(q) S(p_1+k)\gamma_5 V^\alpha(k)\Delta_{\alpha\beta}(k)\,,
\eea
or, in terms of the 
Abelian diagram $(c_1^\nu)$ 
shown in \figB{fig:1ld_green}
and given by \1eq{eq:c1}, 
\be\label{eq:Pd12graph}
-iP_\mu\left(d_1^{\mu\nu}+d_2^{\mu\nu}\right) = c_1^\nu(q,p_1,-q_1)\gamma_5+\gamma_5 c_1^\nu(q,p_2,-q_2) \,.
\ee

Similarly, for $d_3^{\mu\nu}$ 
we obtain 
\bea
-P_\mu d_3^{\mu\nu} & = & ig^2\kappa_b\int_k V^\beta(k') S(p_1+k)V^\alpha(k)\g^{\nu\rho\delta}(q,-k,k')\Delta_{\alpha\rho}(k)\Delta_{\delta\beta}(k')\gamma_5\nonumber\\
\nonumber\\
& +& ig^2\kappa_b\int_k\gamma_5 V^\beta(k') S(p_2+k)V^\alpha(k) \g^{\nu\rho\delta}(q,-k,k')\Delta_{\alpha\rho}(k)\Delta_{\delta\beta}(k')\,,
\label{Pd3}
\eea
where $\kappa_b=iC_A/2$ and $k'=k-q$. As before, 
we may express the r.h.s. 
of \1eq{Pd3}
in terms of the 
non-Abelian diagram $(c_2^\nu)$ 
in \figB{fig:1ld_green},
given by \1eq{eq:c2}, namely 
\be\label{eq:Pd3graph}
-iP_\mu d_3^{\mu\nu} = c_2^\nu(q,p_1,-q_1)\gamma_5+\gamma_5 c_2^\nu(q,p_2,-q_2) \,.
\ee

Adding up the results of \2eqs{eq:Pd12graph}{eq:Pd3graph}, and using the definition of $\g^\mu_{\!\scriptscriptstyle{Q}}$
in \1eq{eq:qgquant}, 
we find
\be
-iP_\mu G_5^{\mu\nu}(P,q,p_2,-q_1)=\g^\mu_{\!\scriptscriptstyle{Q}}(q,p_1,-q_1)\gamma_5+\gamma_5\g^\mu_{\!\scriptscriptstyle{Q}}(q,p_2,-q_2)\,.
\label{qWTI}
\ee
Since $\gamma^\mu\gamma_5+\gamma_5\gamma^\mu=0$, the tree-level terms
of the two quark-gluon vertices may be added for free,  leading  precisely to 
the WTI of \1eq{eq:wtiG5}, with 
the quark-gluon 
vertex defined through the SDE in \1eq{eq:qgsde}.

Given that the vertex $G_5^{\mu\nu}$ satisfies \1eq{qWTI},
the resulting $\ga^\mu(P,p_2,-p_1)$ 
fulfils the axial WTI of  
\1eq{eq:wtig5}, where the quark propagator is given by the gap equation in \figA{fig:summ}. Of course, the 
fully-dressed quark-gluon vertex 
of the gap equation 
is now given by \figB{fig:1ld_green},
\ie the red circle in \figA{fig:summ}
must be replaced by the cyan one.  

\section{Pion BSE with dressed quark-gluon vertices }\label{sec:bse}

In this section we derive the 
BSE of the pion 
by taking the limit $P\to 0$ 
of the SDE for $\ga^\mu(P,p_2,-p_1)$,
and then 
equating the residues of the pole parts appearing on both sides. 

\begin{figure}[!t]
    \hspace*{-1.25cm}
    \includegraphics[scale=1.2]{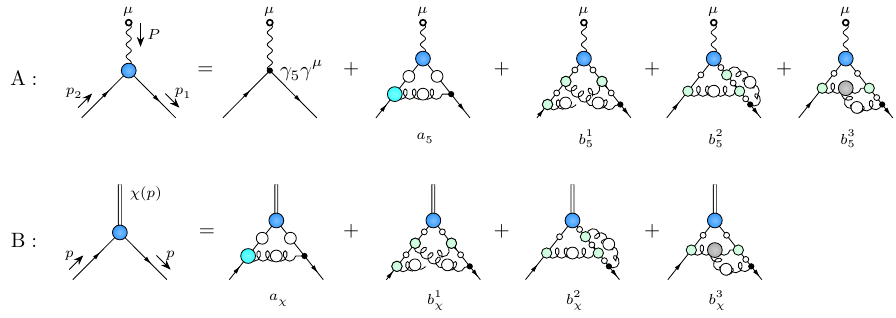}
    \caption{\textbf{\textit{Panel A:}} The SDE for the axial-vector vertex, emerging once the truncation for the $G_5^{\mu\nu}$ shown in Panel A of \fig{fig:1ld_green} has been implemented.
    \textbf{\textit{Panel B:}} The pion BSE, obtained after contracting the SDE in the panel above with $P_\mu$ and then taking the limit $P \to 0$.}
    \label{fig:bsebare}
\end{figure}

The axial-vector vertex $\ga^\mu(P,p_2,-p_1)$ arising from the truncation put forth in the previous section may be determined by substituting the $G_{\!5}^{\mu\nu}$ of \figA{fig:1ld_green}
into 
the SDE of \figA{fig:1ld}, thus leading 
to the diagrammatic expansion shown in 
\figA{fig:bsebare}. In particular,
\begin{equation}\label{eq:SDEg5}
    \Gamma_{\!5\mu}(P,p_2,-p_1)=\gamma_5\gamma_\mu+a_{5\mu}+b_{5\mu}^1+b_{5\mu}^2+b_{5\mu}^3\,,
\end{equation}
with
\bea
a_{5\mu} & = & ~~~c_d\int_q \gamma^\sigma S(q_1)\Gamma_{\!5\mu}(P,q_2,-q_1)\g^\nu(q,p_2,-q_2)\Delta_{\nu\sigma}(q)\nonumber\,,\\
\nonumber\\
b_{5\mu}^i& = & -ic_d\int_q\gamma^\sigma S(q_1) d_{i\mu}^\nu(P,q,p_2,-q_2)\Delta_{\nu\sigma}(q) \,, \qquad i=1,2,3, 
\eea
where we defined $c_d=-ig^2 C_f$ and $q_i=p_i+q$. 
The terms $d_i^{\mu\nu}$ are 
those composing the vertex 
$G_5^{\mu\nu}$, shown in \figA{fig:1ld_green}. 

After contracting the SDE in \1eq{eq:SDEg5} by  $P_\mu$, taking the limit $P\to 0$ to isolate its pole contribution, and finally eliminating $\gamma_5$ from both sides of the resulting equation, the pion BSE shown in \figB{fig:bsebare} emerges. 
Specifically, using that 
$S(p)S(-p) = -c(p)$, with 
$c(p)$ defined in \1eq{eq:bcquark}, 
the BSE assumes the form 
\be\label{eq:bse}
\chi(p)=a_\chi+b^{1}_\chi+b^{2}_\chi+b^{3}_\chi\,,
\ee
with 
\bea
a_\chi & = & ~-c_d\int_q \gamma^\sigma c(q')\chi(p)\left[\ge^\nu(q,p,-q')-\go^\nu(q,p,-q')\right]\Delta_{\nu\sigma}(q)\,,\nonumber\\
\nonumber\\
b_\chi^{1} & = & -c_dc_a\int_{q,k}\gamma^\sigma S(q')  V^\beta(k)c(q'+k)\chi(q'+k)V^\nu(q)S(-k-p)V^\alpha(k)\Delta_{\alpha\beta}(k)\Delta_{\nu\sigma}(q)\,,\nonumber\\
\nonumber\\
b_\chi^{2} & = & c_dc_a\int_{q,k}\gamma^\sigma S(q')V^\beta(k)S(q'+k)V^\nu(q)c(p+k)\chi(p+k)V^\alpha(k)\Delta_{\alpha\beta}(k)\Delta_{\nu\sigma}(q)\,,\nonumber\\
\nonumber\\
b_\chi^{3} & = & c_dc_b\int_{q,k}\gamma^\sigma S(q')V^\beta(k')c(p+k)\chi(p+k)V^\alpha(k)\g^{\nu\rho\delta}(q,-k,k')\Delta_{\alpha\rho}(k)\Delta_{\delta\beta}(k')\Delta_{\nu\sigma}(q)\,,\nonumber\\
\label{abschi}
\eea
where $q'=p+q$, $k'=k-q$.

Clearly, the terms $a_\chi$ and $b_\chi^{(i)}$ in the BSE of \1eq{eq:bse} are in one-to-one correspondence with the diagrams $a_5$ and $b_5^{i}$ of the axial-vector vertex SDE in \1eq{eq:SDEg5}.

In particular, we can now project out the dynamical equation for $\chi_1$, by  simply taking the trace in \1eq{eq:bse}, namely 
\be\label{eq:chi1}
\chi_1(p)=\frac{1}{4}\textrm{Tr}\left[a_\chi+b^{1}_\chi+b^{2}_\chi+b^{3}_\chi\right]\,.
\ee
The traces on the r.h.s. of \1eq{eq:chi1} can be evaluated using the expressions collected in \1eq{abschi}, yielding 
\bea\label{eq:Trbs}
\textrm{Tr}(a_\chi) & = & \gf\int_q c(q')\chi_{1}(q')\textrm{Tr}\left[\gamma^\sigma\go^\nu(q,p,-q')\right]\Delta_{\nu\sigma}(q)\,,\nonumber\\
\nonumber\\
\textrm{Tr}(b^{1}_\chi) & = & -\gf c_a\int_{q,k} c(t_2)\chi_1(t_2) \textrm{Tr}\left[\gamma^\sigma S(q')\gamma^\beta\gamma^\nu S(-t_1)\gamma^\alpha\right]V^2(k)V(q)\Delta_{\alpha\beta}(k)\Delta_{\nu\sigma}(q)\,,\nonumber\\
\nonumber\\
\textrm{Tr}(b^{2}_\chi) & = & \gf c_a\int_{q,k} c(t_1)\chi_1(t_1)\textrm{Tr}\left[\gamma^\sigma S(q')\gamma^\beta S(t_2)\gamma^\nu\gamma^\alpha\right]V^2(k)V(q)\Delta_{\alpha\beta}(k)\Delta_{\nu\sigma}(q)\nonumber\,,\\
\nonumber\\
\textrm{Tr}(b^{3}_\chi) & = & \gf c_b\int_{q,k}c(t_1)\chi_1(t_1)\textrm{Tr}\left[\gamma^\sigma S(q') \gamma^\beta\gamma^\alpha\right]V^2(k)\g^{\nu\rho\delta}(q,-k,k')\Delta_{\alpha\rho}(k)\Delta_{\delta\beta}(k')\Delta_{\nu\sigma}(q)\nonumber\,,\\
\eea
where $t_1=p+k$ and $t_2=t_1+q$.

We note that the 
cyan vertex enters into the 
standard diagram $a_5$, thus effectuating the transition beyond the RL approximation. In that sense, the remaining three diagrams, 
$b_5^{(1)}$, $b_5^{(2)}$, and $b_5^{(3)}$ are required precisely for restoring the symmetry.

\section{Renormalization of the SDE-BSE system}\label{sec:ren}

The final system that one must solve 
consists of three dynamical equations, namely  
(${\it a}$) 
the SDE of the quark propagator (gap equation) given in \1eq{eq:GapEq}, with $m=0$, 
(${\it b}$) the SDE of the quark-gluon vertex in \1eq{eq:qgsde}, and 
(${\it c}$) the pion BSE in \1eq{eq:chi1}.

Before proceeding with the numerical 
treatment, this set of equations must be 
properly renormalized.  
The renormalization proceeds in the standard way, by means of the 
following relations 
\begin{align}
\Delta_{{\rm \s{R}}}(q)&= Z_{3}^{-1} \Delta(q)\,,& S_{{\rm \s{R}}}(q) &= Z_2^{-1}S(q)\,,
\nonumber\\
\g^\mu_{{\rm \s{R}}}(r,p,q)&= Z_1 \g^\mu(r,p,q)\,,&
\chi^{\rm\s{R}}_1(q)&= Z_4 \, \chi_1(q) \,, 
\nonumber\\
g_{{\rm \s{R}}}&= Z_1^{-1} Z_2 Z_3^{1/2} \, g \,, 
\label{Zs_def}
\end{align}
which connect 
bare and unrenormalized quantities by means of the corresponding 
(cutoff-dependent) 
renormalization constants 
$Z_i$, with $i=1,2,3,4$. Note that
the requirement that the fundamental WTI 
of \1eq{eq:wtig5} should 
maintain its form intact after renormalization, imposes the 
constraint $Z_4 = Z_2$. In any case, 
due to the linearity and homogeneity 
of the BSE in \1eq{eq:chi1}, see \figB{fig:bsebare}, the constant 
$Z_4$ drops out automatically from both sides.

For the renormalization scheme, we adopt a variant of momentum subtraction (MOM)~\cite{Celmaster:1979km}, denoted \MOMt{}~\cite{Skullerud:2002ge,Kizilersu:2021jen,Aguilar:2023mam,Aguilar:2024ciu}. This scheme is defined by prescribing that at the renormalization point $\mu$, the propagators reduce to their tree-level forms, 
\be 
\Delta(\mu) = \mu^{-2}\,, \qquad A(\mu) = 1 \,, \label{MOM_prop}
\ee
and that the classical form factor, $\lambda_1(q,r,-p)$, attains its tree-level value in the soft-gluon limit, \ie 
\be
\left. \lambda_1(0,p,-p) \right\vert_{p^2 = \mu^2} = 1 \,. \label{MOM_vert}
\ee
We emphasize that although this is an exceptional kinematic configuration, $\lambda_1(0,p,-p)$ is infrared finite~\cite{Davydychev:2000rt}, thus defining a valid scheme for the subtraction of ultraviolet divergences.

Employing the above relations, it is 
straightforward to show that the 
renormalized gap equation is given by  
\begin{align}
\label{eq:GapEqRen}
S^{-1}_{{\rm \s{R}}}(p)= Z_2  \slashed{p}  
+ ig^2_{{\rm \s{R}}} C_f Z_1 \int_q\gamma^\nu S_{{\rm \s{R}}}(q)\,
\g^\mu_{\rm \s{R}}(q-p,p,-q)\Delta^{\rm \s{R}}_{\mu\nu}(q-p) \,.
\end{align}
Similarly, the renormalized SDE of the quark-gluon vertex reads 
\be\label{eq:qgsderen}
\g^\mu_{\rm \s{R}}(q,r,-p) =
Z_1 \gamma^\mu+c_{1 {\rm \s{R}}}^\mu+ c_{2 {\rm \s{R}}}^\mu\,,
\ee
where the index ``R'' on $c_{1,2}$
indicates that all quantities appearing in the expressions 
given in \2eqs{eq:c1}{eq:c2} have been replaced
by their renormalized counterparts. 
Finally, the renormalized BSE has the 
form 
\be
\label{eq:chi1ren}
\chi^{\rm\s{R}}_1(p)=\frac{1}{4} Z_1 \textrm{Tr}\left[a_{\chi}^{\rm\s{R}}+b^{1,\rm\s{R}}_{\chi}+b^{2,\rm\s{R}}_{\chi}+b^{3,\rm\s{R}}_{\chi}\right]\,,
\ee
where, again, the index ``R'' on the r.h.s. denotes 
that renormalized ingredients have been 
inserted in the relations of \1eq{abschi}.
In what follows we suppress the 
index ``R'', in order to simplify the notation. 

We note that, in both \2eqs{eq:GapEqRen}{eq:chi1ren}, the $Z_1$ appears multiplicatively, 
essentially due to the fact that there is a tree-level quark-gluon vertex in each of the diagrams defining these contributions. The presence of $Z_1$ 
is crucial for the 
cancellation of overlapping divergences 
\cite{Brown:1989hy,Curtis:1990zs,Curtis:1993py,Bloch:2001wz,Bloch:2002eq,Kizilersu:2009kg}.

In what follows we will implement the multiplicative renormalizability by relying on an expedient often utilized in the literature, mainly in the context 
of the quark gap equation~\cite{Fischer:2003rp,Aguilar:2010cn,Aguilar:2018epe}. 
In particular, denoting the 
kernel of the gap equation by 
${\cal K}_{\rm gap}(p,q,t)$, with $t = q-p$,
one carries out the {\it effective} replacement 
\begin{align}
Z_1 {\cal K}_{\rm gap}(p,q,t) \to {\cal C}(t) {\cal K}_{\rm gap}(p,q,t) \,,
\label{effrep}
\end{align}
where,
for large momenta,  
the function ${\cal C}(t)$, 
evolves as the classical  
form factor of the quark-gluon 
vertex [\ie the anomalous dimension of $\lambda_1$ in the 
symmetric limit ($q^2 = r^2 =p^2$)]. 

Adopting this procedure, the gap equation reads 
\be
Z_1 \int_q\gamma^\nu S(q) 
\g^\mu (q-p,p,-q)\Delta^{\rm \s{R}}_{\mu\nu}(q-p) 
\rightarrow 
\int_q\gamma^\nu {\cal C}(q-p)S(q) 
\g^\mu (q-p,p,-q)\Delta^{\rm \s{R}}_{\mu\nu}(q-p) \,.
\ee
Importantly, \2eqs{eq:qgsym}{CtoV} imply that $V(t)$ and ${\cal C}(t)$ scale as the quark-gluon vertex under a change of the renormalization point. Therefore, the combination 
$g^2 {\cal C} S \Gamma \Delta$ 
is guaranteed to be renormalization-group-invariant.

It is clear that if the substitution
in \1eq{effrep}
is carried out at the level of the gap equation, 
the preservation of the 
WTI in \1eq{eq:wtig5} requires  a similar 
replacement 
at the level of the 
SDE for $\ga^\mu(P,p_2,-p_1)$,
see \figA{fig:bsebare}, 
and, finally, 
of the pion BSE derived from it. 
In particular, a factor 
${\cal C} (t)$ 
is introduced in all terms of \1eq{eq:Trbs}, and the 
$Z_1$ is removed from \1eq{eq:chi1ren}. 
Note that the argument of 
${\cal C} (t)$ coincides 
with the momentum of the gluon that enters in the tree-level 
vertex of each diagram. 

While the asymptotic form of 
${\cal C}(t)$ is fixed by 
resorting to renormalization-group arguments, its infrared 
completion remains undetermined.
In the related studies~\cite{Fischer:2003rp,Aguilar:2010cn,Aguilar:2018epe},  appropriate {\it Ans\"atze} 
have been used for  
the infrared part of ${\cal C}(t)$, 
which have the additional effect of 
increasing the 
required strength of the 
gap equation kernel. Given that a function with these characteristics, namely the 
$V(q)$ in \1eq{eq:qgsym}, has already been introduced, in what follows we opt for the 
natural choice 
\be
{\cal C}(t) \rightarrow V(t) \,,
\label{CtoV}
\ee
thus obtaining for the quark gap equation \vspace{0.2cm}
\be
S^{-1}(p) = Z_2  \slashed{p}  
+ ig^2 C_f \int_q\gamma^\nu 
V(q-p) S(q)\,
\g^\mu(q-p,p,-q)\Delta_{\mu\nu}(q-p) \,.
\label{gapV}
\ee

Similarly, the pion BSE 
becomes 
\be
\label{eq:chiwithV}
\chi_1(p)=\frac{1}{4} \textrm{Tr}\left[{\tilde a}_{\chi}+
{\tilde b}^1_{\chi}+
{\tilde b}^2_{\chi}
+{\tilde b}^3_{\chi}\right]\,,
\ee
with
\bea
\label{eq:abchitilde}
\textrm{Tr}(\tilde{a}_\chi) & = & \gf\int_q c(q')\chi_{1}(q')\textrm{Tr}\left[\gamma^\sigma\go^\nu(q,p,-q')\right]V(q)\Delta_{\nu\sigma}(q)\,,\nonumber\\
\nonumber\\
\textrm{Tr}(\tilde{b}^{1}_\chi) & = & -\gf c_a\int_{q,k} c(t_2)\chi_1(t_2) \textrm{Tr}\left[\gamma^\sigma S(q')\gamma^\beta\gamma^\nu S(-t_1)\gamma^\alpha\right]V^2(k)V^2(q)\Delta_{\alpha\beta}(k)\Delta_{\nu\sigma}(q)\,,\nonumber\\
\nonumber\\
\textrm{Tr}(\tilde{b}^{2}_\chi) & = & \gf c_a\int_{q,k} c(t_1)\chi_1(t_1)\textrm{Tr}\left[\gamma^\sigma S(q')\gamma^\beta S(t_2)\gamma^\nu\gamma^\alpha\right]V^2(k)V^2(q)\Delta_{\alpha\beta}(k)\Delta_{\nu\sigma}(q)\nonumber\,,\\
\nonumber\\
\textrm{Tr}(\tilde{b}^{3}_\chi) & = & \gf c_b\int_{q,k}c(t_1)\chi_1(t_1)\textrm{Tr}\left[\gamma^\sigma S(q') \gamma^\beta\gamma^\alpha\right]V^2(k)V(q)\g^{\nu\rho\delta}(q,-k,k')\Delta_{\alpha\rho}(k)\Delta_{\delta\beta}(k')\Delta_{\nu\sigma}(q)\nonumber\,.\\
\eea

\begin{figure}[!t]
    \hspace*{-0.5cm}
    \includegraphics[scale=1.5]{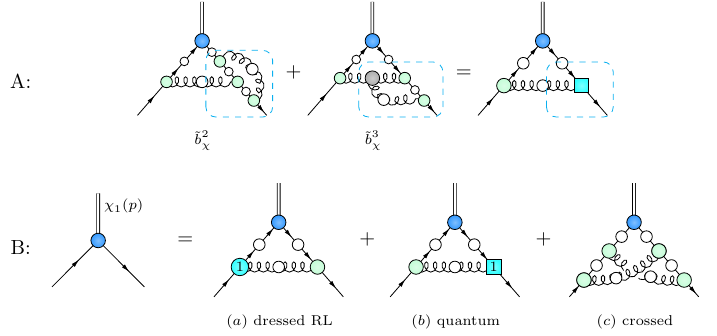}
    \caption{\textbf{\textit{Panel A:}} The diagrammatic formation of $\g_{\!\s Q}$, once the effective renormalization has been carried out. \textbf{\textit{Panel B:}} 
    The final form of the BSE after renormalization. Diagram ($a$) 
    is denominated ``dressed RL'' because 
    it corresponds to the standard RL diagram, but now dressed with a full
    quark-gluon vertex, diagram ($b$) is 
    called ``quantum'' because it  
    contains  
    the quantum part of the same vertex,  
    while diagram ($c$), which is named ``crossed'' due to its geometry, contains only the $V$.}
    \label{fig:bse}
\end{figure}

It turns out that 
the multiplication of 
the terms given in  
\1eq{eq:abchitilde} by $V(t)$ 
leads to a considerable 
simplification in the form of the 
BSE. In particular, as one may 
appreciate in \figA{fig:bse}, the 
sum of $\tilde{b}^{2}_\chi$ and $\tilde{b}^3_\chi$ generates precisely the quantum part 
$\g^\mu_{\!\scriptscriptstyle{Q}}$ 
of the quark-gluon vertex, 
as defined in \1eq{eq:qgquant}
and in \figB{fig:1ld_green}.
Therefore, the BSE assumes the 
final form shown in \figB{fig:bse},
namely 
\be
4\chi_1(p)=(a)+(b)+(c)\,,
\label{eq:BSEabc}
\ee
with
\be
\begin{array}{lcl}
\textrm{dressed RL:} & (a) = & \displaystyle c_d\int_q c(q')\chi_1(q') \textrm{Tr}\left[\gamma^\sigma\go^\nu(q,p,-q')\right]V(q)\Delta_{\nu\sigma}(q)\,,\\
\\
\textrm{quantum:} & (b) = & \displaystyle c_d\int_q c(q')\chi_1(q')\textrm{Tr}\left[\g^\sigma_{\!\s Q,1}(-q,q',-p)\gamma^\nu\right]V(q)\Delta_{\nu\sigma}(q)\,,\\
\\
\textrm{crossed:} & (c) = & \displaystyle -c_d c_a\!\int_{q,k}\!\!\!\!c(t_2)\chi_1(t_2)\textrm{Tr}\left[\gamma^\sigma S(q')\gamma^\beta \gamma^\nu
S(-t_1)\gamma^\alpha\right]V^2(k)V^2(q)\Delta_{\alpha\beta}(k)\Delta_{\nu\sigma}(q) \,,\\
\end{array}
\label{eq:BSE_full}
\ee
where $\g^\sigma_{\!\s Q,1}$
denotes the quantum part of the 
vertex with an odd number of 
Dirac $\gamma$ matrices. We note that the tensorial structures associated with the form factors $\lambda_5$ and $\lambda_7$ drop out from the above equations for the same reason described in \sect{gapnum}, in connection with the equation for $B(p)$; thus the only active components are $\lambda_1$ and $\lambda_6$.

\section{Numerical analysis}\label{num}

In this section we carry out the numerical analysis required for the 
solution of the system of dynamical equations, consisting of the quark-gap equation, the SDE of the quark-gluon vertex, and the BSE of the pion.

\subsection{Inputs}\label{subsec:inputs}

We begin with a presentation of the inputs used in this work, 
namely the gluon propagator, the
leading form factor of the  
three-gluon vertex, the value of the strong coupling, and the form of the function $V(q)$.

\begin{figure}[!t]
    \hspace*{-1.5cm}
    \includegraphics[scale=0.9]{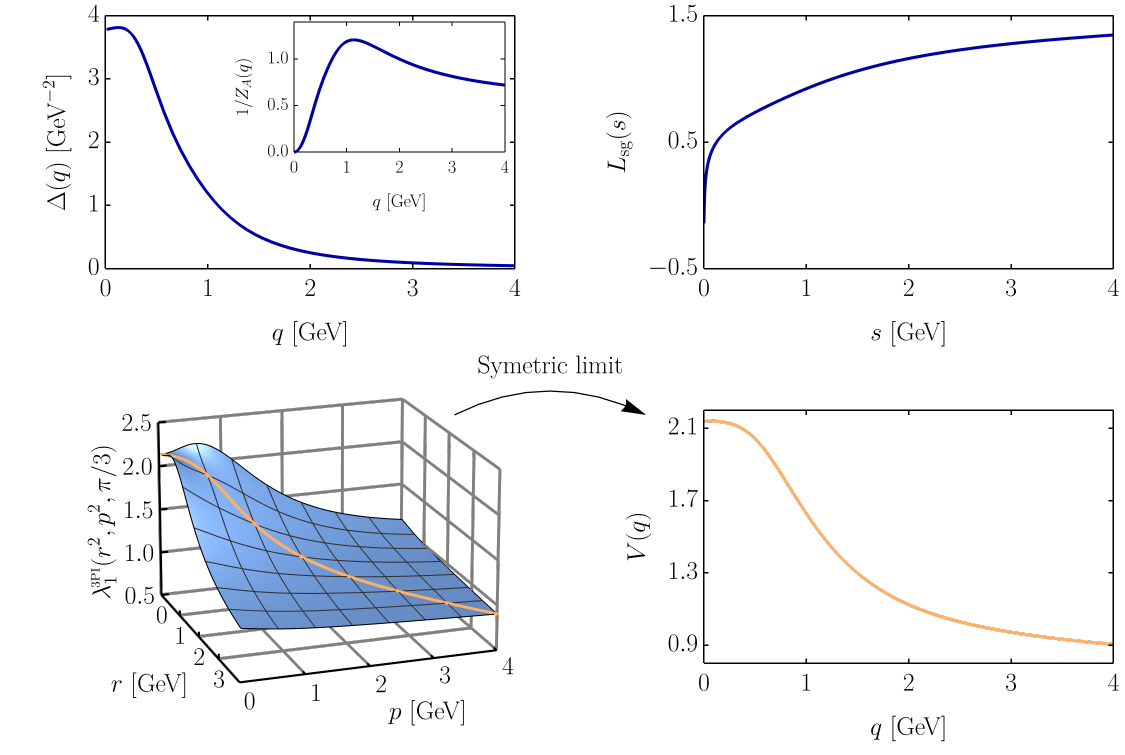}
    \caption{The input functions employed in the numerical solution of the SDE-BSE system. \textbf{\textit{Upper left}}: The gluon propagator, corresponding to the 
    lattice data of \cite{Ayala:2012pb,Binosi:2016xxu}, and the 
    corresponding dressing function (inset);
    \textbf{\textit{Upper right}}: 
    The form factor $L_{sg}(s)$
    of the three-gluon vertex, given in Eq.~(A1) of \cite{Aguilar:2023mam}; \textbf{\textit{Lower left}}: The classical form factor of the quark-gluon vertex obtained in \cite{Aguilar:2024ciu}, for $\theta_{rp}=\pi/3$; the orange line 
    indicates the diagonal $r^2=p^2$, corresponding
    to the symmetric configuration, and \textbf{\textit{Lower right}}: The function $V(q)$, identified with the diagonal of the previous panel, plotted in isolation.}
    \label{fig:inputs}
\end{figure}

\subsubsection{Gluon propagator}\label{glfit}

All integral equations entering in our analysis share 
as common ingredient
the Landau-gauge gluon propagator $\Delta_{\mu\nu}(q)$, 
defined in \1eq{eq:gluoprop}. 
For its scalar part, $\Delta(q)$,
we use a fit to the 
lattice data 
of~\cite{Ayala:2012pb,Binosi:2016xxu}, 
where the gluon propagator was simulated 
with 
two active quark flavours, 
$N_f = 2$.
The  $\Delta(q)$ employed 
is displayed in the upper left panel of \fig{fig:inputs}, 
renormalized at $\mu=2$ GeV. 
The functional form of the fit is described 
in Eq.~(A1) of \cite{Aguilar:2023mam}, and has been employed in the computation of the quark-gluon vertex presented 
in~\cite{Aguilar:2024ciu}. 

Note that the dynamical quarks entering in this propagator 
possess non-vanishing current masses, since the chiral limit may not be simulated on the lattice. Even though our analysis is carried out in the chiral limit, this discrepancy has 
no practical consequences. This may be clearly seen from the work of 
\cite{Cyrol:2017ewj}, where the gluon dressing function, $1/Z_A(q)= q^2 \Delta(q)$
with $N_f =2$ quark flavours 
was obtained for a variety of 
pion masses.
As is evident from Fig.~4a therein, the $1/Z_A(q)$ is essentially 
insensitive to the value of the pion masses, within the range 
\mbox{$m_{\pi} \in [285,60]$ MeV}.
This lack of sensitivity persists 
until the chiral limit, $m_{\pi}=0$, as one may deduce
from the right panel of Fig.~2 in  \cite{Eichmann:2025wgs}. Specifically,   
the lowest pion mass reached in \cite{Cyrol:2017ewj},  
namely \mbox{$m_{\pi} = 60$ MeV}, 
corresponds 
to a current quark mass of approximately \mbox{$m \approx 1$ MeV}.
Between this value and the chiral limit, $m = 0$, the pion mass drops rather abruptly from \mbox{$m_{\pi} = 60$ MeV} down to $m_{\pi}=0$.  However, 
varying $m$ within the narrow interval $[0,1]$ MeV at the level of the quark gap equation leaves the functions 
$A(p)$ and $M(p)$ practically unchanged. 
Given that the dependence of the 
gluon propagator on the quark parameters is fairly mild, of the type $\ln[p^2 + M^2(p)]$, see e.g., \cite{Gao:2021wun}, no appreciable effects are induced into the $1/Z_A(q)$ until the chiral limit has been reached. 

\subsubsection{Three-gluon vertex}\label{3g}

The three-gluon vertex,
$\Gamma_{\alpha\beta\gamma}(q,r,p)$,
enters in the non-Abelian graph, $c_2$, of the quark-gluon vertex SDE, shown in 
\figB{fig:1ld_green},
as well as in the BSE 
of \1eq{eq:BSEabc}, through 
the ``quantum'' diagram shown in  \figA{fig:bse}.
Since, in the Landau gauge, 
the three-gluon vertex 
is contracted by the three projection operators associated with each one of its legs, 
it is natural to introduce 
the transversely-projected 
vertex $\overline{\g}^{\mu\nu\rho}(q,r,p)=P^\mu_\alpha(q)P^\nu_\beta(r)P^\rho_\gamma(p)\g^{\alpha\beta\gamma}(q,r,p)$.
As was shown in a series of works \cite{Blum:2014gna,Eichmann:2014xya,Ferreira:2023fva,Aguilar:2023qqd,Pinto-Gomez:2022brg,Pinto-Gomez:2024mrk}, 
one may capitalize on the 
key property of 
``planar-degeneracy'', 
and achieve
a very accurate description 
of $\overline{\g}^{\mu\nu\rho}(q,r,p)$ by retaining only its tree-level 
structure $\overline{\g}_0^{\mu\nu\rho}(q,r,p)$,  namely 
\be
\overline{\g}^{\mu\nu\rho}(q,r,p)=L_{sg}(s)\overline{\g}_0^{\mu\nu\rho}(q,r,p)\,,\qquad s^2=\frac{1}{2}(q^2+r^2+p^2)\,,
\ee
where
\be
\g_0^{\mu\nu\rho}(q,r,p)=g^{\nu\rho}(r-p)^\mu+g^{\mu\rho}(p-q)^\nu+g^{\mu\nu}(q-r)^\rho\,.
\ee
A fit for the form factor $L_{sg}(s)$ is given in Eq.~(A1) of \cite{Aguilar:2023mam}; the resulting curve is shown in the upper right panel of \fig{fig:inputs}. Note that in the \MOMt{} scheme, defined by \2eqs{MOM_prop}{MOM_vert}, $L_{sg}(\mu) \neq 1$. The conversion between the \MOMt{} scheme and the asymmetric MOM scheme~\cite{Athenodorou:2016oyh,Aguilar:2023qqd}, wherein $L_{sg}^{\s{asym}}(\mu) = 1$, amounts to a finite rescaling $L_{sg}(s) = 1.16 \, L_{sg}^{\s{asym}}(s)$ for $\mu = 2$~GeV (see App.~B of \cite{Aguilar:2023mam}).

\subsubsection{Strong coupling}\label{subsubsec:alpha}

For the value of the coupling in the \MOMt{} scheme, we adopt the estimate $\alpha_s(\mu)=0.55$ at $\mu=2~\textrm{GeV}$. This value was estimated in~\cite{Aguilar:2024ciu} by adjusting the solution of the one-loop dressed SDE for the quark-gluon vertex to the lattice data of~\cite{Kizilersu:2021jen} for the soft-gluon $\lambda_1(0,p,-p)$, computed with $N_f = 2$ dynamical quarks.

\subsubsection{The function \texorpdfstring{$V(q)$}{V(q)}}\label{vparam}

\begin{table}[!t]
    \centering
    \setlength{\tabcolsep}{10pt} 
    \begin{tabular}{cccccccc}
    \toprule
        $d$ &
        $\kappa^2$ &
        $b^2_0$ &
        $b^2_1~[\textrm{GeV}^2]$ &
        $b^2_2~[\textrm{GeV}^2]$ &
        $b^2_3~[\textrm{GeV}^2]$ &
        $e^2_0~[\textrm{GeV}^2]$ &
        $e^2_1~[\textrm{GeV}^2]$ \\
        \midrule
        $1.154$  & $1.774$ & $0.0145$ & $126.113$ & $217.268$ & $4.766$ & $3.916$ & $2.079$ \\
    \bottomrule
    \end{tabular}
    \caption{Best fit parameters for $V(q)$ to reproduce the symmetric slice ($q^2=r^2=p^2$) of $\lambda_1^{\!\rm \s{3PI}}$ as obtained in \cite{Aguilar:2024ciu}.}
    \label{tab:coefsV}
\end{table}

The last ingredient needed for our study is an expression for $V(q)$. In order to maintain a close connection 
to the full quark-gluon vertex, 
we identify $V(q)$ with the 
symmetric limit of the classical form factor 
obtained in the analysis 
of~\cite{Aguilar:2024ciu}.

In that study, the SDE shown 
in~\figB{fig:summ}
was solved 
iteratively, having full
quark-gluon vertices inside the 
defining integrals 
instead of a $V(q)$. 
In Euclidean space, we will express all form factors as functions of $r^2$, $p^2$ and $\theta_{rp}$,
\be 
\lambda_i(q,r,-p) \to \lambda_i(r^2,p^2,\theta_{rp})\,.
\label{conv}
\ee
In particular the form factor associated 
with $\bar{\tau}_1^{\mu}=P^\mu_\nu(q)\gamma^{\nu}$ 
is denoted by $\lambda_i^{\!\rm \s{3PI}}(r^2,p^2,\theta_{rp})$.
The solution for the special value $\theta_{rp} = \pi/3$ 
is shown in the 
 lower left panel of 
\fig{fig:inputs}. 
The orange curve 
marked on that plot 
is the diagonal, $p^2=r^2$, 
which coincides with 
the so-called 
``symmetric limit''.
When expressed in terms of the 
three momenta, this limit
corresponds to $q^2=r^2=p^2$, 
in which case, given that 
$q=p-r$, one finds indeed that $\theta_{rp}= \pi/3$.  
 
In what follows we identify 
$V(q)$ with the orange curve
corresponding to the symmetric limit in \fig{fig:inputs}, 
namely ($r^2,p^2 \to q^2$)
\begin{align}   
\label{Vsl}
V(q) :=  \lambda_1^{\!\rm \s{3PI}} (q^2,q^2,\pi/3) \,.
\end{align} 

For this particular curve 
we employ the fit 
\be
    V(q)=\frac{d}{U(q)^\frac{9}{4\beta_0}+R(q)}\,,
    \label{eq:V_fit}
\ee
where $\beta_0=11-2N_f/3$, and 
\begin{align}
U(q)&=1+\kappa\log\left(\frac{q^2+\eta(q)}{\mu^2+\eta(q)}\right)\,, &R(q)&=\frac{b_0^2+\frac{q^2}{b_1^2}}{1+\frac{q^2}{b_2^2}+\left(\frac{q^2}{b_3^2}\right)^2}\,,& \eta(q)&=\frac{e_0^2}{1+\frac{q^2}{e_1^2}}\,,
\end{align}
are functions adjusted to reproduce the $\lambda_1^{\!\rm \s{3PI}} (q^2,q^2,\pi/3)$ 
from~\cite{Aguilar:2024ciu}, 
at the renormalization scale of 
$\mu=2~\textrm{GeV}$. The optimal values for the fit parameters are collected in \tab{tab:coefsV}, and the resulting curve for $V(q)$ is displayed in the lower right panel of \fig{fig:inputs}. Note that in the \MOMt{} renormalization scheme, $V(\mu) = 1.12$.

\subsection{Quark-gluon vertex}\label{qgnum}

\begin{figure}[!t]
\centering
\includegraphics[scale=0.78]{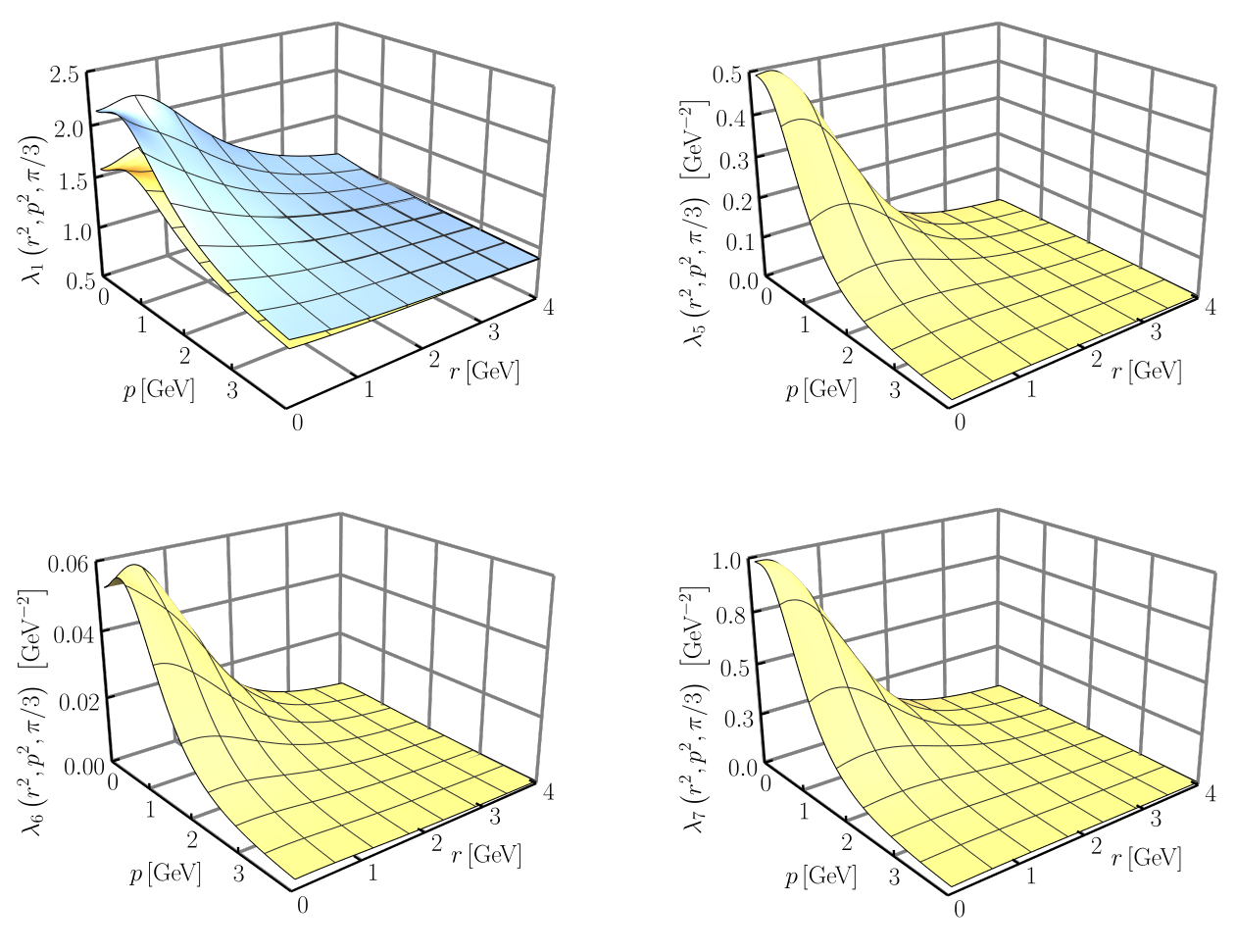}
\caption{The chirally symmetric form factors $\lambda_i$, $i=1,5,6,7$, 
of the 
quark-gluon vertex, plotted as functions of the momenta $r$ and $p$, for a fixed value of 
the angle $\theta_{rp}$ formed between them, $\theta_{rp}=\pi/3$. In the upper left panel we include a comparison with the form factor
$\lambda_1^{\rm\s{3PI}}$ (blue surface),
also shown in the lower left panel of \fig{fig:inputs}.
}
\label{fig:LambdaOdd}
\end{figure}

\begin{figure}[!t]
\centering
\includegraphics[scale=0.78]{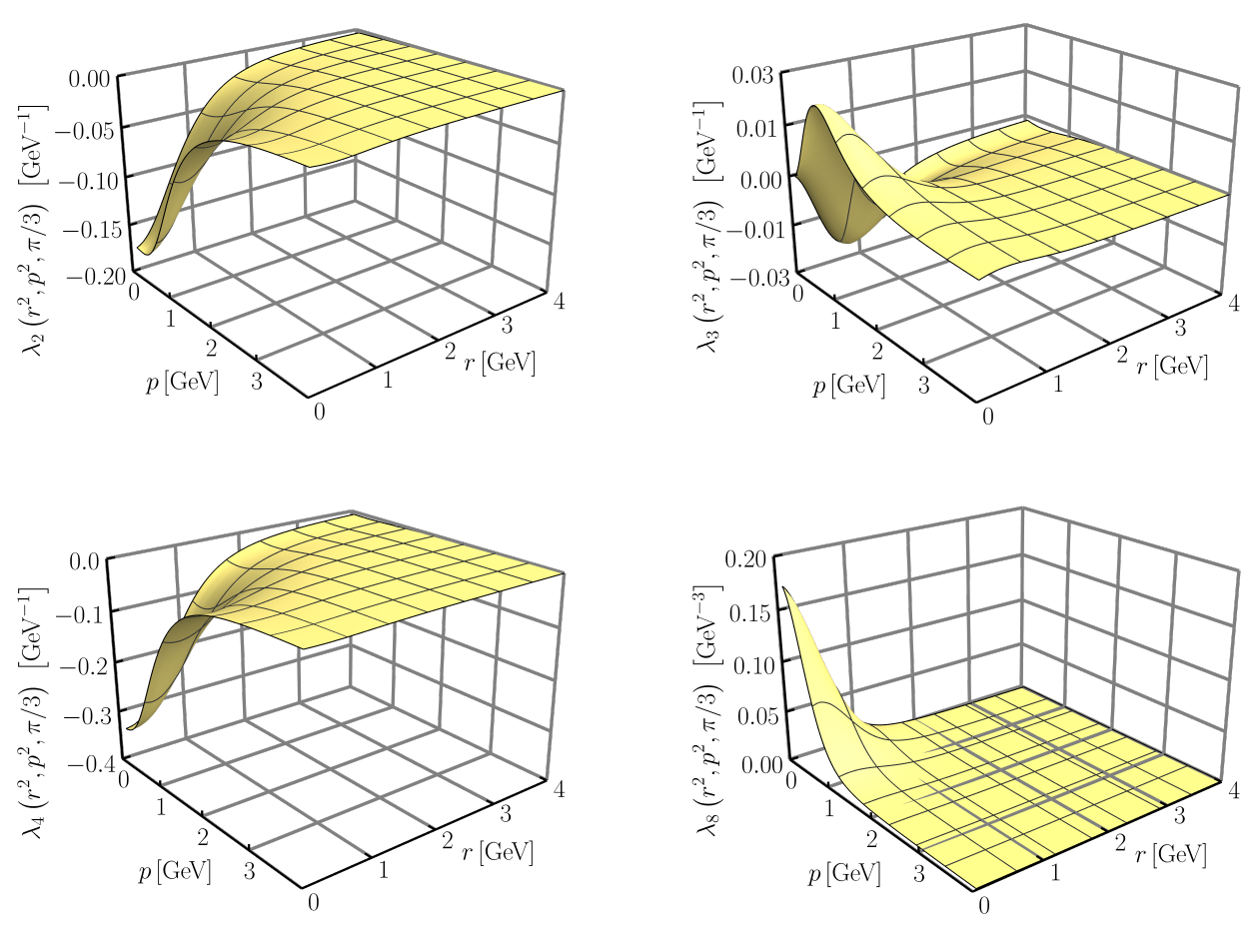}
\caption{The chiral symmetry breaking quark-gluon vertex form factors $\lambda_i$, $i=2,3$ (upper panel) and $i=4,8$ (lower panel) plotted as functions of the quark-momenta $r$ and $p$ for a fixed value $\theta_{rp}=\pi/3$.}
\label{fig:LambdaEven}
\end{figure}

In order to treat the  
SDE of \1eq{eq:qgsde} 
numerically, one must 
first implement the 
transition to Euclidean space;
we do so by  
following the conversion rules 
exposed in detail 
in \cite{Aguilar:2024ciu}, see in 
particular Section~IV.~B and Appendix~A therein. 

Substituting 
into the Euclidean SDE 
the relevant ingredients discussed in \sect{subsec:inputs}, 
one may obtain the 
form factors $\lambda_i$ through simple integration.

The corresponding results 
for the chirally symmetric form factors $\lambda_{1,5,6,7}$, which compose 
the $\got^\nu$, are shown in \fig{fig:LambdaOdd}.
The chiral symmetry breaking 
form factors $\lambda_{2,3,4,8}$,
which compose $\get^\nu$, 
are displayed in \fig{fig:LambdaEven}. 

In the upper left panel of 
\fig{fig:LambdaOdd} we present a 
direct comparison between the 
$\lambda_1$ acquired  
using the $V(q)$
as input and the $\lambda_1^{\!\rm \s{3PI}}$
obtained from the full treatment of~\cite{Aguilar:2024ciu}; the
latter quantity is also shown in 
the lower left panel of 
\fig{fig:inputs}, where its diagonal 
is identified with the $V(q)$. 
We note that the two 
form factors compare rather well 
through the entire kinematic range, showing relatively mild deviations in the infrared region.

\subsection{Gap equation}\label{gapnum}

\begin{figure}[!t]
    \centering
    \includegraphics[scale=0.9]{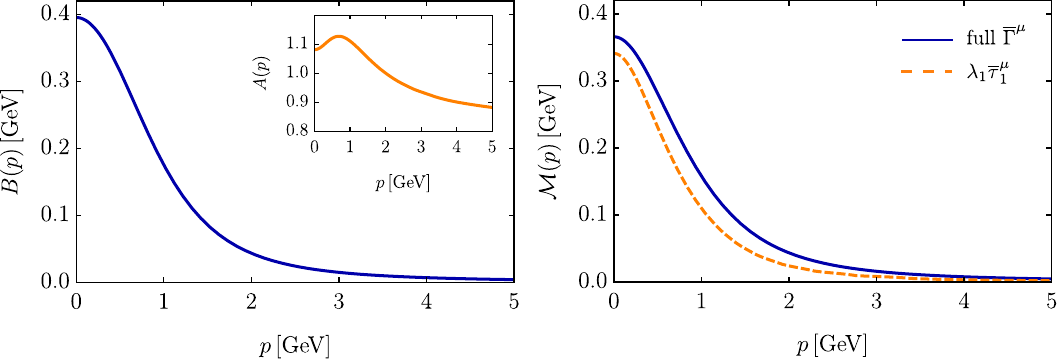}
    \caption{Results of the quark gap–equation solution. 
\textbf{\textit{Left panel}}: The quark dressing function $B(p)$, with the corresponding dressing $A(p)$ displayed in the inset. 
\textbf{\textit{Right panel}}: The constituent quark mass function $\mathcal{M}(p)=B(p)/A(p)$, shown together with the result obtained when retaining only the classical part of the quark–gluon vertex.
}
\label{fig:mass_function}
\end{figure}

After passing to Euclidean space
following standard conventions, 
the discretization of the 
momentum dependence using Gaussian quadratures transforms the integral equations into a system of non-linear algebraic equations. These equations are solved by fixed-point iteration: starting from an initial guess for $A(p)$ and $B(p)$, the loop integrals are evaluated numerically, and the dressing functions are updated until convergence is achieved \cite{Sanchis-Alepuz:2017jjd}. The corresponding results are shown in \fig{fig:mass_function}. In the left panel, we show the result for the function $B(p)$, while 
the function $A(p)$ is displayed in the inset. 

The numerical solution for the corresponding constituent quark mass, $\mathcal{M}(p) =B(p)/A(p)$, is displayed in the right panel of \fig{fig:mass_function} as a continuous blue curve. 
In that figure, we also show, 
as a dashed orange curve, the mass obtained if we solve the gap equation with a quark-gluon vertex that contains only the classical 
component, \ie $\overline{\Gamma}^{\mu} \to \lambda_1\bar{\tau}_1^\mu$. 
As we see, the two results are relatively close, 
which indicates that the bulk of the mass originates from the classical tensor structure. 
The r\^ole of the remaining form factors is to 
enhance the mass by about $7\%$ at the origin and $30\%$ at $p = 1$~GeV. We find that the 
second most important component is the one  
associated with the form factor $\lambda_4$, 
in agreement with earlier observations 
presented in \cite{Gao:2021wun}. 

Lastly, we point out that the form factors $\lambda_5$ and $\lambda_7$ do not contribute to the equation for $B(p)$; their accompanying tensors, $\bar{\tau}_{5}^\mu$ and $\bar{\tau}_{7}^\mu$, are annihilated in the projection appearing in the first line of \1eq{eq:QuarkGapEq}. Nevertheless, these form factors contribute to $\mathcal{M}(p)$ through $A(p)$.

\subsection{BSE}\label{bsenum}
\begin{figure}[!t]
    \centering
    \includegraphics[scale=0.95]{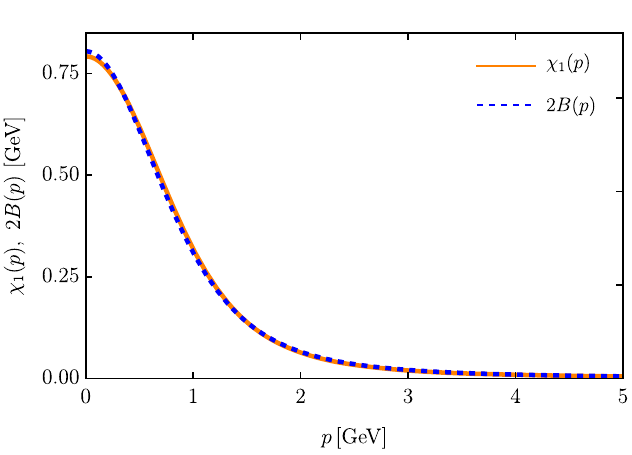}
    \caption{Numerical confirmation of the relation $\chi_1(p)=2B(p)$, \1eq{eq:chi12}, which follows from the axial WTI. The relation is satisfied within the numerical accuracy of the (two-) loop integrals, \ie one percent or better.}
    \label{fig:chi_2B}
\end{figure}
For the numerical solution of the BSE 
given by \2eqs{eq:BSEabc}{eq:BSE_full} we adopt standard techniques, discretizing the momentum integrals using Gauss-Legendre and Gauss-Chebyshev quadratures,  and reformulating the equation as an eigenvalue problem of the form \cite{Eichmann:2016yit, Sanchis-Alepuz:2017jjd, Eichmann:2025wgs, Huber:2025cbd}
\begin{equation}
\lambda(P^2)\, \chi_1(p) = \mathbf{K}(P^2)\, \chi_1(p) \,.
\label{eq:eigenvalue}
\end{equation}
$\mathbf{K}(P^2)$ denotes the interaction kernel, which in our case is composed of the three diagrams shown in \fig{fig:bse}. In general, 
the eigenvalue $\lambda(P^2)$ depends continuously on the total momentum, and the 
mass $M$ of the bound state 
is determined from the condition \mbox{$\lambda(P^2 = -M^2) = 1$}. The dominant eigenvalue and its corresponding eigenvector are obtained numerically using iterative eigensolvers, such as the ``power method'' or the ``Arnoldi iteration'', which ensure stable convergence \cite{doi:10.1137/S0895479895281484, doi:10.1137/1.9780898719628, Sanchis-Alepuz:2017jjd}. 
As already mentioned, in the present analysis we restrict ourselves to the case of the  
chiral limit, where the current quark mass $m$ vanishes, and consequently, the 
pion is exactly massless,  
corresponding to the condition $\lambda(P^2 = 0) = 1$.

As may be seen in \fig{fig:chi_2B},
the pion BSA, $\chi_1(p)$,   
 obtained from this analysis, 
satisfies at a high degree of accuracy (better than $1\%$) the symmetry-induced relation 
given in \1eq{eq:chi12}. 
This result is of central  importance, because it constitutes a clear  
numerical confirmation of the 
symmetry-preserving nature of the 
entire approach. 

\begin{figure}[!t]
    \centering
    \includegraphics[scale=0.6]{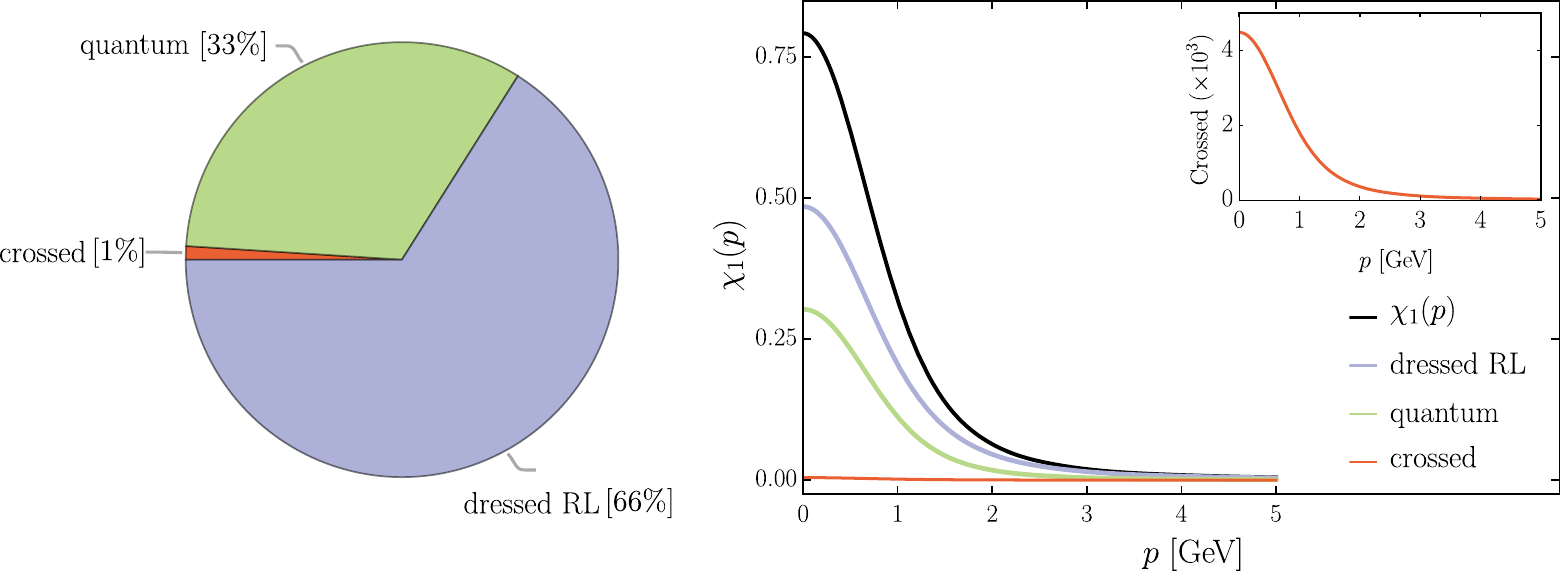}
    \caption{\textbf{\textit{Left panel}}: Individual eigenvalue contributions from each diagram entering into the full BSE solution in \1eq{eq:BSE_full}. 
\textbf{\textit{Right panel}}: Full pion BSA, $\chi_1(p)$, together with the individual contributions stemming from each diagram.
}
\label{fig:pie_chart}
\end{figure}

The individual contributions of the three diagrams to the eigenvalue $\lambda$ are displayed in the left panel of \fig{fig:pie_chart}: the ``dressed RL'' diagram provides approximately $66\%$ of the total value, the ``quantum'' diagram accounts for about $33\%$, while the ``crossed'' diagram contributes around $1\%$. However, even though the ``crossed'' contributes slightly, its r\^ole is crucial in maintaining the exact realization of the chiral solution. In particular, given that the BSE is an eigenvalue problem, even a small deviation from unity, for example, $\lambda(0) = 0.99$ instead of $\lambda(0) = 1$ would correspond to a large mass shift, thus  
 invalidating the masslessness of the pion. 
Therefore, all three diagrams are indispensable: the precise value $\lambda(P^2=0)=1$ arises only when their combined contributions are taken into account. Neglecting any of them distorts this delicate balance, thwarting the proper emergence
of the Nambu-Goldstone boson associated with the dynamical breaking of the chiral symmetry.

As a consistency check, once the amplitude $\chi_1(p)$ is obtained from the eigenvalue problem in \1eq{eq:eigenvalue}, we substitute it back into the full integral equation \1eq{eq:BSE_full}. This allows us to evaluate separately the contribution of each diagram to the total BSA (right panel of \fig{fig:pie_chart}). We find that the sum of these three individual contributions reconstructs the original solution with high numerical accuracy, confirming the internal consistency of our implementation.

\subsection{Varying the \texorpdfstring{$V(q)$}{V(q)}}\label{ernum}

\begin{figure}[!t]
    \centering
    \includegraphics[scale=0.75]{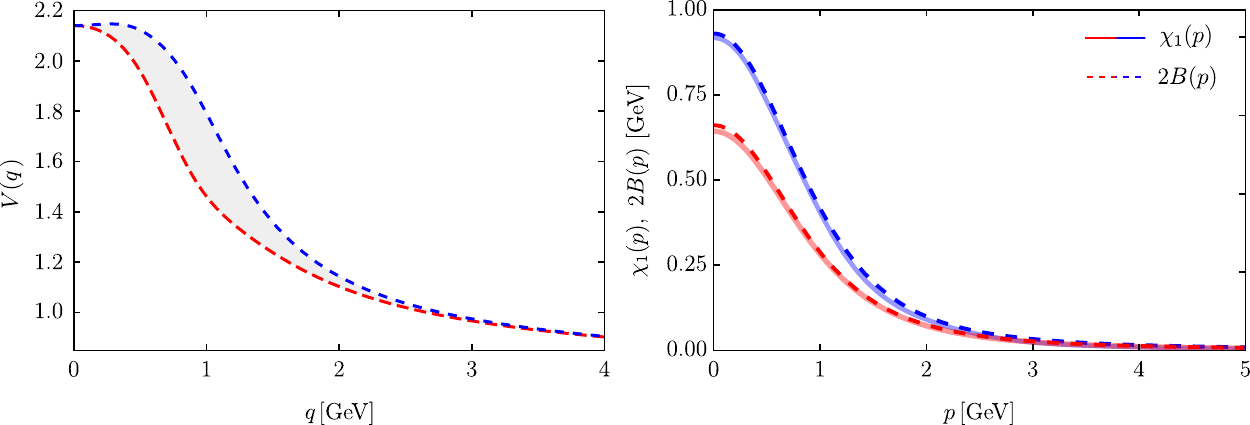}
 \caption{\textbf{\textit{Left panel}}: Variations around the central parametrization of $V(q)$ employed to assess the robustness of the calculation. 
\textbf{\textit{Right panel}}: Resulting pion BSAs obtained from these variations, shown together with $2B(p)$. The blue and red curves correspond to the upper and lower variations of the fit $V(q)$, respectively.
}
\label{fig:V_variations}
\end{figure}

In order to probe 
the robustness of our results and the internal consistency of the truncation underlying the coupled SDE-BSE framework, we have implemented
variations around the central fit 
of $V(q)$, shown in the left panel of \fig{fig:V_variations}, and have tested the 
persistence of the 
key relation $\chi_1(p) = 2 B(p)$,
see \1eq{eq:chi12}. As explained earlier, from the theoretical point of view, the 
validity of this relation does not 
hinge on the specific form of the 
function $V(q)$; the only requirement is that the latter depends on a single variable, which is identified with the gluon momentum 
entering into the specific 
quark-gluon vertex.

The variations implemented on 
$V(q)$ induce corresponding changes
to the vertex form factors 
$\lambda_i$, which then propagate 
to the quark gap equation 
and the BSE. 
Importantly, in all cases considered, the corresponding eigenvalue $\lambda(P^2)$ remains stable and continues to satisfy the condition $\lambda(P^2=0)=1$.
As a result, one obtains 
a sequence of pairs 
$\{\chi_1(p) , B(p)\}$, which, even though they are different from one another, have a key property 
in common: they fulfil 
the WTI-induced constraint  $\chi_1(p) = 2 B(p)$ 
at a high level of accuracy 
(1\% discrepancy), see right panel of \fig{fig:V_variations}. 
We therefore conclude that 
the resulting massless pion is not a contingent feature of a particular parametrization, but a robust consequence of the self-consistent SDE-BSE truncation.

\section{Discussion and conclusions}\label{sec:Disc}

We have presented a detailed numerical  
treatment of the dynamical equations within the recently developed approach to the 
physics of mesons \cite{Miramontes:2025imd}. 
We concentrated on the special case of the chiral limit with a massless pion. The analysis goes significantly beyond the standard RL treatments, employing fully-dressed quark-gluon vertices  
in both the gap equation and the pion BSE.
This quark-gluon vertex is obtained from an 
approximate form of the standard SDE within the 3PI 
formalism, and contains all eight transverse form factors 
with their full momentum dependence. The respective truncation and subsequent renormalization of the BSE kernel leads to a dynamical equation for the 
pion amplitude that consists of three special diagrams, two one-loop dressed, 
denominated as ``dressed RL" and ``quantum", and a two-loop dressed graph, coined ``crossed". The symmetry-preserving character of the truncation was confirmed explicitly by checking of the pivotal relation \mbox{$\chi_1(p) = 2 B(p)$}. It is satisfied exactly within the numerical precision of the respective (two-) loop integrals, that is, a precision of $\lesssim 1\%$ in the entire range of momenta, see \fig{fig:chi_2B}.

The numerical error originates mainly from the treatment of the two-loop contribution (``crossed'' diagram) in the BSE kernel, whose evaluation is considerably more demanding than the one-loop terms.
In particular, the resolution of the integration grid and the need to balance numerical cost 
with stability introduce a controlled uncertainty at very small momenta.  
A refined grid sampling would systematically reduce this effect, and will be implemented in the near future. In any case, the present level of agreement 
already confirms that the truncation preserves the axial symmetry.

In addition, the BSE eigenvalue was found to be equal to unity to four significant figures, and the individual contributions of the aforementioned three graphs to its composition (66$\%$, 33$\%$, and 1$\%$, respectively) was ascertained.
From the technical point of view, 
it is worth stressing that 
the present study deals for the first time with the evaluation of the 
two-loop diagram (``crossed"); in particular, even though the relevance of this graph had been 
briefly alluded in \cite{Sanchis-Alepuz:2015tha}, its numerical 
treatment has never been reported. 

In order to extend the above analysis beyond the 
chiral limit, and obtain the physical masses of the pion and other mesons, the general theoretical framework developed in \cite{Miramontes:2025imd} 
must be generalized to the 
case of nonvanishing quark masses. 
In addition, and more importantly, 
several of the key equations must be 
evaluated for complex values of the 
momenta. The need for such a generalization may already be seen at the level of the meson BSE,
whose incoming momentum must satisfy the 
condition $P^2=-M^2$, where $M$ is the mass of the 
meson in question. Therefore, 
according to the standard 
parametrization \cite{Sanchis-Alepuz:2017jjd}, one sets $P^{\mu} = (0,0,0,iM)$ inside the BSE, and, as a result, knowledge of the BSE ingredients 
beyond the real Euclidean axis is needed.
The need to extend the treatment of the gap equation to accommodate 
complex momenta is already known from the standard 
RL studies, see, \eg \cite{Maris:1999bh, Alkofer:2002bp, Eichmann:2008ae, Qin:2011dd, Hilger:2014nma, Heupel:2012ua, Eichmann:2015cra, Hilger:2015hka, El-Bennich:2016qmb,Eichmann:2016yit,Mojica:2017tvh, Weil:2017knt,Sanchis-Alepuz:2017jjd,Chen:2019otg,Chang:2020iut,Xu:2024fun}; however, the present formulation requires, in addition, the corresponding generalization on the complex plane of the form factors that 
compose the quark-gluon vertex. 
The study of the structure of the quark-gluon vertex 
for complex momenta is naturally integrated in the 
general exploration of QCD correlation functions on the complex plane, which is being implemented using a variety of approaches, such as 
\cite{Alkofer:2003jj,Eichmann:2007nn,Windisch:2012sz,Eichmann:2019dts,Fischer:2020xnb,Huber:2022nzs,Horak:2022myj,Duarte:2022yur,Braun:2022mgx,Eichmann:2023tjk,Horak:2023hkp,Pawlowski:2024kxc,Fischer:2005en,Fischer:2008sp,Krassnigg:2008bob}.
We hope to report progress in this direction in the near future. 

\section*{Acknowledgments}\label{sec:Acknow}

The work of M.N.F. is supported by the National Natural Science Foundation of
China (grants 12135007 and W2433021). A.S.M., J.M.M. and J.P. are funded by the Spanish MICINN grants PID2020-113334GB-I00 and PID2023-151418NB-I00, the Generalitat Valenciana grant CIPROM/2022/66, and CEX2023-001292-S by MCIU/AEI. J.M.P. is funded by the Deutsche Forschungsgemeinschaft (DFG, German Research Foundation) under Germany’s Excellence Strategy EXC 2181/1 - 390900948 (the Heidelberg STRUCTURES Excellence Cluster) and the Collaborative Research Centre SFB 1225 - 273811115 (ISOQUANT).

\appendix

\section{Diagrammatics of \1eq{eq:G5poleG2}.}\label{app:DiagrammaticProof}

In this appendix we 
demonstrate the validity of \1eq{eq:G5poleG2}
when the approximations and truncations described in the main text are implemented; 
the demonstration proceeds 
directly at the 
level of the diagrams that compose both sides of \1eq{eq:G5poleG2}. 

\begin{figure}[t!]
    \centering
    \includegraphics[scale=1.2]{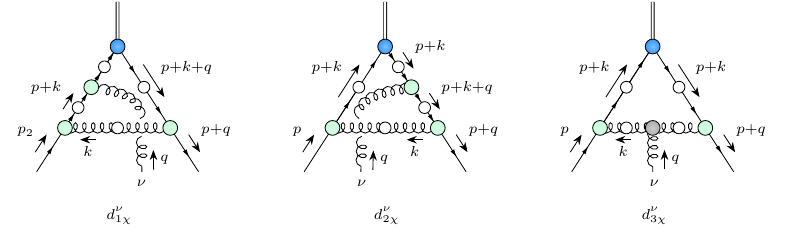}
    \caption{Diagrammatic representation of the three contributions to the pole part of $G_5^{\mu\nu}$.}
    \label{fig:dchi}
\end{figure}

When the  vertex $G_5^{\mu\nu}$ is approximated 
by the sum of the graphs $d_1$, $d_2$, and $d_3$ in \figA{fig:1ld_green}, 
the l.h.s. of the WTI in 
\1eq{eq:G5poleG2}
is given by 
\begin{align} 
\lim_{P\to 0} P_\mu G_5^{\mu\nu}(P,q,p_2,-q_1)  = 
d_{1\chi}^{\,\nu} + d_{2\chi}^{\,\nu} + d_{3\chi}^{\,\nu} \,, 
\label{eq:dxi}
\end{align}
with (see \fig{fig:dchi})
\bea
d_{1\chi}^\nu & = & -i c_a \int_k \gamma^\beta c(p+k+q)\chi_1(p+k+q)\gamma^\nu S(-p-k)\gamma^\alpha \Delta_{\alpha\beta}(k)V^2(k)V(q)\gamma_5\,,\nonumber\\
\nonumber\\
d_{2\chi}^\nu & = & i c_a \int_k \gamma^\beta S(p+k+q)\gamma^\nu c(p+k)\chi_1(p+k)\gamma^\alpha\Delta_{\alpha\beta}(k)V^2(k)V(q)\gamma_5 \,,\nonumber\\
\nonumber\\
d_{3\chi}^\nu & = & ic_b\int_k \gamma^\beta c(p+k)\chi_1(p+k)\gamma^\alpha\g^{\nu\rho\delta}(q,-k,k')\Delta_{\alpha\rho}(k)V(k)\Delta_{\delta\beta}(k')V(k')\gamma_5\,.\nonumber\\
\eea

Since the r.h.s. of  \1eq{eq:G5poleG2}
is given by 
the 
even component of the quark-gluon vertex graphs $c_1^{\nu}$ and $c_2^{\nu}$ in 
\figB{fig:1ld_green}, 
equating both sides yields  
\be 
d_{1\chi}^{\,\nu} + d_{2\chi}^{\,\nu} + d_{3\chi}^{\,\nu} = 
2i \, (c^{\nu}_{1,2} + c^{\nu}_{2,2}) \gamma_5 \,,
\label{eq:insrel}
\ee
where the rightmost subscript ``2'' in $c^{\nu}_{1,2}$ and $c^{\nu}_{2,2}$ indicates the part of the diagrams that contain an even 
number of Dirac $\gamma$ matrices.
In fact, the obvious separation of graphs into Abelian (no three-gluon vertex)  and non-Abelian (three-gluon vertex) reduces \1eq{eq:insrel} to two 
simpler relations, namely  
\be\label{eq:relationsW}
d^{\,\nu}_{1\chi} + d^{\,\nu}_{2\chi} = 2 i c^{\nu}_{1,2} \gamma_5 \,,
\qquad \qquad 
d^{\,\nu}_{3\chi} = 2i c^{\nu}_{2,2} \gamma_5 \,,
\ee 
which we proceed to prove. 

We now introduce the following diagrammatic representation for the components of the quark propagator, see \1eq{eq:bcquark},  
\bea
a(p)\slashed{p}=c(p)A(p)\slashed{p} & = & \includegraphics[scale=1]{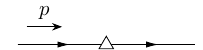}\,,\nonumber\\
b(p)=c(p)B(p) & = & \includegraphics[scale=1]{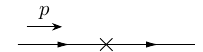}\,,
\label{eq:insertions}
\eea
which is useful in the ensuing analysis.

\begin{figure}
    \centering
    \includegraphics[scale=1.2]{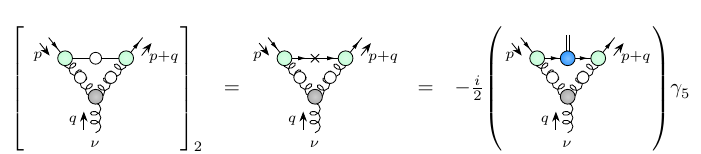}
    \caption{Diagrammatic realization of \1eq{eq:c22d3}.}
    \label{fig:massnonab}
\end{figure}

We start considering the non-Abelian diagram $c_2^{\nu}$, \1eq{eq:c1c2}, and determine its even component, 
\bea
\label{eq:c22d3}
c_{2,2}^\nu & = & c_b\int_k \bigl[\gamma^\beta S(p+k)\gamma^\alpha\bigr]_{\textrm{even}} \g^{\nu\rho\delta}(q,-k,k')\Delta_{\alpha\rho}(k)V(k)\Delta_{\delta\beta}(k')V(k')\nonumber\\
\nonumber\\
& = & c_b\kappa_b\int_k \gamma^\beta b(p+k)\gamma^\alpha \g^{\nu\rho\delta}(q,-k,k')\Delta_{\alpha\rho}(k)V(k)\Delta_{\delta\beta}(k')V(k')\nonumber\\
\nonumber\\
& = & \frac{c_b}{2}\int_k \gamma^\beta c(p+k)\chi_1(p+k)\gamma^\alpha \g^{\nu\rho\delta}(q,-k,k')\Delta_{\alpha\rho}(k)V(k)\Delta_{\delta\beta}(k')V(k')\nonumber\\
\nonumber\\
& = & -\frac{i}{2} d_{3\chi}^\nu \gamma_5\,,
\eea
\ie the relation given in \1eq{eq:relationsW}. 
With the aid of the  symbols introduced in \1eq{eq:insertions}, 
this identity may be expressed diagrammatically as shown in \fig{fig:massnonab}.

\begin{figure}
    \hspace*{-1cm}
    \includegraphics[scale=1]{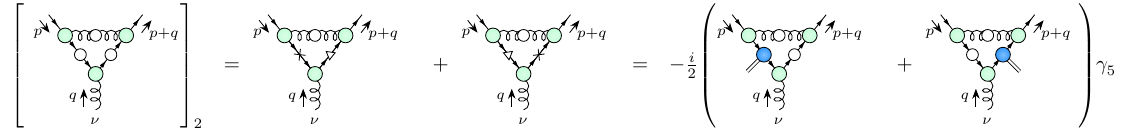}
    \caption{Diagrammatic realization of \1eq{eq:c12d3}.}
    \label{fig:massab}
\end{figure}

The Abelian case proceeds in the same way, but 
is algebraically slightly more involved. 
To begin with, 
an important cancellation takes place 
when 
 $d_{1,\chi}$ and $d_{2,\chi}$ are summed.
In particular,
defining 
\begin{align}
    \mathcal{T}_{\alpha\beta}(k,q) &=\Delta_{\alpha\beta}(k)V^2(k)V(q)\,,& \ell&=p+k+q\,,
\end{align}
we have 
\bea\label{eq:d1d2cancel}
d_{1\chi}+d_{2\chi} & = & ic_a\!\int_k\! \gamma^\beta S(\ell)\gamma^\nu c(p+k)\chi_1(p+k)\gamma^\alpha\mathcal{T}_{\alpha\beta}(k,q)\gamma_5\nonumber\\
\nonumber\\
& - & ic_a\!\int_k\! \gamma^\beta c(\ell)\chi_1(\ell)\gamma^\nu S(-p-k)\gamma^\alpha\mathcal{T}_{\alpha\beta}(k,q)\gamma_5\nonumber\\
\nonumber\\
& = & ic_a\!\int_k\! \gamma^\beta S(\ell)\gamma^\nu c(p+k)\chi_1(p+k)\gamma^\alpha\mathcal{T}_{\alpha\beta}(k,q)\gamma_5\nonumber\\
\nonumber\\
& - & ic_a\!\int_k\! \gamma^\beta c(\ell)\chi_1(\ell)\gamma^\nu S(-p-k)\gamma^\alpha\mathcal{T}_{\alpha\beta}(k,q)\gamma_5\nonumber\\
\nonumber\\
& = & ic_a\int_k \gamma^\beta a(\ell)\slashed{\ell}\gamma^\nu c(p+k)\chi_1(p+k)\gamma^\alpha\mathcal{T}_{\alpha\beta}(k,q)\gamma_5\nonumber\\
\nonumber\\
& + & ic_a\int_k \gamma^\beta c(\ell)\chi_1(\ell)\gamma^\nu a(\ell)\slashed{\ell}\gamma^\alpha\mathcal{T}_{\alpha\beta}(k,q)\gamma_5\nonumber\\
\nonumber\\
& + & ic_a\int_k \gamma^\beta b(\ell)\gamma^\nu c(p+k)\chi_1(p+k)\gamma^\alpha\mathcal{T}_{\alpha\beta}(k,q)\gamma_5\nonumber\\
\nonumber\\
& - & ic_a\int_k \gamma^\beta c(\ell)\chi_1(\ell)\gamma^\nu b(p+k)\gamma^\alpha\mathcal{T}_{\alpha\beta}(k,q)\gamma_5\,.
\eea
The announced cancellation  
takes place between the last two terms, 
once the first relation 
in \1eq{eq:chi12}, 
\mbox{$2b(p)=c(p)\chi_1(p)$}, 
has been invoked, 
leaving us with
\bea
d_{1\chi}+d_{2\chi} & = & ic_a\int_k \gamma^\beta a(\ell)\slashed{\ell}\gamma^\nu c(p+k)\chi_1(p+k)\gamma^\alpha\mathcal{T}_{\alpha\beta}(k,q)\nonumber\\
\nonumber\\
& + & ic_a \int_k \gamma^\beta c(\ell)\chi_1(\ell)\gamma^\nu a(\ell)\slashed{\ell}\gamma^\alpha\mathcal{T}_{\alpha\beta}(k,q)\gamma_5\,.
\label{eq:aftcan}
\eea
Consider now the 
even part of the diagram 
$c_{1}^\nu$, namely  
\bea\label{eq:c12d3}
c_{1,2}^\nu & = & c_a\int_k \bigl[\gamma^\beta S(\ell) \gamma^\nu S(p+k)\gamma^\alpha\bigr]_\textrm{even}\mathcal{T}_{\alpha\beta}(k,q)\nonumber\\
\nonumber\\
& = & c_a\int_k\gamma^\beta a(\ell)\slashed{\ell}\gamma^\nu b(p+k)\gamma^\alpha\mathcal{T}_{\alpha\beta}(k,q)\nonumber\\
\nonumber\\
& + & c_a\int_k\gamma^\beta b(\ell)\gamma^\nu a(p+k)(\slashed{p}+\slashed{k})\gamma^\alpha\mathcal{T}_{\alpha\beta}(k,q)\nonumber\\
\nonumber\\
& = & \frac{c_a}{2}\int_k\gamma^\beta a(\ell)\slashed{\ell}\gamma^\nu c(p+k)\chi_1(p+k)\gamma^\alpha\mathcal{T}_{\alpha\beta}(k,q)\nonumber\\
\nonumber\\
& + & \frac{c_a}{2}\int_k\gamma^\beta c(\ell)\chi_1(\ell)\gamma^\nu a(p+k)(\slashed{p}+\slashed{k})\gamma^\alpha\mathcal{T}_{\alpha\beta}(k,q)\nonumber\\
\nonumber\\
& = & -\frac{i}{2}\left(d_{1\chi}^\nu+d_{2\chi}^\nu\right)\gamma_5\,,
\eea
which is the first relation in 
\1eq{eq:relationsW}; in the last step,
\1eq{eq:aftcan} was used.
The final result may be 
depicted as shown in \fig{fig:massab}.

\section{An interesting identity}\label{app:identity}

In this appendix we derive an approximate integral constraint relating the odd and even components of the quark-gluon vertex, $\go^\mu(q,r,-p)$ and $\ge^\mu(q,r,-p)$, respectively. This constraint stems from an identity relating the odd and even parts of the non-Abelian diagram in the quark-gluon vertex SDE
of \figB{fig:1ld_green}.

\begin{figure}
    \centering
    \includegraphics[scale=1]{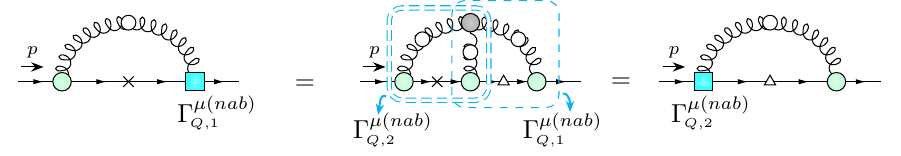}
    \caption{Diagrammatic demonstration of the identity \1eq{id_nab}. }
    \label{fig:id_nab}
\end{figure}

To derive the identity of interest, consider the expression
\be 
i\int_q b(q')\textrm{Tr}\left[\go^{\sigma(nab)}(-q,q',-p)V^\nu(q)\right]\Delta_{\nu\sigma}(q) \,,
\ee
corresponding to the diagram on the l.h.s. of \fig{fig:id_nab}. The superscript ``$nab$'' labels de non-Abelian contribution to the quark-gluon vertex. Note that, up to an overall prefactor, this diagram is a contribution to the renormalized gap equation. Then, using for $\go^{\sigma(nab)}$ the explicit SDE diagram
($c_2$ in \figB{fig:1ld_green}), we obtain the two-loop diagram in the center of \fig{fig:id_nab}. At this point, the loop on the left part of the two-loop diagram, highlighted by a double-dashed line, is immediately recognized as $\ge^{\nu(nab)}$. Thus, we obtain the following identity for the non-Abelian contribution
\be
t_1^{(nab)}(p) = t_2^{(nab)}(p) \,, \label{id_nab}
\ee
where
\begin{align}
t_1^{(nab)}(p):=&\,i\!\int_q b(q')\textrm{Tr}\left[\go^{\sigma(nab)}(-q,q',-p)V^\nu(q)\right]\Delta_{\nu\sigma}(q) \,, \nonumber\\
t_2^{(nab)}(p):=&\,i\!\int_q a(q')\textrm{Tr}\left[V^\sigma(q)\slashed{q}'\ge^{\nu(nab)}(q,p,-q')\right]\Delta_{\nu\sigma}(q)\,. \label{id_terms}
\end{align}

Evidently, repeating this argument but placing $\go^{\sigma(nab)}$ on the quark leg instead of the antiquark in \fig{fig:id_nab}, leads to another identity
\be
{\tilde t}_1^{(nab)}(p) = {\tilde t}_2^{(nab)}(p)\,, \label{id_nab2}
\ee
with
\begin{align}
{\tilde t}_1^{(nab)}(p) :=&\, i\!\int_q b(q')\textrm{Tr}\left[V^\nu(q)\go^{\sigma(nab)}(q,p,-q')\right]\Delta_{\nu\sigma}(q) \,, \nonumber\\
{\tilde t}_2^{(nab)}(p) :=&\, i\!\int_q a(q')\textrm{Tr}\left[\ge^{\nu(nab)}(-q,q',-p)\slashed{q}'V^\sigma(q)\right]\Delta_{\nu\sigma}(q)\,.
\end{align}
Invoking the charge conjugation transformation of the vertex (see, \eg Eq.~(2.7) of \cite{Aguilar:2024ciu}) and elementary properties of the trace, it is straightforward to show that
\be 
{\tilde t}_1^{(nab)}(p) = t_1^{(nab)}(p) \,.
\ee

\begin{figure}
    \centering
    \includegraphics[scale=1]{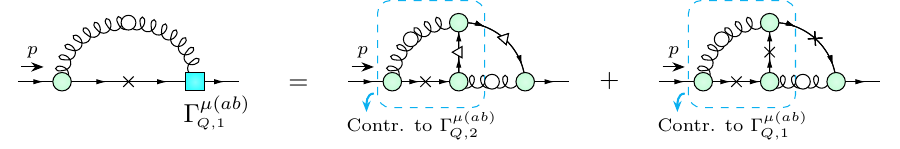}
    \caption{Diagrammatic representation of the integral in \ref{abelian_integral}.}
    \label{fig:not_id_ab}
\end{figure}

Note that a similar result \emph{does not} hold for the Abelian contribution, as shown diagrammatically in \fig{fig:not_id_ab}. In this case, the integral analogous to the l.h.s. of \1eq{id_nab}, namely
\be 
i\int_q b(q')\textrm{Tr}\left[\go^{\sigma(ab)}(-q,q',-p)V^\nu(q)\right]\Delta_{\nu\sigma}(q) \,, \label{abelian_integral}
\ee
with ``$ab$'' labeling the Abelian part of the quark-gluon vertex, corresponds to the diagram on the left of \fig{fig:not_id_ab}. Using the SDE diagrams 
of \figB{fig:1ld_green} 
for $\Gamma_{\!1,{\s Q}}^{\sigma(ab)}$, we get the two-loop diagrams on the right of that figure. There, the first diagram on the right can be seen to furnish a contribution to $\ge^{\nu(ab)}$. However, the second diagram generates a contribution to $\go^{\nu(ab)}$ instead.
Thus, the sum of the two terms enclosed in the 
blue boxes does not furnish the full $\ge^{\nu(ab)}$. 

Nevertheless, the quark-gluon vertex is known to be dominated by its non-Abelian diagram~\cite{Williams:2015cvx,Aguilar:2024ciu}. Consequently, \1eq{id_nab} should hold approximately for the quantum part of the vertex, \ie
\be
t_1(p) \approx  t_2(p) \,, \label{id_approx}
\ee
where
\begin{align} 
t_1(p) :=&\, i\int_q b(q')\textrm{Tr}\left[\Gamma_{\!\s Q,\s1}^\sigma(-q,q',-p)V^\nu(q)\right]\Delta_{\nu\sigma}(q) \,, \nonumber\\
t_2(p) :=&\, i\int_q a(q')\textrm{Tr}\left[V^\sigma(q)\slashed{q}'\Gamma_{\!\s Q,\s 2}^\nu(q,p,-q')\right]\Delta_{\nu\sigma}(q)\,.
\end{align}

\begin{figure}[!t]
    \centering
    \includegraphics[scale=0.75]{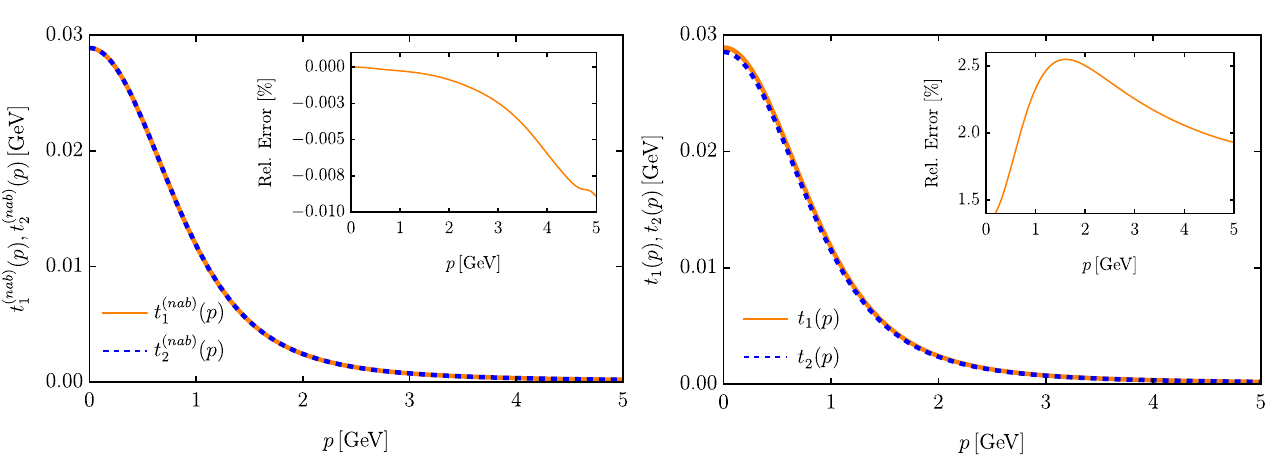}
    \caption{\textbf{\textit{Left panel}}: Integrals $t_1^{(nab)}(p)$ (orange continuous) and $t_2^{(nab)}(p)$ (blue dashed), appearing in the identity of \1eq{id_nab}, and the relative difference between them (inset). \textbf{\textit{Right panel}}: Analogous terms, $t_1(p)$ (orange continuous) and $t_2(p)$ (blue dashed), including the Abelian contribution. In this case, the relative difference is not compatible with zero. Nevertheless, \1eq{id_approx} holds within $2.5\%$.}
    \label{fig:ti}
\end{figure}

Then, using the inputs described in \sect{num} for $a(p)$, $b(p)$, and $V(p)$, we compute the $\Gamma_{\!i}^{\sigma(nab)}$, for $i = 1,2$, and use the latter to evaluate the $t_i^{(nab)}(p)$ through \1eq{id_terms}. The results are shown on the left panel of \fig{fig:ti}. There, the relative error, $[t_1^{(nab)}(p) - t_2^{(nab)}(p)]/t_1^{(nab)}(p)$, which is less than $10^{-4}$, is shown as an inset. Since the numerical precision of our numerical calculations is known to be of that same order, this error is compatible with zero. Thus, \1eq{id_nab} is verified. 

Next, we quantify the precision to which \1eq{id_approx} is satisfied when both Abelian and non-Abelian contributions are included in the vertex SDE. The results for the $t_i(p)$ are shown on the right panel of \fig{fig:ti}, and are barely distinguishable from each other, as well as from the $t_i^{(nab)}$ shown on the left panel, confirming that the non-Abelian contribution dominates the quark-gluon vertex. As a result, \1eq{id_approx} is satisfied within $2.5\%$, as seen on the inset. 

\bibliography{bibliography.bib}

\begin{thebibliography}{114}%
\makeatletter
\providecommand \@ifxundefined [1]{%
 \@ifx{#1\undefined}
}%
\providecommand \@ifnum [1]{%
 \ifnum #1\expandafter \@firstoftwo
 \else \expandafter \@secondoftwo
 \fi
}%
\providecommand \@ifx [1]{%
 \ifx #1\expandafter \@firstoftwo
 \else \expandafter \@secondoftwo
 \fi
}%
\providecommand \natexlab [1]{#1}%
\providecommand \enquote  [1]{``#1''}%
\providecommand \bibnamefont  [1]{#1}%
\providecommand \bibfnamefont [1]{#1}%
\providecommand \citenamefont [1]{#1}%
\providecommand \href@noop [0]{\@secondoftwo}%
\providecommand \href [0]{\begingroup \@sanitize@url \@href}%
\providecommand \@href[1]{\@@startlink{#1}\@@href}%
\providecommand \@@href[1]{\endgroup#1\@@endlink}%
\providecommand \@sanitize@url [0]{\catcode `\\12\catcode `\$12\catcode
  `\&12\catcode `\#12\catcode `\^12\catcode `\_12\catcode `\%12\relax}%
\providecommand \@@startlink[1]{}%
\providecommand \@@endlink[0]{}%
\providecommand \url  [0]{\begingroup\@sanitize@url \@url }%
\providecommand \@url [1]{\endgroup\@href {#1}{\urlprefix }}%
\providecommand \urlprefix  [0]{URL }%
\providecommand \Eprint [0]{\href }%
\providecommand \doibase [0]{https://doi.org/}%
\providecommand \selectlanguage [0]{\@gobble}%
\providecommand \bibinfo  [0]{\@secondoftwo}%
\providecommand \bibfield  [0]{\@secondoftwo}%
\providecommand \translation [1]{[#1]}%
\providecommand \BibitemOpen [0]{}%
\providecommand \bibitemStop [0]{}%
\providecommand \bibitemNoStop [0]{.\EOS\space}%
\providecommand \EOS [0]{\spacefactor3000\relax}%
\providecommand \BibitemShut  [1]{\csname bibitem#1\endcsname}%
\let\auto@bib@innerbib\@empty
\bibitem [{\citenamefont {Miramontes}\ \emph
  {et~al.}(2025{\natexlab{a}})\citenamefont {Miramontes}, \citenamefont
  {Morgado}, \citenamefont {Papavassiliou},\ and\ \citenamefont
  {Pawlowski}}]{Miramontes:2025imd}%
  \BibitemOpen
  \bibfield  {author} {\bibinfo {author} {\bibfnamefont {A.~S.}\ \bibnamefont
  {Miramontes}}, \bibinfo {author} {\bibfnamefont {J.~M.}\ \bibnamefont
  {Morgado}}, \bibinfo {author} {\bibfnamefont {J.}~\bibnamefont
  {Papavassiliou}},\ and\ \bibinfo {author} {\bibfnamefont {J.~M.}\
  \bibnamefont {Pawlowski}},\ }\href
  {https://doi.org/10.1140/epjc/s10052-025-14774-x} {\bibfield  {journal}
  {\bibinfo  {journal} {Eur. Phys. J. C}\ }\textbf {\bibinfo {volume} {85}},\
  \bibinfo {pages} {1055} (\bibinfo {year} {2025}{\natexlab{a}})}\BibitemShut
  {NoStop}%
\bibitem [{\citenamefont {Munczek}(1995)}]{Munczek:1994zz}%
  \BibitemOpen
  \bibfield  {author} {\bibinfo {author} {\bibfnamefont {H.}~\bibnamefont
  {Munczek}},\ }\href {https://doi.org/10.1103/PhysRevD.52.4736} {\bibfield
  {journal} {\bibinfo  {journal} {Phys. Rev.}\ }\textbf {\bibinfo {volume}
  {D52}},\ \bibinfo {pages} {4736} (\bibinfo {year} {1995})}\BibitemShut
  {NoStop}%
\bibitem [{\citenamefont {Matevosyan}\ \emph {et~al.}(2007)\citenamefont
  {Matevosyan}, \citenamefont {Thomas},\ and\ \citenamefont
  {Tandy}}]{Matevosyan:2006bk}%
  \BibitemOpen
  \bibfield  {author} {\bibinfo {author} {\bibfnamefont {H.~H.}\ \bibnamefont
  {Matevosyan}}, \bibinfo {author} {\bibfnamefont {A.~W.}\ \bibnamefont
  {Thomas}},\ and\ \bibinfo {author} {\bibfnamefont {P.~C.}\ \bibnamefont
  {Tandy}},\ }\href {https://doi.org/10.1103/PhysRevC.75.045201} {\bibfield
  {journal} {\bibinfo  {journal} {Phys. Rev.}\ }\textbf {\bibinfo {volume}
  {C75}},\ \bibinfo {pages} {045201} (\bibinfo {year} {2007})}\BibitemShut
  {NoStop}%
\bibitem [{\citenamefont {Fischer}\ \emph {et~al.}(2007)\citenamefont
  {Fischer}, \citenamefont {Nickel},\ and\ \citenamefont
  {Wambach}}]{Fischer:2007ze}%
  \BibitemOpen
  \bibfield  {author} {\bibinfo {author} {\bibfnamefont {C.~S.}\ \bibnamefont
  {Fischer}}, \bibinfo {author} {\bibfnamefont {D.}~\bibnamefont {Nickel}},\
  and\ \bibinfo {author} {\bibfnamefont {J.}~\bibnamefont {Wambach}},\ }\href
  {https://doi.org/10.1103/PhysRevD.76.094009} {\bibfield  {journal} {\bibinfo
  {journal} {Phys. Rev. D}\ }\textbf {\bibinfo {volume} {76}},\ \bibinfo
  {pages} {094009} (\bibinfo {year} {2007})}\BibitemShut {NoStop}%
\bibitem [{\citenamefont {Fischer}\ and\ \citenamefont
  {Williams}(2008)}]{Fischer:2008wy}%
  \BibitemOpen
  \bibfield  {author} {\bibinfo {author} {\bibfnamefont {C.~S.}\ \bibnamefont
  {Fischer}}\ and\ \bibinfo {author} {\bibfnamefont {R.}~\bibnamefont
  {Williams}},\ }\href {https://doi.org/10.1103/PhysRevD.78.074006} {\bibfield
  {journal} {\bibinfo  {journal} {Phys. Rev. D}\ }\textbf {\bibinfo {volume}
  {78}},\ \bibinfo {pages} {074006} (\bibinfo {year} {2008})}\BibitemShut
  {NoStop}%
\bibitem [{\citenamefont {Fischer}\ and\ \citenamefont
  {Williams}(2009)}]{Fischer:2009jm}%
  \BibitemOpen
  \bibfield  {author} {\bibinfo {author} {\bibfnamefont {C.~S.}\ \bibnamefont
  {Fischer}}\ and\ \bibinfo {author} {\bibfnamefont {R.}~\bibnamefont
  {Williams}},\ }\href {https://doi.org/10.1103/PhysRevLett.103.122001}
  {\bibfield  {journal} {\bibinfo  {journal} {Phys. Rev. Lett.}\ }\textbf
  {\bibinfo {volume} {103}},\ \bibinfo {pages} {122001} (\bibinfo {year}
  {2009})}\BibitemShut {NoStop}%
\bibitem [{\citenamefont {Chang}\ and\ \citenamefont
  {Roberts}(2009)}]{Chang:2009zb}%
  \BibitemOpen
  \bibfield  {author} {\bibinfo {author} {\bibfnamefont {L.}~\bibnamefont
  {Chang}}\ and\ \bibinfo {author} {\bibfnamefont {C.~D.}\ \bibnamefont
  {Roberts}},\ }\href {https://doi.org/10.1103/PhysRevLett.103.081601}
  {\bibfield  {journal} {\bibinfo  {journal} {Phys. Rev. Lett.}\ }\textbf
  {\bibinfo {volume} {103}},\ \bibinfo {pages} {081601} (\bibinfo {year}
  {2009})}\BibitemShut {NoStop}%
\bibitem [{\citenamefont {Sanchis-Alepuz}\ and\ \citenamefont
  {Williams}(2015{\natexlab{a}})}]{Sanchis-Alepuz:2015tha}%
  \BibitemOpen
  \bibfield  {author} {\bibinfo {author} {\bibfnamefont {H.}~\bibnamefont
  {Sanchis-Alepuz}}\ and\ \bibinfo {author} {\bibfnamefont {R.}~\bibnamefont
  {Williams}},\ }\href {https://doi.org/10.1088/1742-6596/631/1/012064}
  {\bibfield  {journal} {\bibinfo  {journal} {J. Phys. Conf. Ser.}\ }\textbf
  {\bibinfo {volume} {631}},\ \bibinfo {pages} {012064} (\bibinfo {year}
  {2015}{\natexlab{a}})}\BibitemShut {NoStop}%
\bibitem [{\citenamefont {Williams}(2015)}]{Williams:2014iea}%
  \BibitemOpen
  \bibfield  {author} {\bibinfo {author} {\bibfnamefont {R.}~\bibnamefont
  {Williams}},\ }\href {https://doi.org/10.1140/epja/i2015-15057-4} {\bibfield
  {journal} {\bibinfo  {journal} {Eur. Phys. J.}\ }\textbf {\bibinfo {volume}
  {A51}},\ \bibinfo {pages} {57} (\bibinfo {year} {2015})}\BibitemShut
  {NoStop}%
\bibitem [{\citenamefont {Heupel}\ \emph {et~al.}(2014)\citenamefont {Heupel},
  \citenamefont {Goecke},\ and\ \citenamefont {Fischer}}]{Heupel:2014ina}%
  \BibitemOpen
  \bibfield  {author} {\bibinfo {author} {\bibfnamefont {W.}~\bibnamefont
  {Heupel}}, \bibinfo {author} {\bibfnamefont {T.}~\bibnamefont {Goecke}},\
  and\ \bibinfo {author} {\bibfnamefont {C.~S.}\ \bibnamefont {Fischer}},\
  }\href {https://doi.org/10.1140/epja/i2014-14085-x} {\bibfield  {journal}
  {\bibinfo  {journal} {Eur. Phys. J.}\ }\textbf {\bibinfo {volume} {A50}},\
  \bibinfo {pages} {85} (\bibinfo {year} {2014})}\BibitemShut {NoStop}%
\bibitem [{\citenamefont {Sanchis-Alepuz}\ \emph {et~al.}(2014)\citenamefont
  {Sanchis-Alepuz}, \citenamefont {Fischer},\ and\ \citenamefont
  {Kubrak}}]{Sanchis-Alepuz:2014wea}%
  \BibitemOpen
  \bibfield  {author} {\bibinfo {author} {\bibfnamefont {H.}~\bibnamefont
  {Sanchis-Alepuz}}, \bibinfo {author} {\bibfnamefont {C.~S.}\ \bibnamefont
  {Fischer}},\ and\ \bibinfo {author} {\bibfnamefont {S.}~\bibnamefont
  {Kubrak}},\ }\href {https://doi.org/10.1016/j.physletb.2014.04.031}
  {\bibfield  {journal} {\bibinfo  {journal} {Phys. Lett. B}\ }\textbf
  {\bibinfo {volume} {733}},\ \bibinfo {pages} {151} (\bibinfo {year}
  {2014})}\BibitemShut {NoStop}%
\bibitem [{\citenamefont {Williams}\ \emph {et~al.}(2016)\citenamefont
  {Williams}, \citenamefont {Fischer},\ and\ \citenamefont
  {Heupel}}]{Williams:2015cvx}%
  \BibitemOpen
  \bibfield  {author} {\bibinfo {author} {\bibfnamefont {R.}~\bibnamefont
  {Williams}}, \bibinfo {author} {\bibfnamefont {C.~S.}\ \bibnamefont
  {Fischer}},\ and\ \bibinfo {author} {\bibfnamefont {W.}~\bibnamefont
  {Heupel}},\ }\href {https://doi.org/10.1103/PhysRevD.93.034026} {\bibfield
  {journal} {\bibinfo  {journal} {Phys. Rev.}\ }\textbf {\bibinfo {volume}
  {D93}},\ \bibinfo {pages} {034026} (\bibinfo {year} {2016})}\BibitemShut
  {NoStop}%
\bibitem [{\citenamefont {Sanchis-Alepuz}\ and\ \citenamefont
  {Williams}(2015{\natexlab{b}})}]{Sanchis-Alepuz:2015qra}%
  \BibitemOpen
  \bibfield  {author} {\bibinfo {author} {\bibfnamefont {H.}~\bibnamefont
  {Sanchis-Alepuz}}\ and\ \bibinfo {author} {\bibfnamefont {R.}~\bibnamefont
  {Williams}},\ }\href {https://doi.org/10.1016/j.physletb.2015.08.067}
  {\bibfield  {journal} {\bibinfo  {journal} {Phys. Lett.}\ }\textbf {\bibinfo
  {volume} {B749}},\ \bibinfo {pages} {592} (\bibinfo {year}
  {2015}{\natexlab{b}})}\BibitemShut {NoStop}%
\bibitem [{\citenamefont {Binosi}\ \emph {et~al.}(2016)\citenamefont {Binosi},
  \citenamefont {Chang}, \citenamefont {Papavassiliou}, \citenamefont {Qin},\
  and\ \citenamefont {Roberts}}]{Binosi:2016rxz}%
  \BibitemOpen
  \bibfield  {author} {\bibinfo {author} {\bibfnamefont {D.}~\bibnamefont
  {Binosi}}, \bibinfo {author} {\bibfnamefont {L.}~\bibnamefont {Chang}},
  \bibinfo {author} {\bibfnamefont {J.}~\bibnamefont {Papavassiliou}}, \bibinfo
  {author} {\bibfnamefont {S.-X.}\ \bibnamefont {Qin}},\ and\ \bibinfo {author}
  {\bibfnamefont {C.~D.}\ \bibnamefont {Roberts}},\ }\href
  {https://doi.org/10.1103/PhysRevD.93.096010} {\bibfield  {journal} {\bibinfo
  {journal} {Phys. Rev.}\ }\textbf {\bibinfo {volume} {D93}},\ \bibinfo {pages}
  {096010} (\bibinfo {year} {2016})}\BibitemShut {NoStop}%
\bibitem [{\citenamefont {Williams}(2019)}]{Williams:2018adr}%
  \BibitemOpen
  \bibfield  {author} {\bibinfo {author} {\bibfnamefont {R.}~\bibnamefont
  {Williams}},\ }\href {https://doi.org/10.1016/j.physletb.2019.134943}
  {\bibfield  {journal} {\bibinfo  {journal} {Phys. Lett. B}\ }\textbf
  {\bibinfo {volume} {798}},\ \bibinfo {pages} {134943} (\bibinfo {year}
  {2019})}\BibitemShut {NoStop}%
\bibitem [{\citenamefont {Miramontes}\ \emph {et~al.}(2021)\citenamefont
  {Miramontes}, \citenamefont {Sanchis~Alepuz},\ and\ \citenamefont
  {Alkofer}}]{Miramontes:2021xgn}%
  \BibitemOpen
  \bibfield  {author} {\bibinfo {author} {\bibfnamefont {{\'A}.~S.}\
  \bibnamefont {Miramontes}}, \bibinfo {author} {\bibfnamefont
  {H.}~\bibnamefont {Sanchis~Alepuz}},\ and\ \bibinfo {author} {\bibfnamefont
  {R.}~\bibnamefont {Alkofer}},\ }\href
  {https://doi.org/10.1103/PhysRevD.103.116006} {\bibfield  {journal} {\bibinfo
   {journal} {Phys. Rev. D}\ }\textbf {\bibinfo {volume} {103}},\ \bibinfo
  {pages} {116006} (\bibinfo {year} {2021})}\BibitemShut {NoStop}%
\bibitem [{\citenamefont {Santowsky}\ \emph {et~al.}(2020)\citenamefont
  {Santowsky}, \citenamefont {Eichmann}, \citenamefont {Fischer}, \citenamefont
  {Wallbott},\ and\ \citenamefont {Williams}}]{Santowsky:2020pwd}%
  \BibitemOpen
  \bibfield  {author} {\bibinfo {author} {\bibfnamefont {N.}~\bibnamefont
  {Santowsky}}, \bibinfo {author} {\bibfnamefont {G.}~\bibnamefont {Eichmann}},
  \bibinfo {author} {\bibfnamefont {C.~S.}\ \bibnamefont {Fischer}}, \bibinfo
  {author} {\bibfnamefont {P.~C.}\ \bibnamefont {Wallbott}},\ and\ \bibinfo
  {author} {\bibfnamefont {R.}~\bibnamefont {Williams}},\ }\href
  {https://doi.org/10.1103/PhysRevD.102.056014} {\bibfield  {journal} {\bibinfo
   {journal} {Phys. Rev. D}\ }\textbf {\bibinfo {volume} {102}},\ \bibinfo
  {pages} {056014} (\bibinfo {year} {2020})}\BibitemShut {NoStop}%
\bibitem [{\citenamefont {Miramontes}\ \emph {et~al.}(2022)\citenamefont
  {Miramontes}, \citenamefont {Alkofer}, \citenamefont {Fischer},\ and\
  \citenamefont {Sanchis-Alepuz}}]{Miramontes:2022mex}%
  \BibitemOpen
  \bibfield  {author} {\bibinfo {author} {\bibfnamefont {{\'A}.~S.}\
  \bibnamefont {Miramontes}}, \bibinfo {author} {\bibfnamefont
  {R.}~\bibnamefont {Alkofer}}, \bibinfo {author} {\bibfnamefont {C.~S.}\
  \bibnamefont {Fischer}},\ and\ \bibinfo {author} {\bibfnamefont
  {H.}~\bibnamefont {Sanchis-Alepuz}},\ }\href
  {https://doi.org/10.1016/j.physletb.2022.137291} {\bibfield  {journal}
  {\bibinfo  {journal} {Phys. Lett. B}\ }\textbf {\bibinfo {volume} {833}},\
  \bibinfo {pages} {137291} (\bibinfo {year} {2022})}\BibitemShut {NoStop}%
\bibitem [{\citenamefont {Gao}\ \emph {et~al.}(2025)\citenamefont {Gao},
  \citenamefont {Miramontes}, \citenamefont {Papavassiliou},\ and\
  \citenamefont {Pawlowski}}]{Gao:2024gdj}%
  \BibitemOpen
  \bibfield  {author} {\bibinfo {author} {\bibfnamefont {F.}~\bibnamefont
  {Gao}}, \bibinfo {author} {\bibfnamefont {A.~S.}\ \bibnamefont {Miramontes}},
  \bibinfo {author} {\bibfnamefont {J.}~\bibnamefont {Papavassiliou}},\ and\
  \bibinfo {author} {\bibfnamefont {J.~M.}\ \bibnamefont {Pawlowski}},\ }\href
  {https://doi.org/10.1016/j.physletb.2025.139384} {\bibfield  {journal}
  {\bibinfo  {journal} {Phys. Lett. B}\ }\textbf {\bibinfo {volume} {863}},\
  \bibinfo {pages} {139384} (\bibinfo {year} {2025})}\BibitemShut {NoStop}%
\bibitem [{\citenamefont {Miramontes}\ \emph
  {et~al.}(2025{\natexlab{b}})\citenamefont {Miramontes}, \citenamefont
  {Eichmann},\ and\ \citenamefont {Alkofer}}]{Miramontes:2025ofw}%
  \BibitemOpen
  \bibfield  {author} {\bibinfo {author} {\bibfnamefont {A.~S.}\ \bibnamefont
  {Miramontes}}, \bibinfo {author} {\bibfnamefont {G.}~\bibnamefont
  {Eichmann}},\ and\ \bibinfo {author} {\bibfnamefont {R.}~\bibnamefont
  {Alkofer}},\ }\href {https://doi.org/10.1016/j.physletb.2025.139659}
  {\bibfield  {journal} {\bibinfo  {journal} {Phys. Lett. B}\ }\textbf
  {\bibinfo {volume} {868}},\ \bibinfo {pages} {139659} (\bibinfo {year}
  {2025}{\natexlab{b}})}\BibitemShut {NoStop}%
\bibitem [{\citenamefont {Fu}\ \emph {et~al.}(2025)\citenamefont {Fu},
  \citenamefont {Huang}, \citenamefont {Pawlowski}, \citenamefont {Tan},\ and\
  \citenamefont {Zhou}}]{Fu:2025hcm}%
  \BibitemOpen
  \bibfield  {author} {\bibinfo {author} {\bibfnamefont {W.-j.}\ \bibnamefont
  {Fu}}, \bibinfo {author} {\bibfnamefont {C.}~\bibnamefont {Huang}}, \bibinfo
  {author} {\bibfnamefont {J.~M.}\ \bibnamefont {Pawlowski}}, \bibinfo {author}
  {\bibfnamefont {Y.-y.}\ \bibnamefont {Tan}},\ and\ \bibinfo {author}
  {\bibfnamefont {L.-j.}\ \bibnamefont {Zhou}},\ }\href
  {https://doi.org/10.1103/4sh5-w4yc} {\bibfield  {journal} {\bibinfo
  {journal} {Phys. Rev. D}\ }\textbf {\bibinfo {volume} {112}},\ \bibinfo
  {pages} {054047} (\bibinfo {year} {2025})},\ \Eprint
  {https://arxiv.org/abs/2502.14388} {arXiv:2502.14388 [hep-ph]} \BibitemShut
  {NoStop}%
\bibitem [{\citenamefont {Huber}\ \emph {et~al.}(2025)\citenamefont {Huber},
  \citenamefont {Fischer},\ and\ \citenamefont
  {Sanchis-Alepuz}}]{Huber:2025kwy}%
  \BibitemOpen
  \bibfield  {author} {\bibinfo {author} {\bibfnamefont {M.~Q.}\ \bibnamefont
  {Huber}}, \bibinfo {author} {\bibfnamefont {C.~S.}\ \bibnamefont {Fischer}},\
  and\ \bibinfo {author} {\bibfnamefont {H.}~\bibnamefont {Sanchis-Alepuz}},\
  }\href {https://doi.org/10.1140/epjc/s10052-025-14590-3} {\bibfield
  {journal} {\bibinfo  {journal} {Eur. Phys. J. C}\ }\textbf {\bibinfo {volume}
  {85}},\ \bibinfo {pages} {859} (\bibinfo {year} {2025})},\ \Eprint
  {https://arxiv.org/abs/2503.03821} {arXiv:2503.03821 [hep-ph]} \BibitemShut
  {NoStop}%
\bibitem [{\citenamefont {Miramontes}\ \emph
  {et~al.}(2025{\natexlab{c}})\citenamefont {Miramontes}, \citenamefont
  {Papavassiliou},\ and\ \citenamefont {Pawlowski}}]{Miramontes:2025vzb}%
  \BibitemOpen
  \bibfield  {author} {\bibinfo {author} {\bibfnamefont {A.~S.}\ \bibnamefont
  {Miramontes}}, \bibinfo {author} {\bibfnamefont {J.}~\bibnamefont
  {Papavassiliou}},\ and\ \bibinfo {author} {\bibfnamefont {J.~M.}\
  \bibnamefont {Pawlowski}},\ }\Eprint {https://arxiv.org/abs/2508.20631}
  {arXiv:2508.20631 [hep-ph]}  (\bibinfo {year}
  {2025}{\natexlab{c}})\BibitemShut {NoStop}%
\bibitem [{\citenamefont {Alkofer}\ and\ \citenamefont {von
  Smekal}(2001)}]{Alkofer:2000wg}%
  \BibitemOpen
  \bibfield  {author} {\bibinfo {author} {\bibfnamefont {R.}~\bibnamefont
  {Alkofer}}\ and\ \bibinfo {author} {\bibfnamefont {L.}~\bibnamefont {von
  Smekal}},\ }\href {https://doi.org/10.1016/S0370-1573(01)00010-2} {\bibfield
  {journal} {\bibinfo  {journal} {Phys. Rept.}\ }\textbf {\bibinfo {volume}
  {353}},\ \bibinfo {pages} {281} (\bibinfo {year} {2001})}\BibitemShut
  {NoStop}%
\bibitem [{\citenamefont {Maris}\ and\ \citenamefont
  {Roberts}(2003)}]{Maris:2003vk}%
  \BibitemOpen
  \bibfield  {author} {\bibinfo {author} {\bibfnamefont {P.}~\bibnamefont
  {Maris}}\ and\ \bibinfo {author} {\bibfnamefont {C.~D.}\ \bibnamefont
  {Roberts}},\ }\href {https://doi.org/10.1142/S0218301303001326} {\bibfield
  {journal} {\bibinfo  {journal} {Int. J. Mod. Phys.}\ }\textbf {\bibinfo
  {volume} {E12}},\ \bibinfo {pages} {297} (\bibinfo {year}
  {2003})}\BibitemShut {NoStop}%
\bibitem [{\citenamefont {Fischer}(2006)}]{Fischer:2006ub}%
  \BibitemOpen
  \bibfield  {author} {\bibinfo {author} {\bibfnamefont {C.~S.}\ \bibnamefont
  {Fischer}},\ }\href {https://doi.org/10.1088/0954-3899/32/8/R02} {\bibfield
  {journal} {\bibinfo  {journal} {J. Phys. G}\ }\textbf {\bibinfo {volume}
  {32}},\ \bibinfo {pages} {R253} (\bibinfo {year} {2006})}\BibitemShut
  {NoStop}%
\bibitem [{\citenamefont {Binosi}\ and\ \citenamefont
  {Papavassiliou}(2009)}]{Binosi:2009qm}%
  \BibitemOpen
  \bibfield  {author} {\bibinfo {author} {\bibfnamefont {D.}~\bibnamefont
  {Binosi}}\ and\ \bibinfo {author} {\bibfnamefont {J.}~\bibnamefont
  {Papavassiliou}},\ }\href {https://doi.org/10.1016/j.physrep.2009.05.001}
  {\bibfield  {journal} {\bibinfo  {journal} {Phys. Rept.}\ }\textbf {\bibinfo
  {volume} {479}},\ \bibinfo {pages} {1} (\bibinfo {year} {2009})}\BibitemShut
  {NoStop}%
\bibitem [{\citenamefont {Maas}(2013)}]{Maas:2011se}%
  \BibitemOpen
  \bibfield  {author} {\bibinfo {author} {\bibfnamefont {A.}~\bibnamefont
  {Maas}},\ }\href {https://doi.org/10.1016/j.physrep.2012.11.002} {\bibfield
  {journal} {\bibinfo  {journal} {Phys. Rept.}\ }\textbf {\bibinfo {volume}
  {524}},\ \bibinfo {pages} {203} (\bibinfo {year} {2013})}\BibitemShut
  {NoStop}%
\bibitem [{\citenamefont {Bashir}\ \emph {et~al.}(2012)\citenamefont {Bashir},
  \citenamefont {Chang}, \citenamefont {Cloet}, \citenamefont {El-Bennich},
  \citenamefont {Liu} \emph {et~al.}}]{Bashir:2012fs}%
  \BibitemOpen
  \bibfield  {author} {\bibinfo {author} {\bibfnamefont {A.}~\bibnamefont
  {Bashir}}, \bibinfo {author} {\bibfnamefont {L.}~\bibnamefont {Chang}},
  \bibinfo {author} {\bibfnamefont {I.~C.}\ \bibnamefont {Cloet}}, \bibinfo
  {author} {\bibfnamefont {B.}~\bibnamefont {El-Bennich}}, \bibinfo {author}
  {\bibfnamefont {Y.-X.}\ \bibnamefont {Liu}}, \emph {et~al.},\ }\href
  {https://doi.org/10.1088/0253-6102/58/1/16} {\bibfield  {journal} {\bibinfo
  {journal} {Commun. Theor. Phys.}\ }\textbf {\bibinfo {volume} {58}},\
  \bibinfo {pages} {79} (\bibinfo {year} {2012})}\BibitemShut {NoStop}%
\bibitem [{\citenamefont {Cloet}\ and\ \citenamefont
  {Roberts}(2014)}]{Cloet:2013jya}%
  \BibitemOpen
  \bibfield  {author} {\bibinfo {author} {\bibfnamefont {I.~C.}\ \bibnamefont
  {Cloet}}\ and\ \bibinfo {author} {\bibfnamefont {C.~D.}\ \bibnamefont
  {Roberts}},\ }\href {https://doi.org/10.1016/j.ppnp.2014.02.001} {\bibfield
  {journal} {\bibinfo  {journal} {Prog. Part. Nucl. Phys.}\ }\textbf {\bibinfo
  {volume} {77}},\ \bibinfo {pages} {1} (\bibinfo {year} {2014})}\BibitemShut
  {NoStop}%
\bibitem [{\citenamefont {Eichmann}\ \emph
  {et~al.}(2016{\natexlab{a}})\citenamefont {Eichmann}, \citenamefont
  {Sanchis-Alepuz}, \citenamefont {Williams}, \citenamefont {Alkofer},\ and\
  \citenamefont {Fischer}}]{Eichmann:2016yit}%
  \BibitemOpen
  \bibfield  {author} {\bibinfo {author} {\bibfnamefont {G.}~\bibnamefont
  {Eichmann}}, \bibinfo {author} {\bibfnamefont {H.}~\bibnamefont
  {Sanchis-Alepuz}}, \bibinfo {author} {\bibfnamefont {R.}~\bibnamefont
  {Williams}}, \bibinfo {author} {\bibfnamefont {R.}~\bibnamefont {Alkofer}},\
  and\ \bibinfo {author} {\bibfnamefont {C.~S.}\ \bibnamefont {Fischer}},\
  }\href {https://doi.org/10.1016/j.ppnp.2016.07.001} {\bibfield  {journal}
  {\bibinfo  {journal} {Prog. Part. Nucl. Phys.}\ }\textbf {\bibinfo {volume}
  {91}},\ \bibinfo {pages} {1} (\bibinfo {year}
  {2016}{\natexlab{a}})}\BibitemShut {NoStop}%
\bibitem [{\citenamefont {Fischer}(2019)}]{Fischer:2018sdj}%
  \BibitemOpen
  \bibfield  {author} {\bibinfo {author} {\bibfnamefont {C.~S.}\ \bibnamefont
  {Fischer}},\ }\href {https://doi.org/10.1016/j.ppnp.2019.01.002} {\bibfield
  {journal} {\bibinfo  {journal} {Prog. Part. Nucl. Phys.}\ }\textbf {\bibinfo
  {volume} {105}},\ \bibinfo {pages} {1} (\bibinfo {year} {2019})}\BibitemShut
  {NoStop}%
\bibitem [{\citenamefont {Huber}(2020)}]{Huber:2018ned}%
  \BibitemOpen
  \bibfield  {author} {\bibinfo {author} {\bibfnamefont {M.~Q.}\ \bibnamefont
  {Huber}},\ }\href {https://doi.org/10.1016/j.physrep.2020.04.004} {\bibfield
  {journal} {\bibinfo  {journal} {Phys. Rept.}\ }\textbf {\bibinfo {volume}
  {879}},\ \bibinfo {pages} {1} (\bibinfo {year} {2020})}\BibitemShut {NoStop}%
\bibitem [{\citenamefont {Ferreira}\ and\ \citenamefont
  {Papavassiliou}(2023)}]{Ferreira:2023fva}%
  \BibitemOpen
  \bibfield  {author} {\bibinfo {author} {\bibfnamefont {M.~N.}\ \bibnamefont
  {Ferreira}}\ and\ \bibinfo {author} {\bibfnamefont {J.}~\bibnamefont
  {Papavassiliou}},\ }\href {https://doi.org/10.3390/particles6010017}
  {\bibfield  {journal} {\bibinfo  {journal} {Particles}\ }\textbf {\bibinfo
  {volume} {6}},\ \bibinfo {pages} {312} (\bibinfo {year} {2023})}\BibitemShut
  {NoStop}%
\bibitem [{\citenamefont {Bender}\ \emph {et~al.}(2002)\citenamefont {Bender},
  \citenamefont {Detmold}, \citenamefont {Roberts},\ and\ \citenamefont
  {Thomas}}]{Bender:2002as}%
  \BibitemOpen
  \bibfield  {author} {\bibinfo {author} {\bibfnamefont {A.}~\bibnamefont
  {Bender}}, \bibinfo {author} {\bibfnamefont {W.}~\bibnamefont {Detmold}},
  \bibinfo {author} {\bibfnamefont {C.}~\bibnamefont {Roberts}},\ and\ \bibinfo
  {author} {\bibfnamefont {A.~W.}\ \bibnamefont {Thomas}},\ }\href
  {https://doi.org/10.1103/PhysRevC.65.065203} {\bibfield  {journal} {\bibinfo
  {journal} {Phys. Rev.}\ }\textbf {\bibinfo {volume} {C65}},\ \bibinfo {pages}
  {065203} (\bibinfo {year} {2002})}\BibitemShut {NoStop}%
\bibitem [{\citenamefont {Maris}\ and\ \citenamefont
  {Tandy}(1999)}]{Maris:1999nt}%
  \BibitemOpen
  \bibfield  {author} {\bibinfo {author} {\bibfnamefont {P.}~\bibnamefont
  {Maris}}\ and\ \bibinfo {author} {\bibfnamefont {P.~C.}\ \bibnamefont
  {Tandy}},\ }\href {https://doi.org/10.1103/PhysRevC.60.055214} {\bibfield
  {journal} {\bibinfo  {journal} {Phys. Rev.}\ }\textbf {\bibinfo {volume}
  {C60}},\ \bibinfo {pages} {055214} (\bibinfo {year} {1999})}\BibitemShut
  {NoStop}%
\bibitem [{\citenamefont {Maris}\ and\ \citenamefont
  {Tandy}(2000)}]{Maris:1999bh}%
  \BibitemOpen
  \bibfield  {author} {\bibinfo {author} {\bibfnamefont {P.}~\bibnamefont
  {Maris}}\ and\ \bibinfo {author} {\bibfnamefont {P.~C.}\ \bibnamefont
  {Tandy}},\ }\href {https://doi.org/10.1103/PhysRevC.61.045202} {\bibfield
  {journal} {\bibinfo  {journal} {Phys. Rev. C}\ }\textbf {\bibinfo {volume}
  {61}},\ \bibinfo {pages} {045202} (\bibinfo {year} {2000})}\BibitemShut
  {NoStop}%
\bibitem [{\citenamefont {Alkofer}\ \emph {et~al.}(2002)\citenamefont
  {Alkofer}, \citenamefont {Watson},\ and\ \citenamefont
  {Weigel}}]{Alkofer:2002bp}%
  \BibitemOpen
  \bibfield  {author} {\bibinfo {author} {\bibfnamefont {R.}~\bibnamefont
  {Alkofer}}, \bibinfo {author} {\bibfnamefont {P.}~\bibnamefont {Watson}},\
  and\ \bibinfo {author} {\bibfnamefont {H.}~\bibnamefont {Weigel}},\ }\href
  {https://doi.org/10.1103/PhysRevD.65.094026} {\bibfield  {journal} {\bibinfo
  {journal} {Phys. Rev. D}\ }\textbf {\bibinfo {volume} {65}},\ \bibinfo
  {pages} {094026} (\bibinfo {year} {2002})}\BibitemShut {NoStop}%
\bibitem [{\citenamefont {Eichmann}\ \emph
  {et~al.}(2008{\natexlab{a}})\citenamefont {Eichmann}, \citenamefont
  {Alkofer}, \citenamefont {Cloet}, \citenamefont {Krassnigg},\ and\
  \citenamefont {Roberts}}]{Eichmann:2008ae}%
  \BibitemOpen
  \bibfield  {author} {\bibinfo {author} {\bibfnamefont {G.}~\bibnamefont
  {Eichmann}}, \bibinfo {author} {\bibfnamefont {R.}~\bibnamefont {Alkofer}},
  \bibinfo {author} {\bibfnamefont {I.~C.}\ \bibnamefont {Cloet}}, \bibinfo
  {author} {\bibfnamefont {A.}~\bibnamefont {Krassnigg}},\ and\ \bibinfo
  {author} {\bibfnamefont {C.~D.}\ \bibnamefont {Roberts}},\ }\href
  {https://doi.org/10.1103/PhysRevC.77.042202} {\bibfield  {journal} {\bibinfo
  {journal} {Phys. Rev. C}\ }\textbf {\bibinfo {volume} {77}},\ \bibinfo
  {pages} {042202} (\bibinfo {year} {2008}{\natexlab{a}})}\BibitemShut
  {NoStop}%
\bibitem [{\citenamefont {Qin}\ \emph {et~al.}(2011)\citenamefont {Qin},
  \citenamefont {Chang}, \citenamefont {Liu}, \citenamefont {Roberts},\ and\
  \citenamefont {Wilson}}]{Qin:2011dd}%
  \BibitemOpen
  \bibfield  {author} {\bibinfo {author} {\bibfnamefont {S.-x.}\ \bibnamefont
  {Qin}}, \bibinfo {author} {\bibfnamefont {L.}~\bibnamefont {Chang}}, \bibinfo
  {author} {\bibfnamefont {Y.-x.}\ \bibnamefont {Liu}}, \bibinfo {author}
  {\bibfnamefont {C.~D.}\ \bibnamefont {Roberts}},\ and\ \bibinfo {author}
  {\bibfnamefont {D.~J.}\ \bibnamefont {Wilson}},\ }\href
  {https://doi.org/10.1103/PhysRevC.84.042202} {\bibfield  {journal} {\bibinfo
  {journal} {Phys. Rev.}\ }\textbf {\bibinfo {volume} {C84}},\ \bibinfo {pages}
  {042202} (\bibinfo {year} {2011})}\BibitemShut {NoStop}%
\bibitem [{\citenamefont {Hilger}\ \emph
  {et~al.}(2015{\natexlab{a}})\citenamefont {Hilger}, \citenamefont {Popovici},
  \citenamefont {Gomez-Rocha},\ and\ \citenamefont
  {Krassnigg}}]{Hilger:2014nma}%
  \BibitemOpen
  \bibfield  {author} {\bibinfo {author} {\bibfnamefont {T.}~\bibnamefont
  {Hilger}}, \bibinfo {author} {\bibfnamefont {C.}~\bibnamefont {Popovici}},
  \bibinfo {author} {\bibfnamefont {M.}~\bibnamefont {Gomez-Rocha}},\ and\
  \bibinfo {author} {\bibfnamefont {A.}~\bibnamefont {Krassnigg}},\ }\href
  {https://doi.org/10.1103/PhysRevD.91.034013} {\bibfield  {journal} {\bibinfo
  {journal} {Phys. Rev. D}\ }\textbf {\bibinfo {volume} {91}},\ \bibinfo
  {pages} {034013} (\bibinfo {year} {2015}{\natexlab{a}})}\BibitemShut
  {NoStop}%
\bibitem [{\citenamefont {Heupel}\ \emph {et~al.}(2012)\citenamefont {Heupel},
  \citenamefont {Eichmann},\ and\ \citenamefont {Fischer}}]{Heupel:2012ua}%
  \BibitemOpen
  \bibfield  {author} {\bibinfo {author} {\bibfnamefont {W.}~\bibnamefont
  {Heupel}}, \bibinfo {author} {\bibfnamefont {G.}~\bibnamefont {Eichmann}},\
  and\ \bibinfo {author} {\bibfnamefont {C.~S.}\ \bibnamefont {Fischer}},\
  }\href {https://doi.org/10.1016/j.physletb.2012.11.009} {\bibfield  {journal}
  {\bibinfo  {journal} {Phys. Lett. B}\ }\textbf {\bibinfo {volume} {718}},\
  \bibinfo {pages} {545} (\bibinfo {year} {2012})}\BibitemShut {NoStop}%
\bibitem [{\citenamefont {Eichmann}\ \emph
  {et~al.}(2016{\natexlab{b}})\citenamefont {Eichmann}, \citenamefont
  {Fischer},\ and\ \citenamefont {Heupel}}]{Eichmann:2015cra}%
  \BibitemOpen
  \bibfield  {author} {\bibinfo {author} {\bibfnamefont {G.}~\bibnamefont
  {Eichmann}}, \bibinfo {author} {\bibfnamefont {C.~S.}\ \bibnamefont
  {Fischer}},\ and\ \bibinfo {author} {\bibfnamefont {W.}~\bibnamefont
  {Heupel}},\ }\href {https://doi.org/10.1016/j.physletb.2015.12.036}
  {\bibfield  {journal} {\bibinfo  {journal} {Phys. Lett. B}\ }\textbf
  {\bibinfo {volume} {753}},\ \bibinfo {pages} {282} (\bibinfo {year}
  {2016}{\natexlab{b}})}\BibitemShut {NoStop}%
\bibitem [{\citenamefont {Hilger}\ \emph
  {et~al.}(2015{\natexlab{b}})\citenamefont {Hilger}, \citenamefont
  {Gomez-Rocha},\ and\ \citenamefont {Krassnigg}}]{Hilger:2015hka}%
  \BibitemOpen
  \bibfield  {author} {\bibinfo {author} {\bibfnamefont {T.}~\bibnamefont
  {Hilger}}, \bibinfo {author} {\bibfnamefont {M.}~\bibnamefont
  {Gomez-Rocha}},\ and\ \bibinfo {author} {\bibfnamefont {A.}~\bibnamefont
  {Krassnigg}},\ }\href {https://doi.org/10.1103/PhysRevD.91.114004} {\bibfield
   {journal} {\bibinfo  {journal} {Phys. Rev. D}\ }\textbf {\bibinfo {volume}
  {91}},\ \bibinfo {pages} {114004} (\bibinfo {year}
  {2015}{\natexlab{b}})}\BibitemShut {NoStop}%
\bibitem [{\citenamefont {El-Bennich}\ \emph {et~al.}(2016)\citenamefont
  {El-Bennich}, \citenamefont {Krein}, \citenamefont {Rojas},\ and\
  \citenamefont {Serna}}]{El-Bennich:2016qmb}%
  \BibitemOpen
  \bibfield  {author} {\bibinfo {author} {\bibfnamefont {B.}~\bibnamefont
  {El-Bennich}}, \bibinfo {author} {\bibfnamefont {G.}~\bibnamefont {Krein}},
  \bibinfo {author} {\bibfnamefont {E.}~\bibnamefont {Rojas}},\ and\ \bibinfo
  {author} {\bibfnamefont {F.~E.}\ \bibnamefont {Serna}},\ }\href
  {https://doi.org/10.1007/s00601-016-1133-x} {\bibfield  {journal} {\bibinfo
  {journal} {Few Body Syst.}\ }\textbf {\bibinfo {volume} {57}},\ \bibinfo
  {pages} {955} (\bibinfo {year} {2016})}\BibitemShut {NoStop}%
\bibitem [{\citenamefont {Mojica}\ \emph {et~al.}(2017)\citenamefont {Mojica},
  \citenamefont {Vera}, \citenamefont {Rojas},\ and\ \citenamefont
  {El-Bennich}}]{Mojica:2017tvh}%
  \BibitemOpen
  \bibfield  {author} {\bibinfo {author} {\bibfnamefont {F.~F.}\ \bibnamefont
  {Mojica}}, \bibinfo {author} {\bibfnamefont {C.~E.}\ \bibnamefont {Vera}},
  \bibinfo {author} {\bibfnamefont {E.}~\bibnamefont {Rojas}},\ and\ \bibinfo
  {author} {\bibfnamefont {B.}~\bibnamefont {El-Bennich}},\ }\href
  {https://doi.org/10.1103/PhysRevD.96.014012} {\bibfield  {journal} {\bibinfo
  {journal} {Phys. Rev. D}\ }\textbf {\bibinfo {volume} {96}},\ \bibinfo
  {pages} {014012} (\bibinfo {year} {2017})}\BibitemShut {NoStop}%
\bibitem [{\citenamefont {Raya}\ \emph {et~al.}(2018)\citenamefont {Raya},
  \citenamefont {Bedolla}, \citenamefont {Cobos-Mart{\'\i}nez},\ and\
  \citenamefont {Bashir}}]{Raya:2017ggu}%
  \BibitemOpen
  \bibfield  {author} {\bibinfo {author} {\bibfnamefont {K.}~\bibnamefont
  {Raya}}, \bibinfo {author} {\bibfnamefont {M.~A.}\ \bibnamefont {Bedolla}},
  \bibinfo {author} {\bibfnamefont {J.~J.}\ \bibnamefont
  {Cobos-Mart{\'\i}nez}},\ and\ \bibinfo {author} {\bibfnamefont
  {A.}~\bibnamefont {Bashir}},\ }\href
  {https://doi.org/10.1007/s00601-018-1455-y} {\bibfield  {journal} {\bibinfo
  {journal} {Few Body Syst.}\ }\textbf {\bibinfo {volume} {59}},\ \bibinfo
  {pages} {133} (\bibinfo {year} {2018})}\BibitemShut {NoStop}%
\bibitem [{\citenamefont {Weil}\ \emph {et~al.}(2017)\citenamefont {Weil},
  \citenamefont {Eichmann}, \citenamefont {Fischer},\ and\ \citenamefont
  {Williams}}]{Weil:2017knt}%
  \BibitemOpen
  \bibfield  {author} {\bibinfo {author} {\bibfnamefont {E.}~\bibnamefont
  {Weil}}, \bibinfo {author} {\bibfnamefont {G.}~\bibnamefont {Eichmann}},
  \bibinfo {author} {\bibfnamefont {C.~S.}\ \bibnamefont {Fischer}},\ and\
  \bibinfo {author} {\bibfnamefont {R.}~\bibnamefont {Williams}},\ }\href
  {https://doi.org/10.1103/PhysRevD.96.014021} {\bibfield  {journal} {\bibinfo
  {journal} {Phys. Rev. D}\ }\textbf {\bibinfo {volume} {96}},\ \bibinfo
  {pages} {014021} (\bibinfo {year} {2017})}\BibitemShut {NoStop}%
\bibitem [{\citenamefont {Serna}\ \emph {et~al.}(2017)\citenamefont {Serna},
  \citenamefont {El-Bennich},\ and\ \citenamefont {Krein}}]{Serna:2017nlr}%
  \BibitemOpen
  \bibfield  {author} {\bibinfo {author} {\bibfnamefont {F.~E.}\ \bibnamefont
  {Serna}}, \bibinfo {author} {\bibfnamefont {B.}~\bibnamefont {El-Bennich}},\
  and\ \bibinfo {author} {\bibfnamefont {G.}~\bibnamefont {Krein}},\ }\href
  {https://doi.org/10.1103/PhysRevD.96.014013} {\bibfield  {journal} {\bibinfo
  {journal} {Phys. Rev. D}\ }\textbf {\bibinfo {volume} {96}},\ \bibinfo
  {pages} {014013} (\bibinfo {year} {2017})}\BibitemShut {NoStop}%
\bibitem [{\citenamefont {Guti{\'e}rrez-Guerrero}\ \emph
  {et~al.}(2021)\citenamefont {Guti{\'e}rrez-Guerrero}, \citenamefont
  {Paredes-Torres},\ and\ \citenamefont {Bashir}}]{Gutierrez-Guerrero:2021rsx}%
  \BibitemOpen
  \bibfield  {author} {\bibinfo {author} {\bibfnamefont {L.~X.}\ \bibnamefont
  {Guti{\'e}rrez-Guerrero}}, \bibinfo {author} {\bibfnamefont {G.}~\bibnamefont
  {Paredes-Torres}},\ and\ \bibinfo {author} {\bibfnamefont {A.}~\bibnamefont
  {Bashir}},\ }\href {https://doi.org/10.1103/PhysRevD.104.094013} {\bibfield
  {journal} {\bibinfo  {journal} {Phys. Rev. D}\ }\textbf {\bibinfo {volume}
  {104}},\ \bibinfo {pages} {094013} (\bibinfo {year} {2021})}\BibitemShut
  {NoStop}%
\bibitem [{\citenamefont {Hern{\'a}ndez-Pinto}\ \emph
  {et~al.}(2023)\citenamefont {Hern{\'a}ndez-Pinto}, \citenamefont
  {Guti{\'e}rrez-Guerrero}, \citenamefont {Bashir}, \citenamefont {Bedolla},\
  and\ \citenamefont {Higuera-Angulo}}]{Hernandez-Pinto:2023yin}%
  \BibitemOpen
  \bibfield  {author} {\bibinfo {author} {\bibfnamefont {R.~J.}\ \bibnamefont
  {Hern{\'a}ndez-Pinto}}, \bibinfo {author} {\bibfnamefont {L.~X.}\
  \bibnamefont {Guti{\'e}rrez-Guerrero}}, \bibinfo {author} {\bibfnamefont
  {A.}~\bibnamefont {Bashir}}, \bibinfo {author} {\bibfnamefont {M.~A.}\
  \bibnamefont {Bedolla}},\ and\ \bibinfo {author} {\bibfnamefont {I.~M.}\
  \bibnamefont {Higuera-Angulo}},\ }\href
  {https://doi.org/10.1103/PhysRevD.107.054002} {\bibfield  {journal} {\bibinfo
   {journal} {Phys. Rev. D}\ }\textbf {\bibinfo {volume} {107}},\ \bibinfo
  {pages} {054002} (\bibinfo {year} {2023})}\BibitemShut {NoStop}%
\bibitem [{\citenamefont {Hern{\'a}ndez-Pinto}\ \emph
  {et~al.}(2024)\citenamefont {Hern{\'a}ndez-Pinto}, \citenamefont
  {Guti{\'e}rrez-Guerrero}, \citenamefont {Bedolla},\ and\ \citenamefont
  {Bashir}}]{Hernandez-Pinto:2024kwg}%
  \BibitemOpen
  \bibfield  {author} {\bibinfo {author} {\bibfnamefont {R.~J.}\ \bibnamefont
  {Hern{\'a}ndez-Pinto}}, \bibinfo {author} {\bibfnamefont {L.~X.}\
  \bibnamefont {Guti{\'e}rrez-Guerrero}}, \bibinfo {author} {\bibfnamefont
  {M.~A.}\ \bibnamefont {Bedolla}},\ and\ \bibinfo {author} {\bibfnamefont
  {A.}~\bibnamefont {Bashir}},\ }\href
  {https://doi.org/10.1103/PhysRevD.110.114015} {\bibfield  {journal} {\bibinfo
   {journal} {Phys. Rev. D}\ }\textbf {\bibinfo {volume} {110}},\ \bibinfo
  {pages} {114015} (\bibinfo {year} {2024})}\BibitemShut {NoStop}%
\bibitem [{\citenamefont {Chen}\ and\ \citenamefont
  {Chang}(2019)}]{Chen:2019otg}%
  \BibitemOpen
  \bibfield  {author} {\bibinfo {author} {\bibfnamefont {M.}~\bibnamefont
  {Chen}}\ and\ \bibinfo {author} {\bibfnamefont {L.}~\bibnamefont {Chang}},\
  }\href {https://doi.org/10.1088/1674-1137/43/11/114103} {\bibfield  {journal}
  {\bibinfo  {journal} {Chin. Phys. C}\ }\textbf {\bibinfo {volume} {43}},\
  \bibinfo {pages} {114103} (\bibinfo {year} {2019})}\BibitemShut {NoStop}%
\bibitem [{\citenamefont {Chang}\ and\ \citenamefont
  {Ding}(2021)}]{Chang:2020iut}%
  \BibitemOpen
  \bibfield  {author} {\bibinfo {author} {\bibfnamefont {L.}~\bibnamefont
  {Chang}}\ and\ \bibinfo {author} {\bibfnamefont {M.}~\bibnamefont {Ding}},\
  }\href {https://doi.org/10.1103/PhysRevD.103.074001} {\bibfield  {journal}
  {\bibinfo  {journal} {Phys. Rev. D}\ }\textbf {\bibinfo {volume} {103}},\
  \bibinfo {pages} {074001} (\bibinfo {year} {2021})}\BibitemShut {NoStop}%
\bibitem [{\citenamefont {Xu}(2024)}]{Xu:2024fun}%
  \BibitemOpen
  \bibfield  {author} {\bibinfo {author} {\bibfnamefont {Y.-Z.}\ \bibnamefont
  {Xu}},\ }\href {https://doi.org/10.1007/JHEP07(2024)118} {\bibfield
  {journal} {\bibinfo  {journal} {JHEP}\ }\textbf {\bibinfo {volume}
  {2024}}\bibinfo  {number} { (7)},\ \bibinfo {pages} {118}}\BibitemShut
  {NoStop}%
\bibitem [{\citenamefont {Xu}(2025)}]{Xu:2025hjf}%
  \BibitemOpen
\bibfield  {number} {  }\bibfield  {author} {\bibinfo {author} {\bibfnamefont
  {Y.-Z.}\ \bibnamefont {Xu}},\ }\href {https://doi.org/10.1103/399s-4jrq}
  {\bibfield  {journal} {\bibinfo  {journal} {Phys. Rev. D}\ }\textbf {\bibinfo
  {volume} {111}},\ \bibinfo {pages} {114012} (\bibinfo {year}
  {2025})}\BibitemShut {NoStop}%
\bibitem [{\citenamefont {Xu}\ \emph {et~al.}(2024)\citenamefont {Xu},
  \citenamefont {Raya}, \citenamefont {Rodr\'\i{}guez-Quintero},\ and\
  \citenamefont {Segovia}}]{Xu:2024vkn}%
  \BibitemOpen
  \bibfield  {author} {\bibinfo {author} {\bibfnamefont {Y.~Z.}\ \bibnamefont
  {Xu}}, \bibinfo {author} {\bibfnamefont {K.}~\bibnamefont {Raya}}, \bibinfo
  {author} {\bibfnamefont {J.}~\bibnamefont {Rodr\'\i{}guez-Quintero}},\ and\
  \bibinfo {author} {\bibfnamefont {J.}~\bibnamefont {Segovia}},\ }\href
  {https://doi.org/10.1103/PhysRevD.110.054031} {\bibfield  {journal} {\bibinfo
   {journal} {Phys. Rev. D}\ }\textbf {\bibinfo {volume} {110}},\ \bibinfo
  {pages} {054031} (\bibinfo {year} {2024})}\BibitemShut {NoStop}%
\bibitem [{\citenamefont {Albino}\ \emph {et~al.}(2025)\citenamefont {Albino},
  \citenamefont {Paredes-Torres}, \citenamefont {Raya}, \citenamefont
  {Bashir},\ and\ \citenamefont {Segovia}}]{Albino:2025bnr}%
  \BibitemOpen
  \bibfield  {author} {\bibinfo {author} {\bibfnamefont {L.}~\bibnamefont
  {Albino}}, \bibinfo {author} {\bibfnamefont {G.}~\bibnamefont
  {Paredes-Torres}}, \bibinfo {author} {\bibfnamefont {K.}~\bibnamefont
  {Raya}}, \bibinfo {author} {\bibfnamefont {A.}~\bibnamefont {Bashir}},\ and\
  \bibinfo {author} {\bibfnamefont {J.}~\bibnamefont {Segovia}},\ }\href
  {https://doi.org/10.1103/qpr2-nwtw} {\bibfield  {journal} {\bibinfo
  {journal} {Phys. Rev. D}\ }\textbf {\bibinfo {volume} {112}},\ \bibinfo
  {pages} {074015} (\bibinfo {year} {2025})}\BibitemShut {NoStop}%
\bibitem [{\citenamefont {Alkofer}\ \emph {et~al.}(2009)\citenamefont
  {Alkofer}, \citenamefont {Fischer}, \citenamefont {Llanes-Estrada},\ and\
  \citenamefont {Schwenzer}}]{Alkofer:2008tt}%
  \BibitemOpen
  \bibfield  {author} {\bibinfo {author} {\bibfnamefont {R.}~\bibnamefont
  {Alkofer}}, \bibinfo {author} {\bibfnamefont {C.~S.}\ \bibnamefont
  {Fischer}}, \bibinfo {author} {\bibfnamefont {F.~J.}\ \bibnamefont
  {Llanes-Estrada}},\ and\ \bibinfo {author} {\bibfnamefont {K.}~\bibnamefont
  {Schwenzer}},\ }\href {https://doi.org/10.1016/j.aop.2008.07.001} {\bibfield
  {journal} {\bibinfo  {journal} {Annals Phys.}\ }\textbf {\bibinfo {volume}
  {324}},\ \bibinfo {pages} {106} (\bibinfo {year} {2009})}\BibitemShut
  {NoStop}%
\bibitem [{\citenamefont {Aguilar}\ \emph {et~al.}(2024)\citenamefont
  {Aguilar}, \citenamefont {Ferreira}, \citenamefont {Oliveira}, \citenamefont
  {Papavassiliou},\ and\ \citenamefont {Linhares}}]{Aguilar:2024ciu}%
  \BibitemOpen
  \bibfield  {author} {\bibinfo {author} {\bibfnamefont {A.~C.}\ \bibnamefont
  {Aguilar}}, \bibinfo {author} {\bibfnamefont {M.~N.}\ \bibnamefont
  {Ferreira}}, \bibinfo {author} {\bibfnamefont {B.~M.}\ \bibnamefont
  {Oliveira}}, \bibinfo {author} {\bibfnamefont {J.}~\bibnamefont
  {Papavassiliou}},\ and\ \bibinfo {author} {\bibfnamefont {G.~T.}\
  \bibnamefont {Linhares}},\ }\href
  {https://doi.org/10.1140/epjc/s10052-024-13605-9} {\bibfield  {journal}
  {\bibinfo  {journal} {Eur. Phys. J. C}\ }\textbf {\bibinfo {volume} {84}},\
  \bibinfo {pages} {1231} (\bibinfo {year} {2024})}\BibitemShut {NoStop}%
\bibitem [{\citenamefont {Gao}\ \emph {et~al.}(2021)\citenamefont {Gao},
  \citenamefont {Papavassiliou},\ and\ \citenamefont
  {Pawlowski}}]{Gao:2021wun}%
  \BibitemOpen
  \bibfield  {author} {\bibinfo {author} {\bibfnamefont {F.}~\bibnamefont
  {Gao}}, \bibinfo {author} {\bibfnamefont {J.}~\bibnamefont {Papavassiliou}},\
  and\ \bibinfo {author} {\bibfnamefont {J.~M.}\ \bibnamefont {Pawlowski}},\
  }\href {https://doi.org/10.1103/PhysRevD.103.094013} {\bibfield  {journal}
  {\bibinfo  {journal} {Phys. Rev. D}\ }\textbf {\bibinfo {volume} {103}},\
  \bibinfo {pages} {094013} (\bibinfo {year} {2021})}\BibitemShut {NoStop}%
\bibitem [{\citenamefont {Salpeter}\ and\ \citenamefont
  {Bethe}(1951)}]{Salpeter:1951sz}%
  \BibitemOpen
  \bibfield  {author} {\bibinfo {author} {\bibfnamefont {E.~E.}\ \bibnamefont
  {Salpeter}}\ and\ \bibinfo {author} {\bibfnamefont {H.~A.}\ \bibnamefont
  {Bethe}},\ }\href {https://doi.org/10.1103/PhysRev.84.1232} {\bibfield
  {journal} {\bibinfo  {journal} {Phys. Rev.}\ }\textbf {\bibinfo {volume}
  {84}},\ \bibinfo {pages} {1232} (\bibinfo {year} {1951})}\BibitemShut
  {NoStop}%
\bibitem [{\citenamefont {Gell-Mann}\ and\ \citenamefont
  {Low}(1951)}]{PhysRev.84.350}%
  \BibitemOpen
  \bibfield  {author} {\bibinfo {author} {\bibfnamefont {M.}~\bibnamefont
  {Gell-Mann}}\ and\ \bibinfo {author} {\bibfnamefont {F.}~\bibnamefont
  {Low}},\ }\href {https://doi.org/10.1103/PhysRev.84.350} {\bibfield
  {journal} {\bibinfo  {journal} {Phys. Rev.}\ }\textbf {\bibinfo {volume}
  {84}},\ \bibinfo {pages} {350} (\bibinfo {year} {1951})}\BibitemShut
  {NoStop}%
\bibitem [{\citenamefont {Bethe}\ and\ \citenamefont
  {Salpeter}(1957)}]{Bethe1957}%
  \BibitemOpen
  \bibfield  {author} {\bibinfo {author} {\bibfnamefont {H.~A.}\ \bibnamefont
  {Bethe}}\ and\ \bibinfo {author} {\bibfnamefont {E.~E.}\ \bibnamefont
  {Salpeter}},\ }\bibinfo {title} {Quantum mechanics of one- and two-electron
  systems},\ in\ \href {https://doi.org/10.1007/978-3-642-45869-9_2} {\emph
  {\bibinfo {booktitle} {Atoms I / Atome I}}}\ (\bibinfo  {publisher} {Springer
  Berlin Heidelberg},\ \bibinfo {address} {Berlin, Heidelberg},\ \bibinfo
  {year} {1957})\ pp.\ \bibinfo {pages} {88--436}\BibitemShut {NoStop}%
\bibitem [{\citenamefont {Nakanishi}(1969)}]{Nakanishi:1969ph}%
  \BibitemOpen
  \bibfield  {author} {\bibinfo {author} {\bibfnamefont {N.}~\bibnamefont
  {Nakanishi}},\ }\href {https://doi.org/10.1143/PTPS.43.1} {\bibfield
  {journal} {\bibinfo  {journal} {Prog. Theor. Phys. Suppl.}\ }\textbf
  {\bibinfo {volume} {43}},\ \bibinfo {pages} {1} (\bibinfo {year}
  {1969})}\BibitemShut {NoStop}%
\bibitem [{\citenamefont {Jain}\ and\ \citenamefont
  {Munczek}(1993)}]{Jain:1993qh}%
  \BibitemOpen
  \bibfield  {author} {\bibinfo {author} {\bibfnamefont {P.}~\bibnamefont
  {Jain}}\ and\ \bibinfo {author} {\bibfnamefont {H.~J.}\ \bibnamefont
  {Munczek}},\ }\href {https://doi.org/10.1103/PhysRevD.48.5403} {\bibfield
  {journal} {\bibinfo  {journal} {Phys. Rev. D}\ }\textbf {\bibinfo {volume}
  {48}},\ \bibinfo {pages} {5403} (\bibinfo {year} {1993})}\BibitemShut
  {NoStop}%
\bibitem [{\citenamefont {Fischer}\ and\ \citenamefont
  {Alkofer}(2003)}]{Fischer:2003rp}%
  \BibitemOpen
  \bibfield  {author} {\bibinfo {author} {\bibfnamefont {C.~S.}\ \bibnamefont
  {Fischer}}\ and\ \bibinfo {author} {\bibfnamefont {R.}~\bibnamefont
  {Alkofer}},\ }\href {https://doi.org/10.1103/PhysRevD.67.094020} {\bibfield
  {journal} {\bibinfo  {journal} {Phys. Rev.}\ }\textbf {\bibinfo {volume}
  {D67}},\ \bibinfo {pages} {094020} (\bibinfo {year} {2003})}\BibitemShut
  {NoStop}%
\bibitem [{\citenamefont {Aguilar}\ and\ \citenamefont
  {Papavassiliou}(2011)}]{Aguilar:2010cn}%
  \BibitemOpen
  \bibfield  {author} {\bibinfo {author} {\bibfnamefont {A.~C.}\ \bibnamefont
  {Aguilar}}\ and\ \bibinfo {author} {\bibfnamefont {J.}~\bibnamefont
  {Papavassiliou}},\ }\href {https://doi.org/10.1103/PhysRevD.83.014013}
  {\bibfield  {journal} {\bibinfo  {journal} {Phys. Rev.}\ }\textbf {\bibinfo
  {volume} {D83}},\ \bibinfo {pages} {014013} (\bibinfo {year}
  {2011})}\BibitemShut {NoStop}%
\bibitem [{\citenamefont {Aguilar}\ \emph {et~al.}(2018)\citenamefont
  {Aguilar}, \citenamefont {Cardona}, \citenamefont {Ferreira},\ and\
  \citenamefont {Papavassiliou}}]{Aguilar:2018epe}%
  \BibitemOpen
  \bibfield  {author} {\bibinfo {author} {\bibfnamefont {A.~C.}\ \bibnamefont
  {Aguilar}}, \bibinfo {author} {\bibfnamefont {J.~C.}\ \bibnamefont
  {Cardona}}, \bibinfo {author} {\bibfnamefont {M.~N.}\ \bibnamefont
  {Ferreira}},\ and\ \bibinfo {author} {\bibfnamefont {J.}~\bibnamefont
  {Papavassiliou}},\ }\href {https://doi.org/10.1103/PhysRevD.98.014002}
  {\bibfield  {journal} {\bibinfo  {journal} {Phys. Rev.}\ }\textbf {\bibinfo
  {volume} {D98}},\ \bibinfo {pages} {014002} (\bibinfo {year}
  {2018})}\BibitemShut {NoStop}%
\bibitem [{\citenamefont {Miransky}(1994)}]{Miransky:1994vk}%
  \BibitemOpen
  \bibfield  {author} {\bibinfo {author} {\bibfnamefont {V.~A.}\ \bibnamefont
  {Miransky}},\ }\href {https://doi.org/10.1142/2170} {\emph {\bibinfo {title}
  {{Dynamical symmetry breaking in quantum field theories}}}}\ (\bibinfo
  {publisher} {World Scientific},\ \bibinfo {year} {1994})\BibitemShut
  {NoStop}%
\bibitem [{\citenamefont {Maris}\ \emph {et~al.}(1998)\citenamefont {Maris},
  \citenamefont {Roberts},\ and\ \citenamefont {Tandy}}]{Maris:1997hd}%
  \BibitemOpen
  \bibfield  {author} {\bibinfo {author} {\bibfnamefont {P.}~\bibnamefont
  {Maris}}, \bibinfo {author} {\bibfnamefont {C.~D.}\ \bibnamefont {Roberts}},\
  and\ \bibinfo {author} {\bibfnamefont {P.~C.}\ \bibnamefont {Tandy}},\ }\href
  {https://doi.org/10.1016/S0370-2693(97)01535-9} {\bibfield  {journal}
  {\bibinfo  {journal} {Phys. Lett.}\ }\textbf {\bibinfo {volume} {B420}},\
  \bibinfo {pages} {267} (\bibinfo {year} {1998})}\BibitemShut {NoStop}%
\bibitem [{\citenamefont {Itzykson}\ and\ \citenamefont
  {Zuber}(1980)}]{Itzykson:1980rh}%
  \BibitemOpen
  \bibfield  {author} {\bibinfo {author} {\bibfnamefont {C.}~\bibnamefont
  {Itzykson}}\ and\ \bibinfo {author} {\bibfnamefont {J.~B.}\ \bibnamefont
  {Zuber}},\ }\href@noop {} {\emph {\bibinfo {title} {Quantum Field Theory}}},\
  International Series in Pure and Applied Physics\ (\bibinfo  {publisher} {New
  York, USA: Mcgraw-Hill (1980) 705 p.},\ \bibinfo {year} {1980})\BibitemShut
  {NoStop}%
\bibitem [{\citenamefont {Mitter}\ \emph {et~al.}(2015)\citenamefont {Mitter},
  \citenamefont {Pawlowski},\ and\ \citenamefont
  {Strodthoff}}]{Mitter:2014wpa}%
  \BibitemOpen
  \bibfield  {author} {\bibinfo {author} {\bibfnamefont {M.}~\bibnamefont
  {Mitter}}, \bibinfo {author} {\bibfnamefont {J.~M.}\ \bibnamefont
  {Pawlowski}},\ and\ \bibinfo {author} {\bibfnamefont {N.}~\bibnamefont
  {Strodthoff}},\ }\href {https://doi.org/10.1103/PhysRevD.91.054035}
  {\bibfield  {journal} {\bibinfo  {journal} {Phys. Rev.}\ }\textbf {\bibinfo
  {volume} {D91}},\ \bibinfo {pages} {054035} (\bibinfo {year}
  {2015})}\BibitemShut {NoStop}%
\bibitem [{\citenamefont {Cyrol}\ \emph {et~al.}(2018)\citenamefont {Cyrol},
  \citenamefont {Mitter}, \citenamefont {Pawlowski},\ and\ \citenamefont
  {Strodthoff}}]{Cyrol:2017ewj}%
  \BibitemOpen
  \bibfield  {author} {\bibinfo {author} {\bibfnamefont {A.~K.}\ \bibnamefont
  {Cyrol}}, \bibinfo {author} {\bibfnamefont {M.}~\bibnamefont {Mitter}},
  \bibinfo {author} {\bibfnamefont {J.~M.}\ \bibnamefont {Pawlowski}},\ and\
  \bibinfo {author} {\bibfnamefont {N.}~\bibnamefont {Strodthoff}},\ }\href
  {https://doi.org/10.1103/PhysRevD.97.054006} {\bibfield  {journal} {\bibinfo
  {journal} {Phys. Rev.}\ }\textbf {\bibinfo {volume} {D97}},\ \bibinfo {pages}
  {054006} (\bibinfo {year} {2018})}\BibitemShut {NoStop}%
\bibitem [{\citenamefont {Ihssen}\ \emph {et~al.}(2024)\citenamefont {Ihssen},
  \citenamefont {Pawlowski}, \citenamefont {Sattler},\ and\ \citenamefont
  {Wink}}]{Ihssen:2024miv}%
  \BibitemOpen
  \bibfield  {author} {\bibinfo {author} {\bibfnamefont {F.}~\bibnamefont
  {Ihssen}}, \bibinfo {author} {\bibfnamefont {J.~M.}\ \bibnamefont
  {Pawlowski}}, \bibinfo {author} {\bibfnamefont {F.~R.}\ \bibnamefont
  {Sattler}},\ and\ \bibinfo {author} {\bibfnamefont {N.}~\bibnamefont
  {Wink}},\ }\Eprint {https://arxiv.org/abs/2408.08413} {arXiv:2408.08413
  [hep-ph]}  (\bibinfo {year} {2024})\BibitemShut {NoStop}%
\bibitem [{\citenamefont {Celmaster}\ and\ \citenamefont
  {Gonsalves}(1979)}]{Celmaster:1979km}%
  \BibitemOpen
  \bibfield  {author} {\bibinfo {author} {\bibfnamefont {W.}~\bibnamefont
  {Celmaster}}\ and\ \bibinfo {author} {\bibfnamefont {R.~J.}\ \bibnamefont
  {Gonsalves}},\ }\href {https://doi.org/10.1103/PhysRevD.20.1420} {\bibfield
  {journal} {\bibinfo  {journal} {Phys. Rev. D}\ }\textbf {\bibinfo {volume}
  {20}},\ \bibinfo {pages} {1420} (\bibinfo {year} {1979})}\BibitemShut
  {NoStop}%
\bibitem [{\citenamefont {Skullerud}\ and\ \citenamefont
  {Kizilersu}(2002)}]{Skullerud:2002ge}%
  \BibitemOpen
  \bibfield  {author} {\bibinfo {author} {\bibfnamefont {J.}~\bibnamefont
  {Skullerud}}\ and\ \bibinfo {author} {\bibfnamefont {A.}~\bibnamefont
  {Kizilersu}},\ }\href {https://doi.org/10.1088/1126-6708/2002/09/013}
  {\bibfield  {journal} {\bibinfo  {journal} {J. High Energy Phys.}\ }\textbf
  {\bibinfo {volume} {2002}}\bibinfo  {number} { (09)},\ \bibinfo {pages}
  {013}}\BibitemShut {NoStop}%
\bibitem [{\citenamefont {K\i{}z\i{}lers\"u}\ \emph {et~al.}(2021)\citenamefont
  {K\i{}z\i{}lers\"u}, \citenamefont {Oliveira}, \citenamefont {Silva},
  \citenamefont {Skullerud},\ and\ \citenamefont
  {Sternbeck}}]{Kizilersu:2021jen}%
  \BibitemOpen
\bibfield  {number} {  }\bibfield  {author} {\bibinfo {author} {\bibfnamefont
  {A.}~\bibnamefont {K\i{}z\i{}lers\"u}}, \bibinfo {author} {\bibfnamefont
  {O.}~\bibnamefont {Oliveira}}, \bibinfo {author} {\bibfnamefont {P.~J.}\
  \bibnamefont {Silva}}, \bibinfo {author} {\bibfnamefont {J.-I.}\ \bibnamefont
  {Skullerud}},\ and\ \bibinfo {author} {\bibfnamefont {A.}~\bibnamefont
  {Sternbeck}},\ }\href {https://doi.org/10.1103/PhysRevD.103.114515}
  {\bibfield  {journal} {\bibinfo  {journal} {Phys. Rev. D}\ }\textbf {\bibinfo
  {volume} {103}},\ \bibinfo {pages} {114515} (\bibinfo {year}
  {2021})}\BibitemShut {NoStop}%
\bibitem [{\citenamefont {Aguilar}\ \emph
  {et~al.}(2023{\natexlab{a}})\citenamefont {Aguilar}, \citenamefont
  {Ferreira}, \citenamefont {Iba\~nez},\ and\ \citenamefont
  {Papavassiliou}}]{Aguilar:2023mam}%
  \BibitemOpen
  \bibfield  {author} {\bibinfo {author} {\bibfnamefont {A.~C.}\ \bibnamefont
  {Aguilar}}, \bibinfo {author} {\bibfnamefont {M.~N.}\ \bibnamefont
  {Ferreira}}, \bibinfo {author} {\bibfnamefont {D.}~\bibnamefont {Iba\~nez}},\
  and\ \bibinfo {author} {\bibfnamefont {J.}~\bibnamefont {Papavassiliou}},\
  }\href {https://doi.org/10.1140/epjc/s10052-023-12103-8} {\bibfield
  {journal} {\bibinfo  {journal} {Eur. Phys. J. C}\ }\textbf {\bibinfo {volume}
  {83}},\ \bibinfo {pages} {967} (\bibinfo {year}
  {2023}{\natexlab{a}})}\BibitemShut {NoStop}%
\bibitem [{\citenamefont {Davydychev}\ \emph {et~al.}(2001)\citenamefont
  {Davydychev}, \citenamefont {Osland},\ and\ \citenamefont
  {Saks}}]{Davydychev:2000rt}%
  \BibitemOpen
  \bibfield  {author} {\bibinfo {author} {\bibfnamefont {A.~I.}\ \bibnamefont
  {Davydychev}}, \bibinfo {author} {\bibfnamefont {P.}~\bibnamefont {Osland}},\
  and\ \bibinfo {author} {\bibfnamefont {L.}~\bibnamefont {Saks}},\ }\href
  {https://doi.org/10.1103/PhysRevD.63.014022} {\bibfield  {journal} {\bibinfo
  {journal} {Phys. Rev.}\ }\textbf {\bibinfo {volume} {D63}},\ \bibinfo {pages}
  {014022} (\bibinfo {year} {2001})}\BibitemShut {NoStop}%
\bibitem [{\citenamefont {Brown}\ and\ \citenamefont
  {Dorey}(1991)}]{Brown:1989hy}%
  \BibitemOpen
  \bibfield  {author} {\bibinfo {author} {\bibfnamefont {N.}~\bibnamefont
  {Brown}}\ and\ \bibinfo {author} {\bibfnamefont {N.}~\bibnamefont {Dorey}},\
  }\href {https://doi.org/10.1142/S0217732391000294} {\bibfield  {journal}
  {\bibinfo  {journal} {Mod. Phys. Lett.}\ }\textbf {\bibinfo {volume} {A6}},\
  \bibinfo {pages} {317} (\bibinfo {year} {1991})}\BibitemShut {NoStop}%
\bibitem [{\citenamefont {Curtis}\ and\ \citenamefont
  {Pennington}(1990)}]{Curtis:1990zs}%
  \BibitemOpen
  \bibfield  {author} {\bibinfo {author} {\bibfnamefont {D.~C.}\ \bibnamefont
  {Curtis}}\ and\ \bibinfo {author} {\bibfnamefont {M.~R.}\ \bibnamefont
  {Pennington}},\ }\href {https://doi.org/10.1103/PhysRevD.42.4165} {\bibfield
  {journal} {\bibinfo  {journal} {Phys. Rev. D}\ }\textbf {\bibinfo {volume}
  {42}},\ \bibinfo {pages} {4165} (\bibinfo {year} {1990})}\BibitemShut
  {NoStop}%
\bibitem [{\citenamefont {Curtis}\ and\ \citenamefont
  {Pennington}(1993)}]{Curtis:1993py}%
  \BibitemOpen
  \bibfield  {author} {\bibinfo {author} {\bibfnamefont {D.~C.}\ \bibnamefont
  {Curtis}}\ and\ \bibinfo {author} {\bibfnamefont {M.~R.}\ \bibnamefont
  {Pennington}},\ }\href {https://doi.org/10.1103/PhysRevD.48.4933} {\bibfield
  {journal} {\bibinfo  {journal} {Phys. Rev. D}\ }\textbf {\bibinfo {volume}
  {48}},\ \bibinfo {pages} {4933} (\bibinfo {year} {1993})}\BibitemShut
  {NoStop}%
\bibitem [{\citenamefont {Bloch}(2001)}]{Bloch:2001wz}%
  \BibitemOpen
  \bibfield  {author} {\bibinfo {author} {\bibfnamefont {J.~C.~R.}\
  \bibnamefont {Bloch}},\ }\href {https://doi.org/10.1103/PhysRevD.64.116011}
  {\bibfield  {journal} {\bibinfo  {journal} {Phys. Rev. D}\ }\textbf {\bibinfo
  {volume} {64}},\ \bibinfo {pages} {116011} (\bibinfo {year}
  {2001})}\BibitemShut {NoStop}%
\bibitem [{\citenamefont {Bloch}(2002)}]{Bloch:2002eq}%
  \BibitemOpen
  \bibfield  {author} {\bibinfo {author} {\bibfnamefont {J.~C.~R.}\
  \bibnamefont {Bloch}},\ }\href {https://doi.org/10.1103/PhysRevD.66.034032}
  {\bibfield  {journal} {\bibinfo  {journal} {Phys. Rev. D}\ }\textbf {\bibinfo
  {volume} {66}},\ \bibinfo {pages} {034032} (\bibinfo {year}
  {2002})}\BibitemShut {NoStop}%
\bibitem [{\citenamefont {Kizilersu}\ and\ \citenamefont
  {Pennington}(2009)}]{Kizilersu:2009kg}%
  \BibitemOpen
  \bibfield  {author} {\bibinfo {author} {\bibfnamefont {A.}~\bibnamefont
  {Kizilersu}}\ and\ \bibinfo {author} {\bibfnamefont {M.}~\bibnamefont
  {Pennington}},\ }\href {https://doi.org/10.1103/PhysRevD.79.125020}
  {\bibfield  {journal} {\bibinfo  {journal} {Phys. Rev.}\ }\textbf {\bibinfo
  {volume} {D79}},\ \bibinfo {pages} {125020} (\bibinfo {year}
  {2009})}\BibitemShut {NoStop}%
\bibitem [{\citenamefont {Ayala}\ \emph {et~al.}(2012)\citenamefont {Ayala},
  \citenamefont {Bashir}, \citenamefont {Binosi}, \citenamefont
  {Cristoforetti},\ and\ \citenamefont {Rodriguez-Quintero}}]{Ayala:2012pb}%
  \BibitemOpen
  \bibfield  {author} {\bibinfo {author} {\bibfnamefont {A.}~\bibnamefont
  {Ayala}}, \bibinfo {author} {\bibfnamefont {A.}~\bibnamefont {Bashir}},
  \bibinfo {author} {\bibfnamefont {D.}~\bibnamefont {Binosi}}, \bibinfo
  {author} {\bibfnamefont {M.}~\bibnamefont {Cristoforetti}},\ and\ \bibinfo
  {author} {\bibfnamefont {J.}~\bibnamefont {Rodriguez-Quintero}},\ }\href
  {https://doi.org/10.1103/PhysRevD.86.074512} {\bibfield  {journal} {\bibinfo
  {journal} {Phys. Rev.}\ }\textbf {\bibinfo {volume} {D86}},\ \bibinfo {pages}
  {074512} (\bibinfo {year} {2012})}\BibitemShut {NoStop}%
\bibitem [{\citenamefont {Binosi}\ \emph {et~al.}(2017)\citenamefont {Binosi},
  \citenamefont {Roberts},\ and\ \citenamefont
  {Rodriguez-Quintero}}]{Binosi:2016xxu}%
  \BibitemOpen
  \bibfield  {author} {\bibinfo {author} {\bibfnamefont {D.}~\bibnamefont
  {Binosi}}, \bibinfo {author} {\bibfnamefont {C.~D.}\ \bibnamefont
  {Roberts}},\ and\ \bibinfo {author} {\bibfnamefont {J.}~\bibnamefont
  {Rodriguez-Quintero}},\ }\href {https://doi.org/10.1103/PhysRevD.95.114009}
  {\bibfield  {journal} {\bibinfo  {journal} {Phys. Rev. D}\ }\textbf {\bibinfo
  {volume} {95}},\ \bibinfo {pages} {114009} (\bibinfo {year}
  {2017})}\BibitemShut {NoStop}%
\bibitem [{\citenamefont {Eichmann}(2025)}]{Eichmann:2025wgs}%
  \BibitemOpen
  \bibfield  {author} {\bibinfo {author} {\bibfnamefont {G.}~\bibnamefont
  {Eichmann}},\ }\Eprint {https://arxiv.org/abs/2503.10397} {arXiv:2503.10397
  [hep-ph]}  (\bibinfo {year} {2025})\BibitemShut {NoStop}%
\bibitem [{\citenamefont {Blum}\ \emph {et~al.}(2014)\citenamefont {Blum},
  \citenamefont {Huber}, \citenamefont {Mitter},\ and\ \citenamefont {von
  Smekal}}]{Blum:2014gna}%
  \BibitemOpen
  \bibfield  {author} {\bibinfo {author} {\bibfnamefont {A.}~\bibnamefont
  {Blum}}, \bibinfo {author} {\bibfnamefont {M.~Q.}\ \bibnamefont {Huber}},
  \bibinfo {author} {\bibfnamefont {M.}~\bibnamefont {Mitter}},\ and\ \bibinfo
  {author} {\bibfnamefont {L.}~\bibnamefont {von Smekal}},\ }\href
  {https://doi.org/10.1103/PhysRevD.89.061703} {\bibfield  {journal} {\bibinfo
  {journal} {Phys. Rev.}\ }\textbf {\bibinfo {volume} {D89}},\ \bibinfo {pages}
  {061703} (\bibinfo {year} {2014})}\BibitemShut {NoStop}%
\bibitem [{\citenamefont {Eichmann}\ \emph {et~al.}(2014)\citenamefont
  {Eichmann}, \citenamefont {Williams}, \citenamefont {Alkofer},\ and\
  \citenamefont {Vujinovic}}]{Eichmann:2014xya}%
  \BibitemOpen
  \bibfield  {author} {\bibinfo {author} {\bibfnamefont {G.}~\bibnamefont
  {Eichmann}}, \bibinfo {author} {\bibfnamefont {R.}~\bibnamefont {Williams}},
  \bibinfo {author} {\bibfnamefont {R.}~\bibnamefont {Alkofer}},\ and\ \bibinfo
  {author} {\bibfnamefont {M.}~\bibnamefont {Vujinovic}},\ }\href
  {https://doi.org/10.1103/PhysRevD.89.105014} {\bibfield  {journal} {\bibinfo
  {journal} {Phys. Rev.}\ }\textbf {\bibinfo {volume} {D89}},\ \bibinfo {pages}
  {105014} (\bibinfo {year} {2014})}\BibitemShut {NoStop}%
\bibitem [{\citenamefont {Aguilar}\ \emph
  {et~al.}(2023{\natexlab{b}})\citenamefont {Aguilar}, \citenamefont
  {Ferreira}, \citenamefont {Papavassiliou},\ and\ \citenamefont
  {Santos}}]{Aguilar:2023qqd}%
  \BibitemOpen
  \bibfield  {author} {\bibinfo {author} {\bibfnamefont {A.~C.}\ \bibnamefont
  {Aguilar}}, \bibinfo {author} {\bibfnamefont {M.~N.}\ \bibnamefont
  {Ferreira}}, \bibinfo {author} {\bibfnamefont {J.}~\bibnamefont
  {Papavassiliou}},\ and\ \bibinfo {author} {\bibfnamefont {L.~R.}\
  \bibnamefont {Santos}},\ }\href
  {https://doi.org/10.1140/epjc/s10052-023-11732-3} {\bibfield  {journal}
  {\bibinfo  {journal} {Eur. Phys. J. C}\ }\textbf {\bibinfo {volume} {83}},\
  \bibinfo {pages} {549} (\bibinfo {year} {2023}{\natexlab{b}})}\BibitemShut
  {NoStop}%
\bibitem [{\citenamefont {Pinto-G\'omez}\ \emph {et~al.}(2023)\citenamefont
  {Pinto-G\'omez}, \citenamefont {De~Soto}, \citenamefont {Ferreira},
  \citenamefont {Papavassiliou},\ and\ \citenamefont
  {Rodr\'\i{}guez-Quintero}}]{Pinto-Gomez:2022brg}%
  \BibitemOpen
  \bibfield  {author} {\bibinfo {author} {\bibfnamefont {F.}~\bibnamefont
  {Pinto-G\'omez}}, \bibinfo {author} {\bibfnamefont {F.}~\bibnamefont
  {De~Soto}}, \bibinfo {author} {\bibfnamefont {M.~N.}\ \bibnamefont
  {Ferreira}}, \bibinfo {author} {\bibfnamefont {J.}~\bibnamefont
  {Papavassiliou}},\ and\ \bibinfo {author} {\bibfnamefont {J.}~\bibnamefont
  {Rodr\'\i{}guez-Quintero}},\ }\href
  {https://doi.org/10.1016/j.physletb.2023.137737} {\bibfield  {journal}
  {\bibinfo  {journal} {Phys. Lett. B}\ }\textbf {\bibinfo {volume} {838}},\
  \bibinfo {pages} {137737} (\bibinfo {year} {2023})}\BibitemShut {NoStop}%
\bibitem [{\citenamefont {Pinto-G\'omez}\ \emph {et~al.}(2024)\citenamefont
  {Pinto-G\'omez}, \citenamefont {De~Soto},\ and\ \citenamefont
  {Rodr\'\i{}guez-Quintero}}]{Pinto-Gomez:2024mrk}%
  \BibitemOpen
  \bibfield  {author} {\bibinfo {author} {\bibfnamefont {F.}~\bibnamefont
  {Pinto-G\'omez}}, \bibinfo {author} {\bibfnamefont {F.}~\bibnamefont
  {De~Soto}},\ and\ \bibinfo {author} {\bibfnamefont {J.}~\bibnamefont
  {Rodr\'\i{}guez-Quintero}},\ }\href
  {https://doi.org/10.1103/PhysRevD.110.014005} {\bibfield  {journal} {\bibinfo
   {journal} {Phys. Rev. D}\ }\textbf {\bibinfo {volume} {110}},\ \bibinfo
  {pages} {014005} (\bibinfo {year} {2024})}\BibitemShut {NoStop}%
\bibitem [{\citenamefont {Athenodorou}\ \emph {et~al.}(2016)\citenamefont
  {Athenodorou}, \citenamefont {Binosi}, \citenamefont {Boucaud}, \citenamefont
  {De~Soto}, \citenamefont {Papavassiliou}, \citenamefont
  {Rodriguez-Quintero},\ and\ \citenamefont
  {Zafeiropoulos}}]{Athenodorou:2016oyh}%
  \BibitemOpen
  \bibfield  {author} {\bibinfo {author} {\bibfnamefont {A.}~\bibnamefont
  {Athenodorou}}, \bibinfo {author} {\bibfnamefont {D.}~\bibnamefont {Binosi}},
  \bibinfo {author} {\bibfnamefont {P.}~\bibnamefont {Boucaud}}, \bibinfo
  {author} {\bibfnamefont {F.}~\bibnamefont {De~Soto}}, \bibinfo {author}
  {\bibfnamefont {J.}~\bibnamefont {Papavassiliou}}, \bibinfo {author}
  {\bibfnamefont {J.}~\bibnamefont {Rodriguez-Quintero}},\ and\ \bibinfo
  {author} {\bibfnamefont {S.}~\bibnamefont {Zafeiropoulos}},\ }\href
  {https://doi.org/10.1016/j.physletb.2016.08.065} {\bibfield  {journal}
  {\bibinfo  {journal} {Phys. Lett.}\ }\textbf {\bibinfo {volume} {B761}},\
  \bibinfo {pages} {444} (\bibinfo {year} {2016})}\BibitemShut {NoStop}%
\bibitem [{\citenamefont {Sanchis-Alepuz}\ and\ \citenamefont
  {Williams}(2018)}]{Sanchis-Alepuz:2017jjd}%
  \BibitemOpen
  \bibfield  {author} {\bibinfo {author} {\bibfnamefont {H.}~\bibnamefont
  {Sanchis-Alepuz}}\ and\ \bibinfo {author} {\bibfnamefont {R.}~\bibnamefont
  {Williams}},\ }\href {https://doi.org/10.1016/j.cpc.2018.05.020} {\bibfield
  {journal} {\bibinfo  {journal} {Comput. Phys. Commun.}\ }\textbf {\bibinfo
  {volume} {232}},\ \bibinfo {pages} {1} (\bibinfo {year} {2018})}\BibitemShut
  {NoStop}%
\bibitem [{\citenamefont {Huber}(2025)}]{Huber:2025cbd}%
  \BibitemOpen
  \bibfield  {author} {\bibinfo {author} {\bibfnamefont {M.~Q.}\ \bibnamefont
  {Huber}},\ }\Eprint {https://arxiv.org/abs/2510.18960} {arXiv:2510.18960
  [hep-ph]}  (\bibinfo {year} {2025})\BibitemShut {NoStop}%
\bibitem [{\citenamefont {Lehoucq}\ and\ \citenamefont
  {Sorensen}(1996)}]{doi:10.1137/S0895479895281484}%
  \BibitemOpen
  \bibfield  {author} {\bibinfo {author} {\bibfnamefont {R.~B.}\ \bibnamefont
  {Lehoucq}}\ and\ \bibinfo {author} {\bibfnamefont {D.~C.}\ \bibnamefont
  {Sorensen}},\ }\href {https://doi.org/10.1137/S0895479895281484} {\bibfield
  {journal} {\bibinfo  {journal} {SIAM Journal on Matrix Analysis and
  Applications}\ }\textbf {\bibinfo {volume} {17}},\ \bibinfo {pages} {789}
  (\bibinfo {year} {1996})},\ \Eprint
  {https://arxiv.org/abs/https://doi.org/10.1137/S0895479895281484}
  {https://doi.org/10.1137/S0895479895281484} \BibitemShut {NoStop}%
\bibitem [{\citenamefont {Lehoucq}\ \emph {et~al.}(1998)\citenamefont
  {Lehoucq}, \citenamefont {Sorensen},\ and\ \citenamefont
  {Yang}}]{doi:10.1137/1.9780898719628}%
  \BibitemOpen
  \bibfield  {author} {\bibinfo {author} {\bibfnamefont {R.~B.}\ \bibnamefont
  {Lehoucq}}, \bibinfo {author} {\bibfnamefont {D.~C.}\ \bibnamefont
  {Sorensen}},\ and\ \bibinfo {author} {\bibfnamefont {C.}~\bibnamefont
  {Yang}},\ }\href {https://doi.org/10.1137/1.9780898719628} {\emph {\bibinfo
  {title} {ARPACK Users' Guide}}}\ (\bibinfo  {publisher} {Society for
  Industrial and Applied Mathematics},\ \bibinfo {year} {1998})\ \Eprint
  {https://arxiv.org/abs/https://epubs.siam.org/doi/pdf/10.1137/1.9780898719628}
  {https://epubs.siam.org/doi/pdf/10.1137/1.9780898719628} \BibitemShut
  {NoStop}%
\bibitem [{\citenamefont {Alkofer}\ \emph {et~al.}(2004)\citenamefont
  {Alkofer}, \citenamefont {Detmold}, \citenamefont {Fischer},\ and\
  \citenamefont {Maris}}]{Alkofer:2003jj}%
  \BibitemOpen
  \bibfield  {author} {\bibinfo {author} {\bibfnamefont {R.}~\bibnamefont
  {Alkofer}}, \bibinfo {author} {\bibfnamefont {W.}~\bibnamefont {Detmold}},
  \bibinfo {author} {\bibfnamefont {C.~S.}\ \bibnamefont {Fischer}},\ and\
  \bibinfo {author} {\bibfnamefont {P.}~\bibnamefont {Maris}},\ }\href
  {https://doi.org/10.1103/PhysRevD.70.014014} {\bibfield  {journal} {\bibinfo
  {journal} {Phys. Rev. D}\ }\textbf {\bibinfo {volume} {70}},\ \bibinfo
  {pages} {014014} (\bibinfo {year} {2004})}\BibitemShut {NoStop}%
\bibitem [{\citenamefont {Eichmann}\ \emph
  {et~al.}(2008{\natexlab{b}})\citenamefont {Eichmann}, \citenamefont
  {Krassnigg}, \citenamefont {Schwinzerl},\ and\ \citenamefont
  {Alkofer}}]{Eichmann:2007nn}%
  \BibitemOpen
  \bibfield  {author} {\bibinfo {author} {\bibfnamefont {G.}~\bibnamefont
  {Eichmann}}, \bibinfo {author} {\bibfnamefont {A.}~\bibnamefont {Krassnigg}},
  \bibinfo {author} {\bibfnamefont {M.}~\bibnamefont {Schwinzerl}},\ and\
  \bibinfo {author} {\bibfnamefont {R.}~\bibnamefont {Alkofer}},\ }\href
  {https://doi.org/10.1016/j.aop.2008.02.007} {\bibfield  {journal} {\bibinfo
  {journal} {Annals Phys.}\ }\textbf {\bibinfo {volume} {323}},\ \bibinfo
  {pages} {2505} (\bibinfo {year} {2008}{\natexlab{b}})}\BibitemShut {NoStop}%
\bibitem [{\citenamefont {Windisch}\ \emph {et~al.}(2013)\citenamefont
  {Windisch}, \citenamefont {Huber},\ and\ \citenamefont
  {Alkofer}}]{Windisch:2012sz}%
  \BibitemOpen
  \bibfield  {author} {\bibinfo {author} {\bibfnamefont {A.}~\bibnamefont
  {Windisch}}, \bibinfo {author} {\bibfnamefont {M.~Q.}\ \bibnamefont
  {Huber}},\ and\ \bibinfo {author} {\bibfnamefont {R.}~\bibnamefont
  {Alkofer}},\ }\href {https://doi.org/10.1103/PhysRevD.87.065005} {\bibfield
  {journal} {\bibinfo  {journal} {Phys. Rev. D}\ }\textbf {\bibinfo {volume}
  {87}},\ \bibinfo {pages} {065005} (\bibinfo {year} {2013})}\BibitemShut
  {NoStop}%
\bibitem [{\citenamefont {Eichmann}\ \emph {et~al.}(2019)\citenamefont
  {Eichmann}, \citenamefont {Duarte}, \citenamefont {Pe{\~n}a},\ and\
  \citenamefont {Stadler}}]{Eichmann:2019dts}%
  \BibitemOpen
  \bibfield  {author} {\bibinfo {author} {\bibfnamefont {G.}~\bibnamefont
  {Eichmann}}, \bibinfo {author} {\bibfnamefont {P.}~\bibnamefont {Duarte}},
  \bibinfo {author} {\bibfnamefont {M.~T.}\ \bibnamefont {Pe{\~n}a}},\ and\
  \bibinfo {author} {\bibfnamefont {A.}~\bibnamefont {Stadler}},\ }\href
  {https://doi.org/10.1103/PhysRevD.100.094001} {\bibfield  {journal} {\bibinfo
   {journal} {Phys. Rev. D}\ }\textbf {\bibinfo {volume} {100}},\ \bibinfo
  {pages} {094001} (\bibinfo {year} {2019})}\BibitemShut {NoStop}%
\bibitem [{\citenamefont {Fischer}\ and\ \citenamefont
  {Huber}(2020)}]{Fischer:2020xnb}%
  \BibitemOpen
  \bibfield  {author} {\bibinfo {author} {\bibfnamefont {C.~S.}\ \bibnamefont
  {Fischer}}\ and\ \bibinfo {author} {\bibfnamefont {M.~Q.}\ \bibnamefont
  {Huber}},\ }\href {https://doi.org/10.1103/PhysRevD.102.094005} {\bibfield
  {journal} {\bibinfo  {journal} {Phys. Rev. D}\ }\textbf {\bibinfo {volume}
  {102}},\ \bibinfo {pages} {094005} (\bibinfo {year} {2020})}\BibitemShut
  {NoStop}%
\bibitem [{\citenamefont {Huber}\ \emph {et~al.}(2023)\citenamefont {Huber},
  \citenamefont {Kern},\ and\ \citenamefont {Alkofer}}]{Huber:2022nzs}%
  \BibitemOpen
  \bibfield  {author} {\bibinfo {author} {\bibfnamefont {M.~Q.}\ \bibnamefont
  {Huber}}, \bibinfo {author} {\bibfnamefont {W.~J.}\ \bibnamefont {Kern}},\
  and\ \bibinfo {author} {\bibfnamefont {R.}~\bibnamefont {Alkofer}},\ }\href
  {https://doi.org/10.1103/PhysRevD.107.074026} {\bibfield  {journal} {\bibinfo
   {journal} {Phys. Rev. D}\ }\textbf {\bibinfo {volume} {107}},\ \bibinfo
  {pages} {074026} (\bibinfo {year} {2023})}\BibitemShut {NoStop}%
\bibitem [{\citenamefont {Horak}\ \emph {et~al.}(2025)\citenamefont {Horak},
  \citenamefont {Pawlowski},\ and\ \citenamefont {Wink}}]{Horak:2022myj}%
  \BibitemOpen
  \bibfield  {author} {\bibinfo {author} {\bibfnamefont {J.}~\bibnamefont
  {Horak}}, \bibinfo {author} {\bibfnamefont {J.~M.}\ \bibnamefont
  {Pawlowski}},\ and\ \bibinfo {author} {\bibfnamefont {N.}~\bibnamefont
  {Wink}},\ }\href {https://doi.org/10.21468/SciPostPhysCore.8.3.048}
  {\bibfield  {journal} {\bibinfo  {journal} {SciPost Phys. Core}\ }\textbf
  {\bibinfo {volume} {8}},\ \bibinfo {pages} {048} (\bibinfo {year}
  {2025})}\BibitemShut {NoStop}%
\bibitem [{\citenamefont {Duarte}\ \emph {et~al.}(2022)\citenamefont {Duarte},
  \citenamefont {Frederico}, \citenamefont {de~Paula},\ and\ \citenamefont
  {Ydrefors}}]{Duarte:2022yur}%
  \BibitemOpen
  \bibfield  {author} {\bibinfo {author} {\bibfnamefont {D.~C.}\ \bibnamefont
  {Duarte}}, \bibinfo {author} {\bibfnamefont {T.}~\bibnamefont {Frederico}},
  \bibinfo {author} {\bibfnamefont {W.}~\bibnamefont {de~Paula}},\ and\
  \bibinfo {author} {\bibfnamefont {E.}~\bibnamefont {Ydrefors}},\ }\href
  {https://doi.org/10.1103/PhysRevD.105.114055} {\bibfield  {journal} {\bibinfo
   {journal} {Phys. Rev. D}\ }\textbf {\bibinfo {volume} {105}},\ \bibinfo
  {pages} {114055} (\bibinfo {year} {2022})}\BibitemShut {NoStop}%
\bibitem [{\citenamefont {Braun}\ \emph {et~al.}(2023)\citenamefont {Braun}
  \emph {et~al.}}]{Braun:2022mgx}%
  \BibitemOpen
  \bibfield  {author} {\bibinfo {author} {\bibfnamefont {J.}~\bibnamefont
  {Braun}} \emph {et~al.},\ }\href
  {https://doi.org/10.21468/SciPostPhysCore.6.3.061} {\bibfield  {journal}
  {\bibinfo  {journal} {SciPost Phys. Core}\ }\textbf {\bibinfo {volume} {6}},\
  \bibinfo {pages} {061} (\bibinfo {year} {2023})}\BibitemShut {NoStop}%
\bibitem [{\citenamefont {Eichmann}\ \emph {et~al.}(2024)\citenamefont
  {Eichmann}, \citenamefont {G{\'o}mez}, \citenamefont {Horak}, \citenamefont
  {Pawlowski}, \citenamefont {Wessely},\ and\ \citenamefont
  {Wink}}]{Eichmann:2023tjk}%
  \BibitemOpen
  \bibfield  {author} {\bibinfo {author} {\bibfnamefont {G.}~\bibnamefont
  {Eichmann}}, \bibinfo {author} {\bibfnamefont {A.}~\bibnamefont {G{\'o}mez}},
  \bibinfo {author} {\bibfnamefont {J.}~\bibnamefont {Horak}}, \bibinfo
  {author} {\bibfnamefont {J.~M.}\ \bibnamefont {Pawlowski}}, \bibinfo {author}
  {\bibfnamefont {J.}~\bibnamefont {Wessely}},\ and\ \bibinfo {author}
  {\bibfnamefont {N.}~\bibnamefont {Wink}},\ }\href
  {https://doi.org/10.1103/PhysRevD.109.096024} {\bibfield  {journal} {\bibinfo
   {journal} {Phys. Rev. D}\ }\textbf {\bibinfo {volume} {109}},\ \bibinfo
  {pages} {096024} (\bibinfo {year} {2024})}\BibitemShut {NoStop}%
\bibitem [{\citenamefont {Horak}\ \emph {et~al.}(2024)\citenamefont {Horak},
  \citenamefont {Ihssen}, \citenamefont {Pawlowski}, \citenamefont {Wessely},\
  and\ \citenamefont {Wink}}]{Horak:2023hkp}%
  \BibitemOpen
  \bibfield  {author} {\bibinfo {author} {\bibfnamefont {J.}~\bibnamefont
  {Horak}}, \bibinfo {author} {\bibfnamefont {F.}~\bibnamefont {Ihssen}},
  \bibinfo {author} {\bibfnamefont {J.~M.}\ \bibnamefont {Pawlowski}}, \bibinfo
  {author} {\bibfnamefont {J.}~\bibnamefont {Wessely}},\ and\ \bibinfo {author}
  {\bibfnamefont {N.}~\bibnamefont {Wink}},\ }\href
  {https://doi.org/10.1103/PhysRevD.110.056009} {\bibfield  {journal} {\bibinfo
   {journal} {Phys. Rev. D}\ }\textbf {\bibinfo {volume} {110}},\ \bibinfo
  {pages} {056009} (\bibinfo {year} {2024})}\BibitemShut {NoStop}%
\bibitem [{\citenamefont {Pawlowski}\ and\ \citenamefont
  {Wessely}(2025)}]{Pawlowski:2024kxc}%
  \BibitemOpen
  \bibfield  {author} {\bibinfo {author} {\bibfnamefont {J.~M.}\ \bibnamefont
  {Pawlowski}}\ and\ \bibinfo {author} {\bibfnamefont {J.}~\bibnamefont
  {Wessely}},\ }\href {https://doi.org/10.1140/epjc/s10052-025-14683-z}
  {\bibfield  {journal} {\bibinfo  {journal} {Eur. Phys. J. C}\ }\textbf
  {\bibinfo {volume} {85}},\ \bibinfo {pages} {970} (\bibinfo {year}
  {2025})}\BibitemShut {NoStop}%
\bibitem [{\citenamefont {Fischer}\ \emph {et~al.}(2005)\citenamefont
  {Fischer}, \citenamefont {Watson},\ and\ \citenamefont
  {Cassing}}]{Fischer:2005en}%
  \BibitemOpen
  \bibfield  {author} {\bibinfo {author} {\bibfnamefont {C.~S.}\ \bibnamefont
  {Fischer}}, \bibinfo {author} {\bibfnamefont {P.}~\bibnamefont {Watson}},\
  and\ \bibinfo {author} {\bibfnamefont {W.}~\bibnamefont {Cassing}},\ }\href
  {https://doi.org/10.1103/PhysRevD.72.094025} {\bibfield  {journal} {\bibinfo
  {journal} {Phys. Rev. D}\ }\textbf {\bibinfo {volume} {72}},\ \bibinfo
  {pages} {094025} (\bibinfo {year} {2005})}\BibitemShut {NoStop}%
\bibitem [{\citenamefont {Fischer}\ \emph {et~al.}(2009)\citenamefont
  {Fischer}, \citenamefont {Nickel},\ and\ \citenamefont
  {Williams}}]{Fischer:2008sp}%
  \BibitemOpen
  \bibfield  {author} {\bibinfo {author} {\bibfnamefont {C.~S.}\ \bibnamefont
  {Fischer}}, \bibinfo {author} {\bibfnamefont {D.}~\bibnamefont {Nickel}},\
  and\ \bibinfo {author} {\bibfnamefont {R.}~\bibnamefont {Williams}},\ }\href
  {https://doi.org/10.1140/epjc/s10052-008-0821-1} {\bibfield  {journal}
  {\bibinfo  {journal} {Eur. Phys. J. C}\ }\textbf {\bibinfo {volume} {60}},\
  \bibinfo {pages} {47} (\bibinfo {year} {2009})}\BibitemShut {NoStop}%
\bibitem [{\citenamefont {Krassnigg}(2008)}]{Krassnigg:2008bob}%
  \BibitemOpen
  \bibfield  {author} {\bibinfo {author} {\bibfnamefont {A.}~\bibnamefont
  {Krassnigg}},\ }\href {https://doi.org/10.22323/1.077.0075} {\bibfield
  {journal} {\bibinfo  {journal} {PoS}\ }\textbf {\bibinfo {volume}
  {CONFINEMENT8}},\ \bibinfo {pages} {075} (\bibinfo {year}
  {2008})}\BibitemShut {NoStop}%
\end{thebibliography}%

\end{document}